\documentclass{aa}
\usepackage{epsfig,times}
\def\mincir{\raise -2.truept\hbox{\rlap{\hbox{$\sim$}}\raise5.truept \hbox{$<$}\ }}
\def\mincireq{\hbox{\raise0.5ex\hbox{$<\lower1.06ex\hbox{$\kern-1.07em{\sim}$}$}}}
\def\magcir{\raise-2.truept\hbox{\rlap{\hbox{$\sim$}}\raise5.truept \hbox{$>$}\ }}
\def\gr{\kern 2pt\hbox{}^\circ{\kern -2pt K}} 

\def\_{\thinspace}

\begin{document}

\title{Galactic star-formation rates 
	gauged by stellar end-products}

\author{Massimo Persic\inst{1}, and 
        Yoel Rephaeli\inst{2,3}}

\offprints{M.P.; e-mail: {\tt persic@ts.astro.it}}

\institute{
INAF/Osservatorio Astronomico di Trieste, via G.B.Tiepolo 11, 34143 Trieste, Italy
        \and
School of Physics and Astronomy, Tel Aviv University, Tel Aviv 69978,
Israel
        \and
CASS, University of California, San Diego, La Jolla, CA 92093, USA
}
\date{Received ..................; accepted ...................}

\abstract{
Young galactic X-ray point sources (XPs) closely trace the ongoing star 
formation in galaxies. From measured XP number counts we extract the 
collective 2-10 keV luminosity of young XPs, $L_{x}^{\rm yXP}$, which we 
use to gauge the current star-formation rate (SFR) in galaxies. We find 
that, for a sample of local star-forming galaxies (i.e., normal spirals 
and mild starbursts), $L_{x}^{\rm yXP}$ correlates linearly with the SFR 
over three decades in luminosity. A separate, high-SFR sample of starburst 
ULIRGs can be used to check the calibration of the relation. Using their 
(presumably SF-related) total 2-10 keV luminosities we find that these 
sources satisfy the SFR--$L_{x}^{\rm yXP}$ relation, as defined by the 
weaker sample, and extend it to span $\sim$$5$ decades in luminosity. 
The SFR--$L_{x}^{\rm yXP}$ relation is likely to hold also for distant 
($z$$\sim$$1$) {\it Hubble} Deep Field North galaxies, especially so if 
these high-SFR objects are similar to the (more nearby) ULIRGs. It is 
argued that the SFR--$L_{x}^{\rm yXP}$ relation provides the most adequate 
X-ray estimator of instantaneous SFR by the phenomena characterizing 
massive stars from their birth (FIR emission from placental dust clouds) 
through their death as compact remnants (emitting X-rays by accreting 
from a close donor). For local, low/intermediate-SFR galaxies, the 
simultaneous existence of a correlation of the instantaneous SFR with 
the total 2-10 keV luminosity, $L_x$, which traces the SFR integrated 
over the last $\sim$$10^9$ yr, suggests that during such epoch the SF in 
these galaxies has been proceeding at a relatively constant rate.
\keywords{
galaxies: X-rays -- 
galaxies: spiral -- 
galaxies: starburst --
galaxies: stellar content --
galaxies: evolution --
infrared: galaxies --
radio continuum: galaxies --
stars: formation -- 
X-rays: binaries --
X-rays: galaxies}}

\maketitle
\markboth{Persic \& Rephaeli: Stellar birth-rates from XPs}{}

\section{Introduction}

Star formation (SF) leads to X-ray emission on various spatial and temporal 
scales, including emission from O stars, X-ray binaries, supernovae (SN) 
and their remnants (SNRs), galactic-scale emission from diffuse hot gas, and 
Compton scattering of FIR \& CMB photons by SN-accelerated electrons. 
Integrated spectra of star-forming galaxies (SFGs) are expected to show all 
these components, as well as emission from an active nucleus (Rephaeli et 
al. 1991, 1995). Based on lower-resolution X-ray data (mainly from {\it 
Einstein}, {\it ASCA}, {\it BeppoSAX}, and {\it RXTE}), most SFGs have 
remarkably similar integrated spectra that include a soft (single- or 
multiple-temperature) sub-keV thermal component which dominates at energies 
$\epsilon$$\mincir$$2$ keV, plus a hard power law (PL) which dominates at 
$\epsilon$$\magcir$$2$ keV (e.g., Dahlem et al. 1998).

By quantitatively assessing the spectral components of the various X-ray 
emission mechanisms in SFGs, Persic \& Rephaeli (2002) concluded that the 
2-10 keV emission is dominated by compact X-ray binaries, specifically HMXBs 
if the star-formation rate (SFR) is very high.

\begin{table*}
\caption[] {
Data Ia: local normal and starburst galaxies (SFG sample).}
\begin{flushleft}
\begin{tabular}{ l  l  l  l  l  l  l  l  l  l  l  l }
\noalign{\smallskip}
\hline
\hline
\noalign{\smallskip}
Object & D$^{(a)}$ & $L_{x}^{(b)}$ & $L_{\rm XP}^{(c)}$ & $L_{\rm min},\, L_{\rm max}^{(d)}$ & 
$\gamma^{(e)}$ & 
$B_{\rm T}^{0\,(f)}$ & $f_{12\mu}^{(g)}$ & $f_{25\mu}^{(g)}$ & $f_{60\mu}^{(g)}$ & $f_{100\mu}^{(g)}$ & 
$f_{1.4}^{(h)}$ \\
\noalign{\smallskip}
\hline
\noalign{\smallskip}
NGC~~253 &   3.0& 39.73& 39.22& 36.82, ~38.42& 0.75&  7.09&  41.04& 154.67&  967.81& 1288.15& 6.18 \\
NGC~~628 &   9.7& 39.45& 39.42& 37.32, ~38.42& 1.11&  9.76&   2.45&   2.87&   21.54&   54.45& 0.173 \\
NGC~~891 &   9.6& 40.32& 39.57& 37.42, ~39.12& 0.76&  9.37&   5.27&   7.00&   66.46&  172.23& 0.658 \\
IC~342   &   3.9& 40.30& 40.02& 37.44, ~40.44& 0.55&  6.04&  14.92&  34.48&  180.80&  391.66& 2.48 \\
NGC~1569 &   1.6& 37.83& 37.72& 35.82, ~37.32& 0.44&  9.42&   1.24&   9.03&   54.36&   55.29& 0.339 \\
NGC~2146 &  17.2& 40.59& 40.44& 37.85, ~39.55& 0.71& 10.58&   6.83&  18.81&  146.69&  194.05& 1.07 \\
NGC~2403 &   4.2& 39.29& 38.45& 36.02, ~37.75& 0.65&  8.43&   2.82&   3.57&   41.47&   99.13& 0.387 \\
NGC~3034 &   5.2& 40.70& 39.82& 37.32, ~39.32& 0.57&  5.58&  79.43& 332.63& 1480.42& 1373.69& 8.36 \\
NGC~3077 &   2.1& 37.91& 37.58&              &     & 10.24&   0.76&   1.88&   15.90&   26.53&      \\
NGC~3079 &  20.4& 40.59& 39.69& 37.62, ~39.32& 0.80& 10.41&   2.54&   3.61&   50.67&  104.69& 0.845 \\
NGC~3628 &   7.7& 39.84& 39.32& 36.82, ~39.12& 0.82&  9.31&   3.13&   4.85&   54.80&  105.76& 0.402 \\
Arp~299  &  41.6& 41.35& 40.82& 38.86, ~40.26& 0.50& 11.85&   3.97&  24.51&  113.05&  111.42& 0.977 \\
NGC~4038/9& 25.4& 40.61& 40.75& 37.95, ~39.95& 0.63& 10.62&   1.94&   6.54&   45.16&   87.09&      \\
NGC~4214 &  3.48& 38.68& 38.65& 36.36, ~38.44& 0.82& 10.14&   0.58&   2.46&   17.57&   29.08&      \\
NGC~4449 &   3.0& 38.72& 38.82& 36.82, ~38.22& 0.70&  9.94&   2.1 &   4.7 &    36.0&    73.0& 0.600 \\
NGC~4631 &   6.9& 39.82& 39.32& 36.82, ~39.02& 0.69&  8.61&   5.16&   8.97&   85.40&  160.08& 1.12 \\
NGC~4945 &   5.2& 41.22& 39.62& 37.42, ~38.92& 0.70&  7.43&  27.74&  42.34&  625.46& 1329.70& 6.60 \\
NGC~5236 &   4.7& 40.09& 39.62& 37.32, ~38.82& 0.91&  7.98&  21.46&  43.57&  265.84&  524.09& 2.60 \\
NGC~5457 &   5.4& 39.38& 39.62& 36.82, ~39.02& 0.85&  8.21&   6.20&  11.78&   88.04&  252.84& 0.808 \\
NGC~6946 &   5.5& 39.64& 39.66& 36.76, ~39.76& 0.64&  7.78&  12.11&  20.70&  129.78&  290.69& 1.43 \\
\noalign{\smallskip}
\hline
\hline
\end{tabular}
\end{flushleft}
\smallskip

$^{(a)}$ Distances (in Mpc) are from Tully (1988) if $cz \leq 3000$ 
km s$^{-1}$, or consistently assume H$_0=75$ km s$^{-1}$ Mpc$^{-1}$ otherwise.

$^{(b)}$ Integrated 2-10 keV luminosities (in erg s$^{-1}$; given in log 
form). For references see Persic et al. (2004a). For NGC~3077 and NGC~4214 see below. For 
NGC~4945 see Guainazzi et al. (2000).

$^{(c)}$ Cumulative XP luminosities in the 2-10 keV band (in erg s$^{-1}$; 
in log form). The quoted values result from converting the values measured 
in several different bands [0.3-10 keV for NGC~2403 (Schlegel \& Pannuti 
2003); 0.5-5 keV for NGC~6946 (Holt et al. 2003); 
0.2-12 keV for IC~342 (Kong 2003); 
0.5-8 keV for Arp~299 (Zezas et al. 2003); 
2-8 keV for NGC~891 (Temple et al. 2005);
0.3-8 keV for NGC~4214 (Hartwell et al. 2004) and 
for all other objects (Colbert et al. 2004)] 
into the 2-10 keV band through an assumed photon spectrum 
$\phi \propto \epsilon^{-\alpha}$ [with $\alpha=2$ in Kong (2003), $1.8$ in 
Hartwell et al. (2004) and Temple et al. (2005), and $1.7$ elsewhere]. 

\noindent
In a few cases (NGC~4038/9, NGC~4449, NGC~5457, NGC~6946) it turns out that 
${\rm log}L_{\rm XP} > {\rm log}L_{x}$ by $0.10 <\delta{\rm log}L \mincir 
0.30$: assuming such unphysical situation to descend from non-optimal fits 
being performed on most XP spectra (except for those few XPs in each galaxy 
which had enough counts to allow detailed fits to be performed), in our 
computations we set $L_{\rm XP}$=$L_x$.

\noindent
For NGC~4214 the XPLF slope is taken slightly steeper than quoted in 
Hartwell et al. (2004), to account for incompleteness that admittedly 
affects the XPLF shape below log$L= 36.6$.

\noindent
For NGC~2403 the reported value of $L_{\rm XP}$ is based on the mean flux 
from its XPs (Schlegel \& Pannuti 2003), excluding the 4 brightest sources and 
12 presumed interlopers. The reported $\gamma$ and $\eta$ refer to 
the same set of sources represented by $L_{\rm XP}$. See discussion in 
Section 4.1. 

\noindent
For NGC~2146 the 2-10 keV XP data are taken from Inui et al. (2005).

\noindent
For NGC~3077 the luminosities of its 6 XPs and of the diffuse thermal plasma, 
measured in the 0.3-6 keV band (Ott et al. 2003), have been converted to 2-10 
keV luminosities, using the individual published spectral models of Ott et 
al. (2003), and combined to give $L_x$. The reported $L_{\rm XP}$ refers 
to the estimated young-XP luminosity. No published XPLF is available. See 
discussion in Section 4.1. 

$^{(d)}$ Minimum and maximum XP luminosities (in erg s$^{-1}$; given in 
log form), between which the XPLF index, $\gamma$, was derived. For spectral 
band information, see previous point.

$^{(e)}$ Index of the cumulative XPLF, of the form $N(>L)$$\propto$$L^{-\gamma}$. 
For references, see point $(c)$. 

$^{(f)}$ Blue apparent magnitudes, corrected to face-on and for Galactic 
absorption, from RC3 (de Vaucouleurs et al. 1991).

$^{(g)}$ {\it IRAS} flux densities at $12\mu$m, $25\mu$m, 
$60\mu$m, and $100\mu$m (in Jy), from Sanders et al. (2003; except for 
NGC~4449: Hunter et al. 1986). The FIR flux is accordingly defined (Helou 
et al. 1985) as $f_{\rm FIR} \equiv 1.26 \times 10^{-11} (2.58\, f_{60} + 
f_{100})$ erg s$^{-1}$ cm$^{-1}$.

$^{(h)}$ 1.4 GHz flux density (in Jy). Data are from K\"uhr et al. 1981 (NGC~253), 
White \& Becker 1992 (NGC~628, NGC~891, IC~342, NGC~2403, NGC~3034, NGC~3079, 
NGC~3628, Arp~299, NGC~4631, NGC~5457, NGC~6946), Condon et al. 1998 (NGC~1569, 
NGC~2146), Heeschen \& Wade 1964 (NGC~4449), Wright \& Otrupcek 1990 (NGC~4945, 
NGC~5236). No entry means that no data, or only upper limits to $f_{1.4}$, were 
found in the literature.

{\it Notes.} {\it (i)} Compared with Persic et al. (2004a), several objects 
could not find their way into the present analysis. These systems and 
the reason for their exclusion are: NGC~55, NGC~891, NGC~1808, NGC~2276, 
NGC~5782, NGC~2903, NGC~3256$^{[l]}$, NGC~3310$^{[l]}$, NGC~3367, NGC~3556, 
NGC~4654, NGC~4666$^{[l]}$, NGC~7552: no published XPLF available ($[l]$: 
observed with {\it Chandra} and/or {\it XMM-Newton} (Lira et al. 2002; 
Jenkins et al. 2004; Persic et al. 2004b). NGC~5253: no integrated 2-10 
keV luminosity available (see Summers et al. 2004). NGC~7469, NGC~7679: 
AGN-dominated 2-10 keV emission. {\it (ii)} Other commonly used names for 
some of the objects are: M~74 for NGC~628, M~82 for NGC~3034, 
NGC~3690~+~IC~694 for Arp~299, Antennae for NGC~4038/9, M~83 for NGC~5236, 
and M~101 for NGC~5457.

\end{table*}

\begin{table}
\caption[] {
Data Ib: local normal and starburst galaxies (SFG sample).}
\begin{flushleft}
\begin{tabular}{ l  l  l  l  l  l }
\noalign{\smallskip}
\hline
\hline
\noalign{\smallskip}
Object & $\eta^{(a)}$ & $L_{\rm FIR}^{(b)}$ & $L_{\rm FIR}^{{\rm corr}\, (c)}$ & 
SFR$^{(d)}$ &  SFR$^{{\rm corr} \, (e)}$\\
\noalign{\smallskip}
\hline
\noalign{\smallskip}
NGC~~253 &  0.34 & 43.71 & 43.66 &~~3.9  &~~3.4\\
NGC~~628 &  0.03 & 43.19 & 43.03 &~~1.2  &~~0.8\\
NGC~~891 &  0.47 & 43.68 & 43.61 &~~3.6  &~~3.1\\
IC~342   &  0.95 & 43.29 &       &~~1.5  &     \\
NGC~1569 &  1.00 & 41.88 & 41.76 &~~0.06 &~~0.04\\
NGC~2146 &  0.65 & 44.41 & 44.39 & 19.2  & 18.6\\
NGC~2403 &  0.33 & 42.74 & 42.37 &~~0.41 &~~0.17\\
NGC~3034 &  0.85 & 44.33 & 44.16 & 15.9  & 10.9\\
NGC~3077 &       & 41.65 & 41.48 &~~0.03 &~~0.02\\
NGC~3079 &  0.46 & 44.17 & 44.13 & 11.1  & 10.2\\
NGC~3628 &  0.29 & 43.34 & 43.24 &~~1.7  &~~1.3\\
Arp~299  &  0.96 & 45.02 & 45.02 & 78.9  & 77.9\\
NGC~4038/9& 0.85 & 44.30 & 44.26 & 14.9  & 13.7\\
NGC~4214 &  0.19 & 42.13 & 41.96 &~~0.10 &~~0.07\\
NGC~4449 &  0.33 & 42.35 & 42.27 &~~0.17 &~~0.14\\
NGC~4631 &  0.53 & 43.44 & 43.30 &~~2.0  &~~1.5\\
NGC~4945 &  0.52 & 44.08 & 44.03 &~~9.0  &~~8.1\\
NGC~5236 &  0.15 & 43.61 & 43.54 &~~3.0  &~~2.6\\
NGC~5457 &  0.23 & 43.32 & 43.17 &~~1.6  &~~1.1\\
NGC~6946 &  0.73 & 43.46 & 43.27 &~~2.1  &~~1.4\\
\noalign{\smallskip}
\hline
\hline
\end{tabular}
\end{flushleft}
\smallskip

$^{(a)}$ Derived 2-10 keV luminosity fraction of young XPs.

$^{(b)}$ Cirrus-uncorrected FIR luminosities (in erg s$^{-1}$), 
in log form.

$^{(c)}$ Cirrus-corrected FIR luminosities (in erg s$^{-1}$), 
in log form.

$^{(d)}$ Star-formation rates, in $M_\odot$ yr$^{-1}$, uncorrected 
for cirrus emission. They are derived from IR luminosities using 
eq.(3) and setting $L_{\rm IR}$$=$$1.65$$\times$$L_{\rm FIR}$. 

$^{(e)}$ Star-formation rates, in $M_\odot$ yr$^{-1}$, corrected 
for cirrus emission. They are derived as described in point $(d)$.

{\it Note.} For objects located at low Galactic latitudes ($|b| < 15$ 
degrees) the corrections for foreground Galactic absorption tend to be 
large and uncertain. An extreme case is IC~342 ($b$$=$$10.58$ deg), for 
which the $B$-band absorption is estimated to be as large as 3.360 mag 
(Burstein \& Heiles 1982) or 2.407 mag (Schlegel et al. 1998). As the 
adopted statistical correction for cirrus emission turns out to be 
unphysical in this case, we will drop IC~342 from further analyses. None 
of the results reported in this paper will depend on the exclusion of 
this object. 

\end{table}

Stellar-related X-ray emission can be used as an indicator of SFR (David et al. 
1992; Bauer et al. 2002; Grimm et al. 2003; Franceschini et al. 2003; Ranalli 
et al. 2003; Gilfanov et al. 2004a). The basic notion is that the ongoing SFR 
can be measured based on stellar end-products which are both sufficiently 
X-ray-bright for their collective emission to be unambiguously identified, and 
sufficiently short-lived so that they trace the 'instantaneous' SFR. Of the 
three main types of Galactic stellar endproduct X-ray sources (LMXBs, SNRs, and 
HMXBs), the latter two provide a suitable combination of short delay between 
star formation and onset of X-ray emission and significant X-ray brightness. 

Persic et al. (2004a) examined ways in which the 2-10 keV luminosity (hereafter: 
$L_{x}$) can be used as an estimator of ongoing galactic SFR. They concluded that 
the collective 2-10 keV emission from HMXBs, $L_{x}^{\rm HMXB}$, was a reliable 
SFR estimator (supporting independent suggestions by, e.g., Grimm et al. 2003). 
Given the diverse sources of X-ray emission in SFGs, Persic et al. (2004a) argued 
that the level of HMXB emission could be evaluated by modelling SFG spectra with 
Persic \& Rephaeli's (2002) template -- in which the HMXB component was represented 
as a $\Gamma$=$1.2$ power law. Upon analyzing suitable {\it ASCA}, {\it BeppoSAX}, 
and {\it Newton/XMM} spectra, Persic et al. (2004a) suggested that $L_{x}^{\rm 
HMXB}$$\sim 0.2\,L_x$ in moderately star-forming galaxies and $L_{x}^{\rm HMXB}
$$\sim$$L_x$ in intensely star-forming ones. 

However, Persic et al.'s (2004a) approach was potentially limited. The template 
SFG spectrum of Persic \& Rephaeli (2002), on which their spectral analysis was 
based, explicitly assumed a Galactic stellar population: in principle that feature 
could bias Persic et al.'s (2004a) procedure because any SF-tracing population 
of X-ray point sources (XPs), not represented in our own Galaxy but present 
elsewhere (and spectrally different from Galactic HMXBs), would be missed. This 
actually turns out to be the case for the Ultra-Luminous X-ray sources (ULXs) 
which dominate the XP luminosity ($L_{\rm XP}$) of most SFGs. Furthermore, few 
integrated spectra were of sufficiently good quality to be fitted with the 
multi-component template of Persic \& Rephaeli (2002).

It is important then to further scrutinize the role and effectiveness of the 
collective 2-10 keV emission of young XPs, $L_{x}^{\rm yXP}$, in measuring the 
ongoing SFR in galaxies -- as well as ways to evaluate it. In this paper we 
propose to evaluate $L_{x}^{\rm yXP}$ by modelling the luminosity functions 
of galactic XPs (hereafter: XPLFs) as linear combinations of the 'universal' 
young- and old-XPLFs. To demostrate the effectiveness of the proposed technique, 
we will apply it to a sample of galaxies with available {\it Chandra} XPLFs. 
For the same sample, we will then check how the derived values of $L_{x}^{\rm 
yXP}$ correlate with the ongoing SFR as deduced from the galaxies' (thermal) 
FIR and (non-thermal) radio emission.

The plan of this paper is as follows. In section 2 we review the properties of 
XPLFs in galaxies. The two local and nearby samples of star-forming galaxies 
are described in section 3. In section 4 we review the FIR- and radio-based SFR 
indicators. In section 5 we apply our proposed young/old-XP decomposition 
technique to a sample of SFGs and evaluate $L_{x}^{\rm yXP}$, which are then 
contrasted with FIR/radio-based SFRs: the results are discussed in section 6.
In section 7 we extend our investigation to high-SFR galaxies, and discuss SF 
in the nearby universe. In section 8 we discuss X-ray SFR indicators at high 
redshift. X-ray and radio SFR indicators are compared in section 9. Section 10 
closes with a short summary of our main results. Throughout this paper we 
assume $\Omega_0=1$, H$_0$$=$$75$ km s$^{-1}$Mpc$^{-1}$. (No result in this 
paper is substantially affected by choosing this one particular cosmology.)

\section{X-ray point sources in galaxies}

Observations of nearby galaxies led to the detection of XPs and their 
luminosity functions (XPLFs) down to limiting luminosities of $\sim$10$^
{36}$ erg s$^{-1}$ (see Fabbiano 2005 for a recent comprehensive review.) 
Such XPs include SNRs, close binary systems in which the accreting object 
can be either a pulsar ($L_x \leq 2 \times 10^{38}$ erg s$^{-1}$) or a 
black hole (BH) ($2 \times 10^{38} \leq L_x/({\rm erg~ s}^{-1}) \leq 10^
{39}$), and ULXs ($L_x$$>$$10^{39}$ erg s$^{-1}$). The quoted limits 
correspond to Eddington luminosities for spherical accretion onto 
$\sim$$2\,M_\odot$ and $\sim$$8\,M_\odot$ BHs, respectively, which 
is the approximate range of BH masses that are thought to be therein
attainable via ordinary stellar evolution. 

Young SNRs, which result from the explosion of short-lived $\magcir$5$M_
\odot$ progenitor stars and are X-ray bright for $\sim$$10^3$ yr, trace 
current SF. X-ray binaries, in which X-ray emission results from mass 
accretion onto a compact stellar remnant (NS or BH) from a main-sequence 
donor, are of the high-mass (HMXB) type or the low-mass (LMXB) type 
according to whether the donor mass is $M$$\magcir$8$\,M_\odot$ or 
$M$$\mincir$1$M_\odot$ (see Persic \& Rephaeli 2002 and references therein). 
HMXBs and LMXBs therefore trace, respectively, the current or average past SFR. 
The accreting object in ULXs is presumed to be a stellar (super-stellar) 
mass BH, accreting at super- (sub-) Eddington rates. Association with 
intense SF activity suggests that most ULXs are of the HMXB type (e.g., 
Zezas et al. 2002) -- their optical counterparts having sometimes been 
positively identified as O stars (e.g., Liu et al. 2002). When observed 
in E/S0 galaxies (only occasionally, and limited to $L_x \mincir 2 \times 
10^{39}$ erg s$^{-1}$), ULXs are likely of the LMXB type (Sarazin et al. 
2000; Kim \& Fabbiano 2004). In both cases, ULXs appear to extrapolate 
X-ray binaries to higher accretor masses (e.g., Swartz et al. 2004). 

The association of XPs with their host environment proves useful to 
measure XPLFs for uniform source population. Functions measured in starburst  
environments yield the young-XPLF, and XPLFs measured in 'sterile' 
environments (e.g., in elliptical galaxies) reproduce the old-XPLF. 
As measured in systems with high SFR (e.g., in NGC~4038/9: Zezas \& 
Fabbiano 2002; see discussion in section 4.3), the differential 
young-XPLFs can be described by 
\begin{eqnarray}
{{\rm d}N_{\rm y} \over {\rm d}L} ~=~ n_{\rm y,0}~ L^{-\beta_{\rm y}}, ~~~~\beta_{\rm y} \sim 1.5
\end{eqnarray}
with the cumulative counts $N_{\rm y}(>$$L)$$\propto$$L^{-(\beta-1)}$. In the 
following we shall take eq.(1) to represent the 'universal' young-XPLF. 
The statistically superposed functions from a sample of E/S0 galaxies 
result in a completeness-corrected differential old-XPLF which is given 
by 
\begin{eqnarray}
{{\rm d}N_{\rm v} \over {\rm d}L} ~=~ n_{\rm v,0}~ \times ~
\left\{
\begin{array}{ll}
L^{-\beta_{\rm v,1} } & \mbox{~~~~~~~~~~~$L_{\rm 1}  \leq L <    L_{\rm 
br}$} \\
L^{-\beta_{\rm v,2} } & \mbox{~~~~~~~~~~~$L_{\rm br} \leq L \leq L_{\rm 
2}$ }
\end{array}
\right.
\end{eqnarray}
with $\beta_{\rm v,1}$$\sim$$1.8$ and $\beta_{\rm v,2}$$\sim$$2.8$
the faint-end slope and bright-end slope,
$L_{\rm 1}$$\sim$$5$$\times$$10^{37}$ erg s$^{-1}$ and $L_{\rm 2} 
$$\sim$$2$$\times$$10^{39}$ erg s$^{-1}$ the limiting luminosities, and 
$L_{\rm br}$$\sim$$5$$\times$$10^{38}$ erg s$^{-1}$ the break luminosity 
(Kim \& Fabbiano 2004; see also Gilfanov 2004). The corresponding 
cumulative function has a low-$L$ slope $\simeq$1 and a high-$L$ 
slope 1.8. In the following we shall take eq.(2) to represent the 
'universal' old-XPLF (see discussion in section 4.3). The break may 
highlight a change in the nature of the XP population: as its value 
approaches the Eddington limit for an accreting NS, the break may 
signal the NS to BH transition in the LMXB population. 

Galaxies with mild ongoing SF, like our own Galaxy, have both young and 
old XPs. A direct separation of young and old XPs has been possible only 
in few cases (e.g., in M~81: Tennant et al. 2001). Usually, measured XPLFs 
reflect galaxy-integrated counts and hence result from a combination of 
young-XPLF and old-XPLF, whose relative normalization depends on the current 
to average-past SFR (see Grimm et al. 2003; Gilfanov 2004). Indeed, based on 
1441 XPs detected in 32 nearby galaxies with different levels of SF activity, 
Colbert et al. (2004) concluded that XPs are linked to both the old and young 
stellar populations or, equivalently, to both the past and present SF activity 
(confirming the earlier suggestions of Fabbiano \& Trinchieri 1985 and David 
et al. 1992). To estimate the current SFR, young XPs have to be culled out 
from the total population.

\section{Galaxy samples}

Sample 1 (see Tables 1,2) consists of local SFGs with available {\it IRAS} 
FIR fluxes, {\it ASCA}/{\it BeppoSAX}/{\it RossiXTE} 2-10 keV fluxes, and {\it 
Chandra} XPLFs. It spans $\sim$3 decades in SFR (estimated from $L_{\rm FIR}$, 
see section 4), from levels typical of quiescent isolated spirals ($\mincir$1$
\,M_\odot$yr$^{-1}$) all the way up to strong merging starbursts ($\sim$50$\,
M_\odot$yr$^{-1}$). As such, sample 1 is fairly representative of the range of 
SF activity in the local Universe.

Sample 2 (see Table 3) comprises a set of nearby Ultra-Luminous Infra-Red 
Galaxies (ULIRGs) with available {\it IRAS} flux densities and {\it XMM} 2-10 
keV fluxes. As suggested by their $L_{\rm FIR}$, these galaxies are sites 
of very intense star formation (SFR$>$$100$$\,M_\odot$yr$^{-1}$). Their 2-10 
keV spectra show no evidence of AGN emission and are reminiscent, in their 
shapes, of Galactic HMXBs: this suggests that the entire $L_x$ of these 
galaxies may be related to ongoing SF (see Franceschini et al. 2003). This 
set of starburst ULIRGs probes the peak of SF activity in the nearby Universe.

It is instructive to check the evolutionary stage of the SFGs represented in 
samples 1 and 2. As the spectral region $\sim$8-120 $\mu$m (sampled by the 
{\it IRAS} broad-band filters), which is characterized by continuum emission 
from hot dust, is very strongly affected by heating processes associated with 
SF, the {\it IRAS} color-color plot can be interpreted in terms of evolution 
of the SF activity. Based on advanced spectro-photometric modeling, Vega et 
al. (2005) suggested that the evolution of a starburst can be described as a 
sequence of four main phases (see Fig.1): 
{\it (a)} an early-starburst phase, when the newly formed stars are still 
deep inside their placental clouds, and the escaping radiation field has not 
reached its peak emission; 
{\it (b)} a peak-starburst phase, when most massive stars are produced and 
are still embedded in their placental clouds, the SED is dominated by hot 
dust emission, and the starburst reaches its hottest colors; 
{\it (c)} an evolved-starburst phase, when the current SFR has decreased 
dramatically, the young hot stars have emerged from their progenitor clouds, 
and the cirrus emission is important; 
and {\it (d)} a post-starburst phase, when the current SF is mainly due to 
the quiescent disk and the colors are evolving towards those of normal spirals. 
According to this scheme, the location of our sample objects in Fig.1 suggests 
that the ULIRGs are dominated by a starburst in its peak, while most local SFGs 
represent later phases, from evolved- through post-starburst to quiescent. 

\begin{figure}
\vspace{6.1cm}
\includegraphics{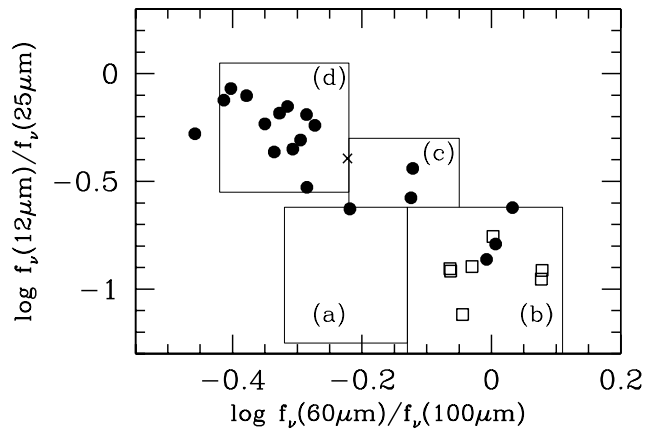}
\caption{ {\it IRAS} color-color diagram of the galaxies of samples 1 (filled 
dots) and 2 (empty squares). Areas (a) through (d) denote successive phases 
of starburst evolution according to Vega et al. (2005; respectively: early-, peak-, 
evolved-, post-starburst). }
\end{figure}

\section{FIR, radio galactic-SFR indicators}

A significant fraction of the bolometric luminosity of an actively star-forming 
galaxy is absorbed by interstellar dust and re-emitted in the FIR band. As the 
absorption cross-section of dust is strongly peaked in the UV which inherently 
traces massive SF, the FIR emission can be a sensitive tracer of the current SFR. 
As discussed by Kennicutt (1998a,b), there probably is no single calibration 
that applies to all galaxy types. However, based on the main characteristics of our 
composite sample and guided by the general principle that the FIR emission should 
provide an excellent measure of the SFR in dusty starbursts, we shall adopt the 
most appropriate FIR SFR indicator for our current purposes.

A wide range of conversion relations between SFR and $L_{\rm FIR}$ are found in 
the literature, based on either a starburst model or an observational analysis 
(e.g., Hunter et al. 1986; Meurer et al. 1997; Kennicutt 1998a,b). Using the 
continuous-burst model of 10-100 Myr duration of Leitherer \& Heckman (1995) 
and the Salpeter (1955) stellar initial mass function (IMF) with mass limits 
0.1-100 $M_\odot$, Kennicutt (1998a,b) derived a conversion relation appropriate 
for starbursts:
\begin{eqnarray} 
{\rm SFR} ~=~ {L_{\rm IR} \over 2.2 \times 10^{43} {\rm erg \,s}^{-1}} ~ 
M_\odot {\rm yr}^{-1}
\end{eqnarray}
with $\sim$30$\%$ uncertainty. Here $L_{\rm IR}$ refers to the full 8-1000 
$\mu$m band: for starbursts with typical dust temperatures and emissivities, 
however, most of the emission falls in the FIR ($\sim$40-120 $\mu$m) band, 
$L_{\rm FIR}/L_{\rm IR}$$\sim$0.6 (Helou et al. 1988). Strictly speaking, the 
relation in eq.(3) applies only to young starbursts embedded in optically 
thick dust clouds. In more quiescent SFGs the assumptions underlying the 
relation in eq.(3) are not verified. Among these, the dust optical depth is 
lower, and a colder "cirrus" component, originating from the heating of the 
ISM by the galactic UV emission that is powered mostly by old stars, will 
contribute to the total FIR emission. 

In this paper we adopt Kennicutt's conversion in eq.(3) as a FIR-based
SFR indicator. The insight on the evolutionary status of the 
galaxies in samples 1 and 2, discussed in section 3, suggests some more 
accurate way of applying eq.(3). As suggested by their FIR colors (see 
Fig.1), all the ULIRGs of sample 2 and some of the SFGs of sample 1 are 
young, dusty, optically-thick starbursts that presumably meet the 
assumptions underlying eq.(3) and hence can be straightforwardly treated 
with it and their SFR can then be directly estimated from the observed 
$L_{\rm IR}$. For the remaining, milder SFGs of sample 1, which have 
evolved past the peak starburst phase, some of the assumption underlying 
eq.(3) are not valid, so in principle their SFR can not be directly 
estimated from the observed $L_{\rm IR}$. In particular, the FIR 
emission of these galaxies should be corrected for cirrus emission before 
being used as a SFR indicator. Since a detailed spectro-photometric 
modeling (e.g., Vega et al. 2005) of our sample galaxies is beyond our 
immediate scope, in this paper we follow David et al. (1992) in adopting 
Devereux \& Eales's (1989) {\it statistical} cirrus correction. The basic 
assumption is that the strong empirical FIR-radio correlation holding for 
SFGs (Helou et al. 1985), interpreted as a consequence of massive SF (see 
below), is blurred by the SF-unrelated cirrus FIR emission, $L_{\rm FIR}^
{\rm cir}$, most notably so at low luminosities where in fact some 
nonlinearity occurs (see Condon 1992; Bell 2003). Setting $L_{\rm FIR}^
{\rm cir}$$\propto$$L_{\rm B}$ (the blue luminosity being a proxy for the 
galactic stellar content), then the SF-related FIR component is then $L_{\rm 
FIR}^{\rm SF}$$=$$L_{\rm FIR}$$-$$x \,L_{\rm B}$: the FIR--radio correlation 
is linearized and optimized if $x=0.14$. The FIR luminosities in sample 1, 
although corrected according to this recipe for SFR-computing purposes, 
for simplicity will still be called $L_{\rm FIR}$ (see Table 2). Finally, 
the FIR ($\sim$$40$$-$$120$$\mu$m) luminosities need a correction for the 
wider bandwidth ($\sim$8-1000 $\mu$m) required by Kennicutt's formula. 
Although such correction clearly depends on the detailed spectral energy 
distribution of each object, we here adopt a statistical correction and set 
\begin{eqnarray}
L_{\rm IR} ~=~ f ~ L_{\rm FIR} ~~~~~~~~~~~ f=1.65
\end{eqnarray}
to be used in eq.(3). This bandwidth correction is, strictly speaking, valid 
for starbursts with $f_{60}/f_{100}=1$ and dust emissivity index equal to 0 
(Helou et al. 1988) and hence it may apply only to the ULIRGs of sample 2 and 
to some strong SFGs of sample 1; however we assume that, after removal of the 
cirrus component, it is sensible also for the remaining, more mildly star-forming 
objects of Tables 1 and 2. In substantial agreement with our choice, Hopkins 
et al. (2003), Bell (2003), Kewley et al. (2002), and Calzetti et al. (2000) 
use a factor of 1.75 to convert from FIR to 1-1000 $\mu$m, the contribution 
to the 1-8 $\mu$m being in fact of the order of a few percent (see also Dale 
et al. 2001). 

\begin{figure*}
\vspace{3.0cm}
\includegraphics{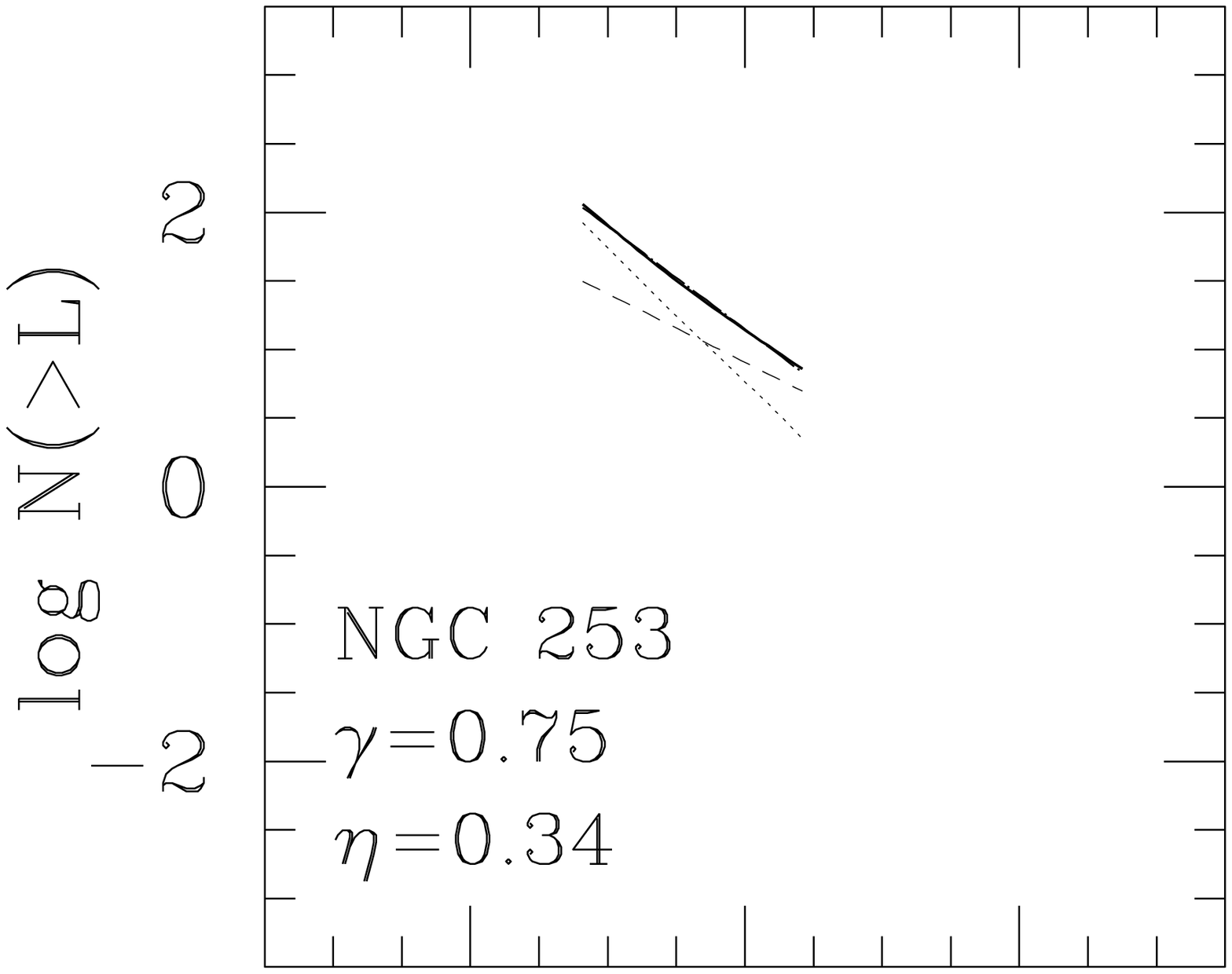}
\includegraphics{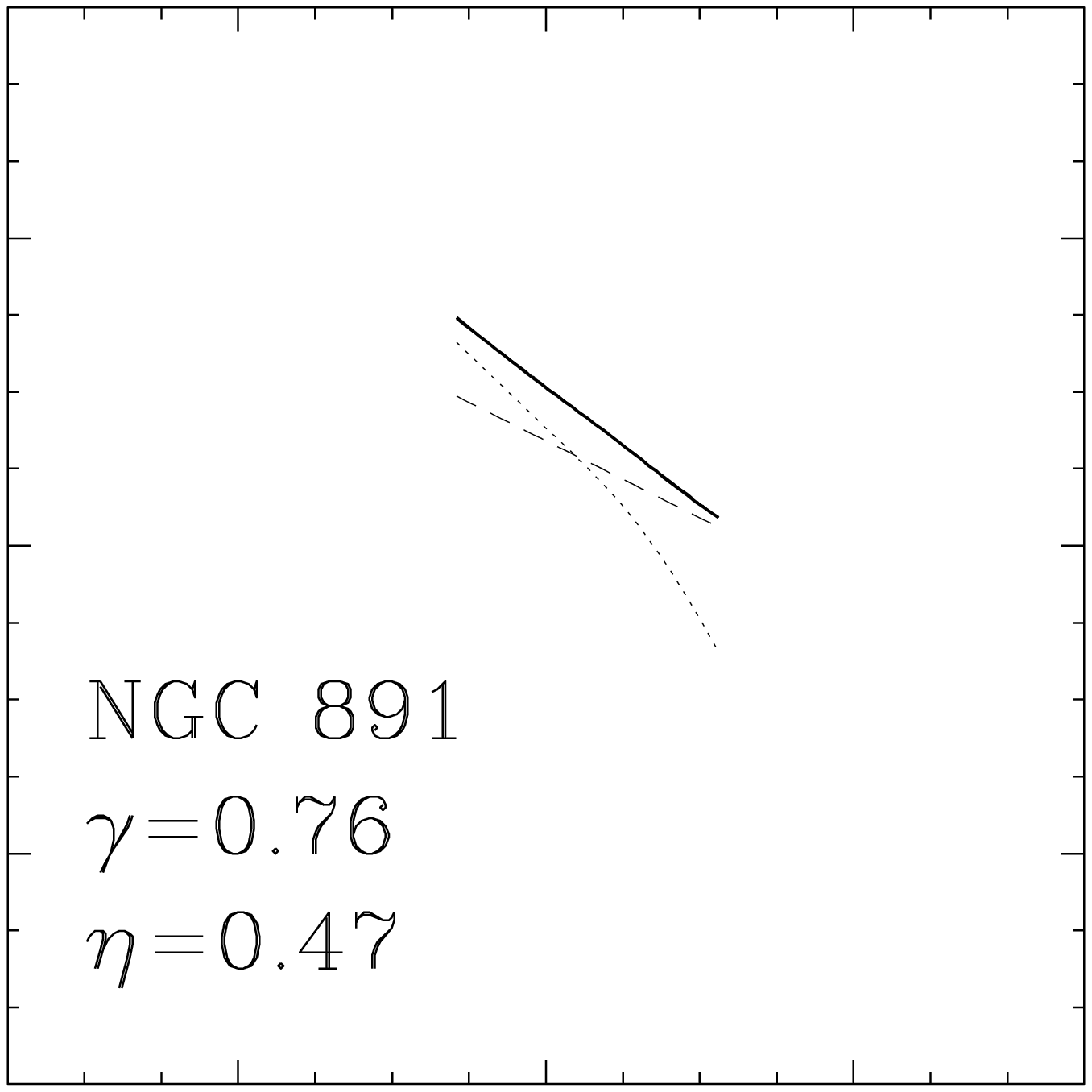}
\includegraphics{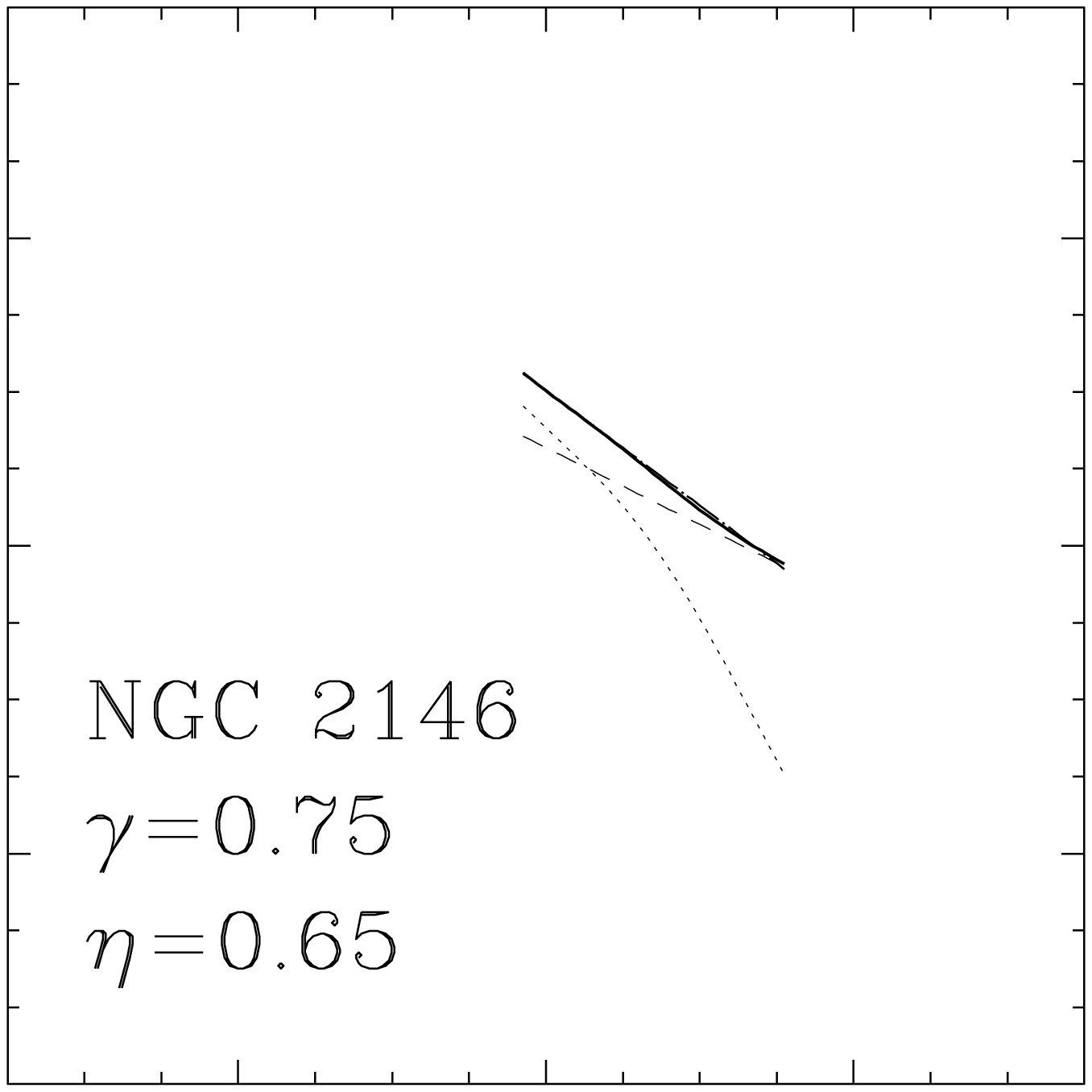}
\includegraphics{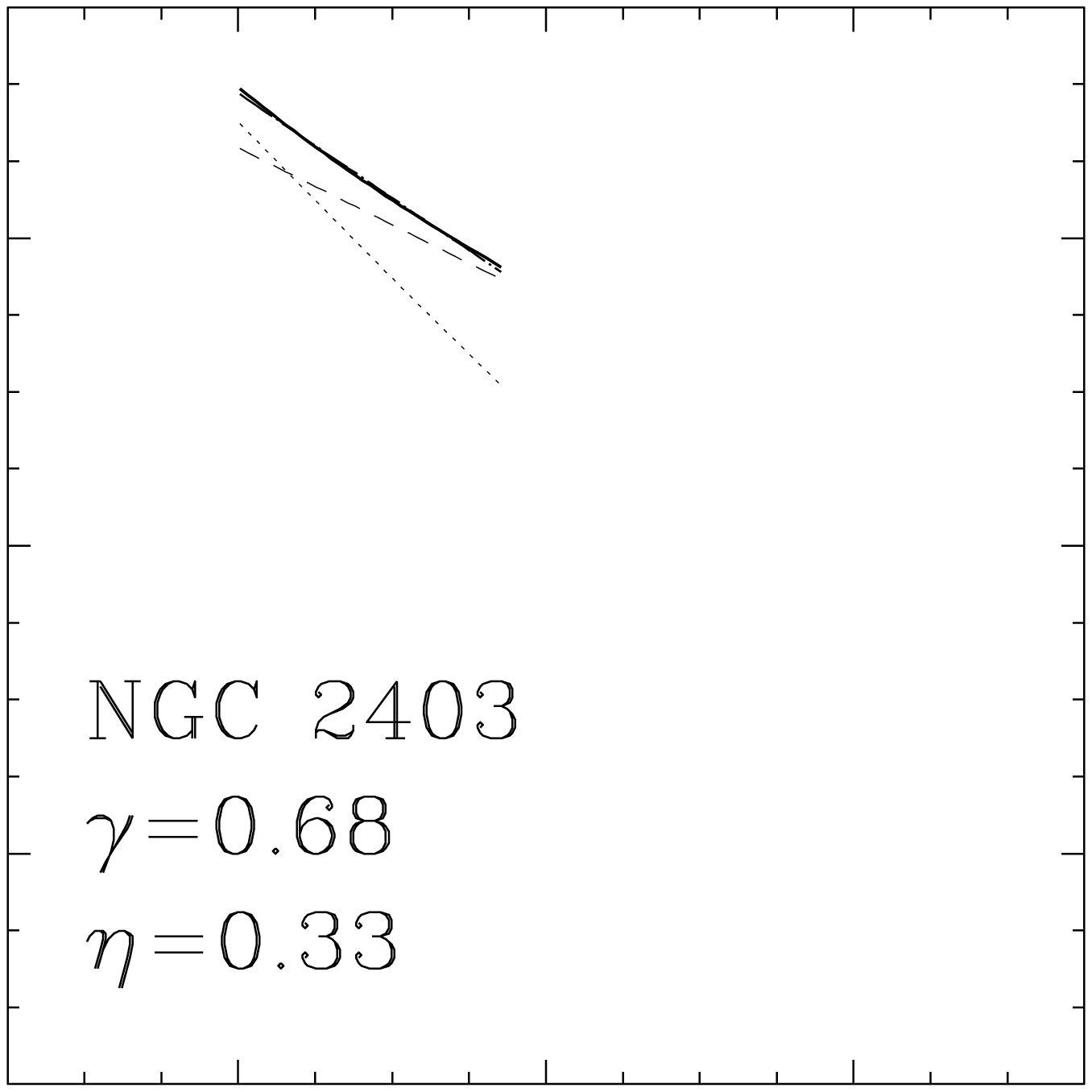}
\includegraphics{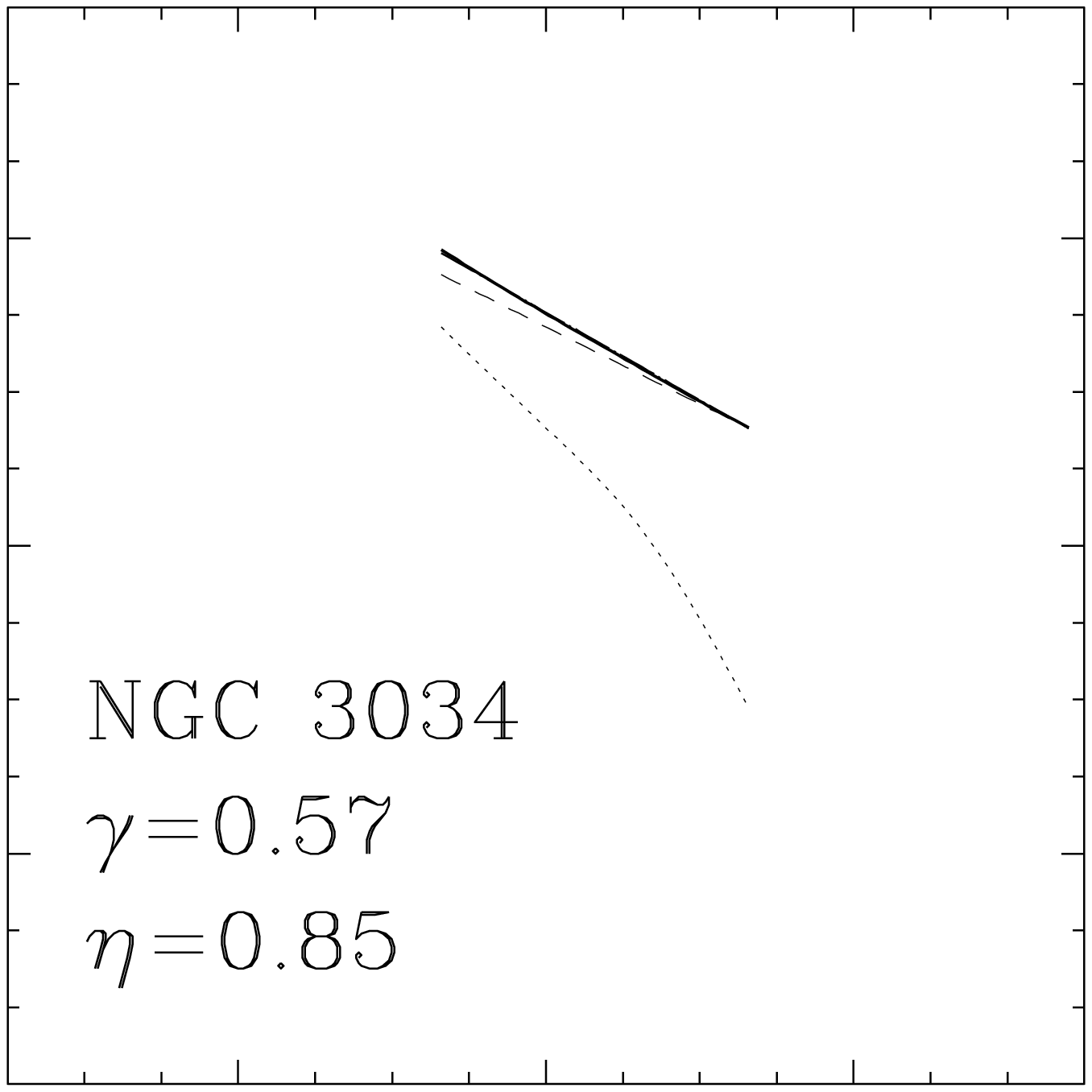}
\includegraphics{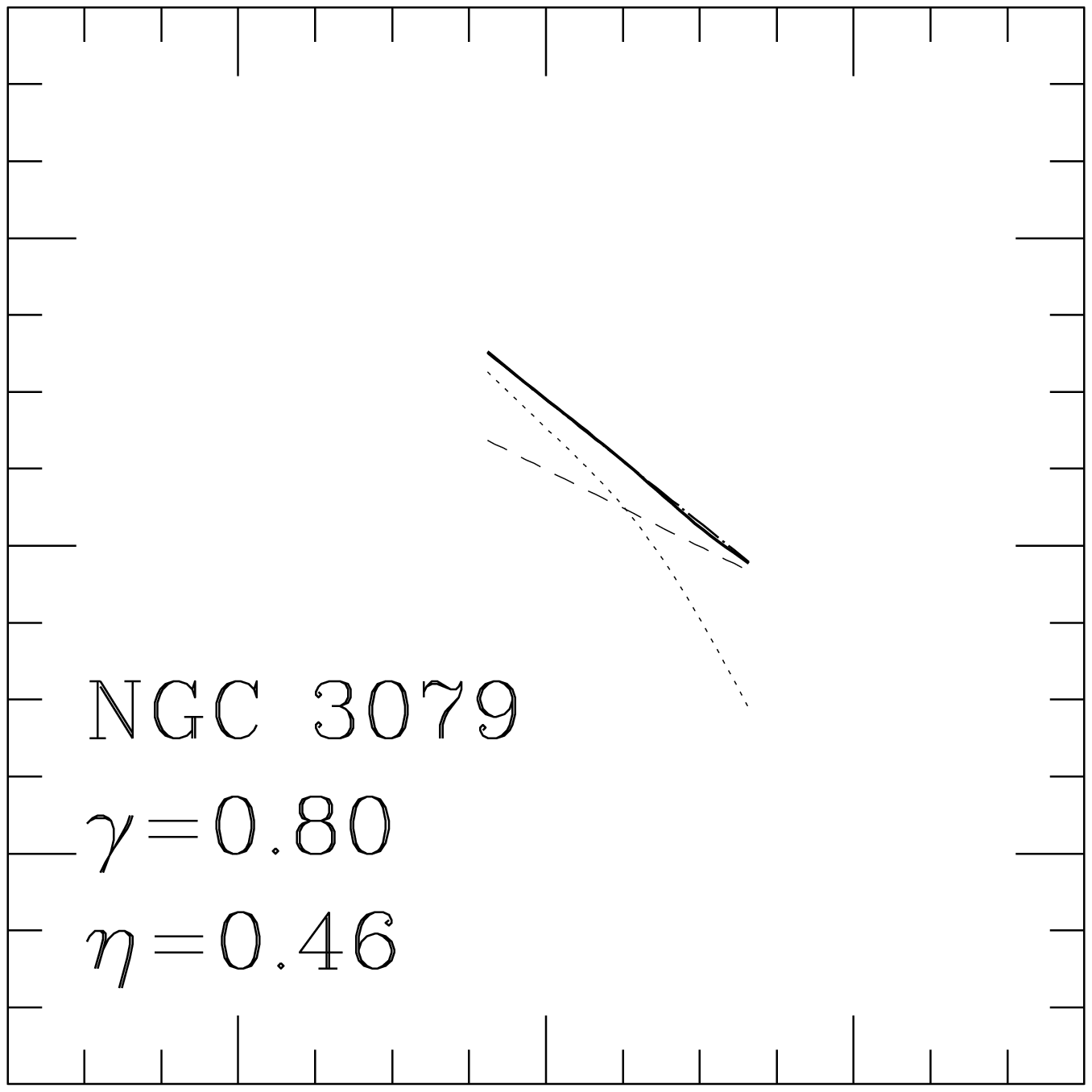}
\vspace{0.1cm}
\vspace{3.0cm}
\includegraphics{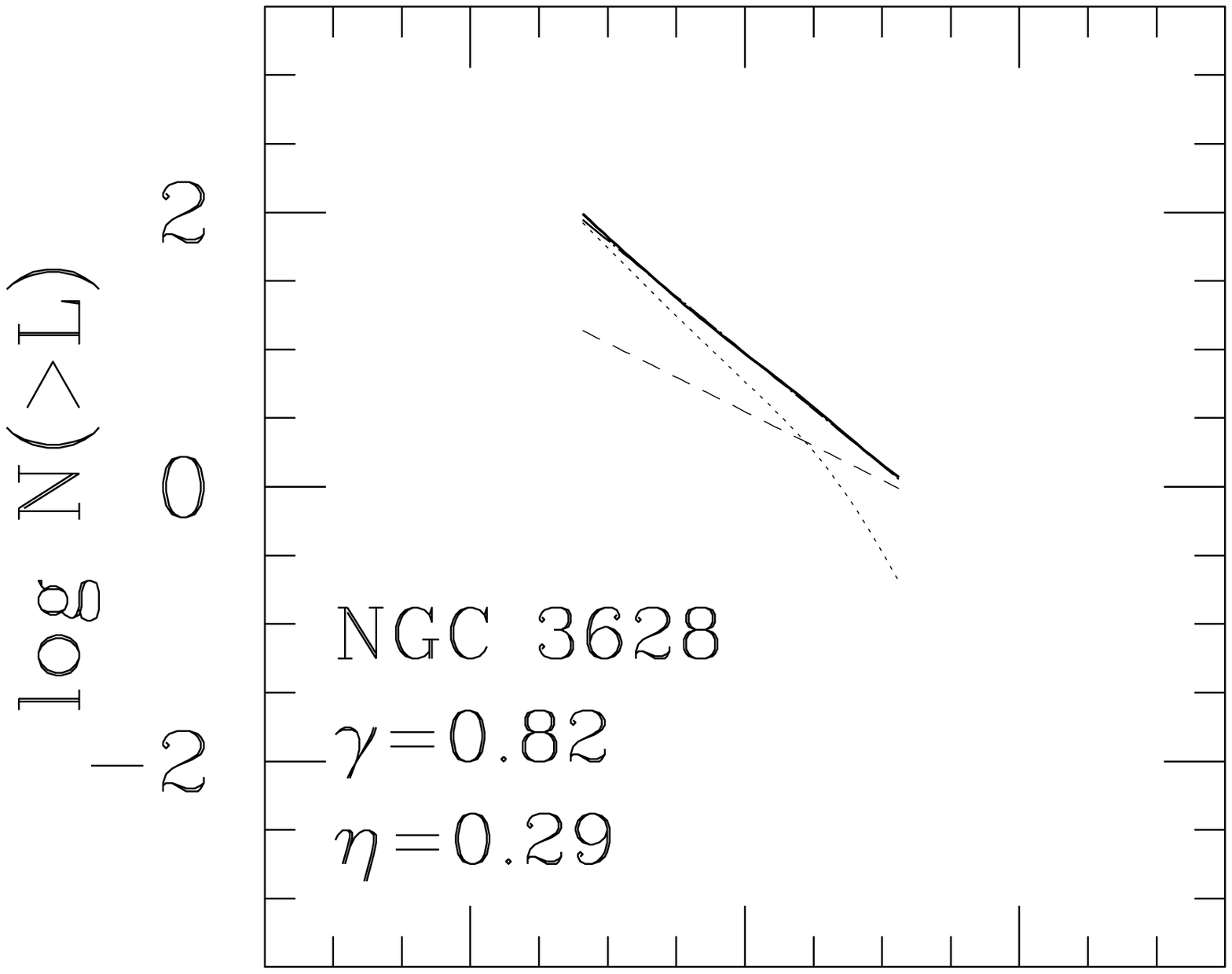}
\includegraphics{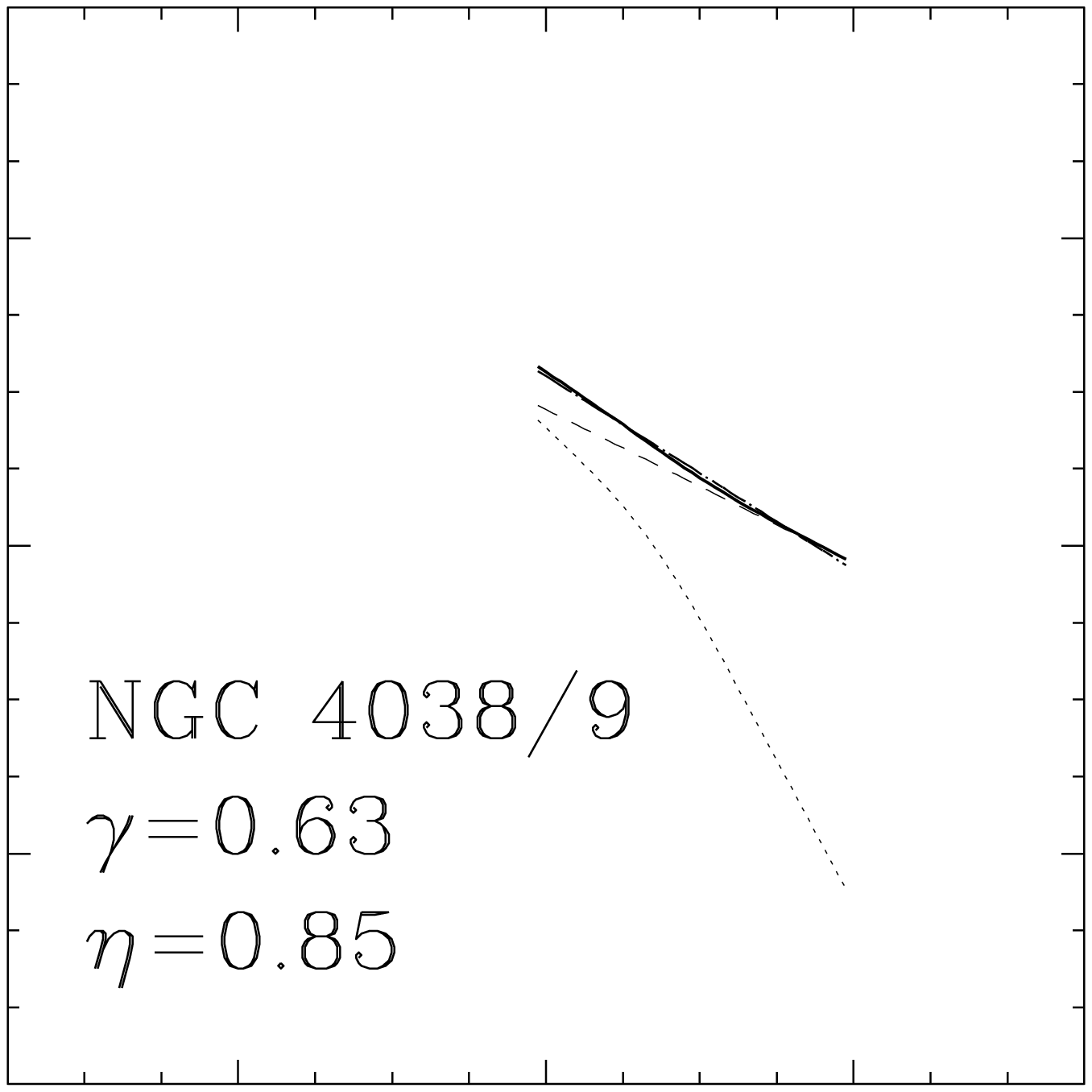}
\includegraphics{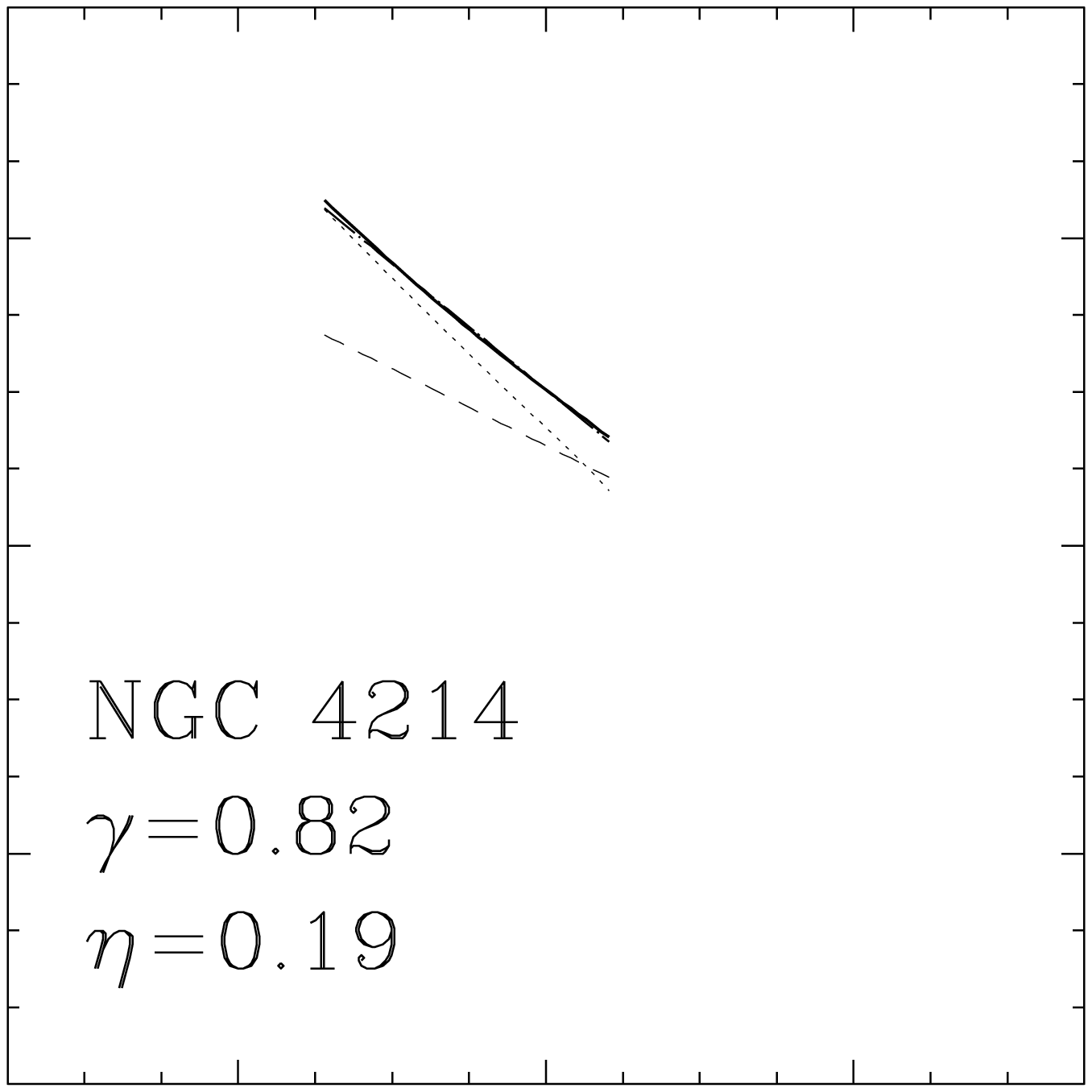}
\includegraphics{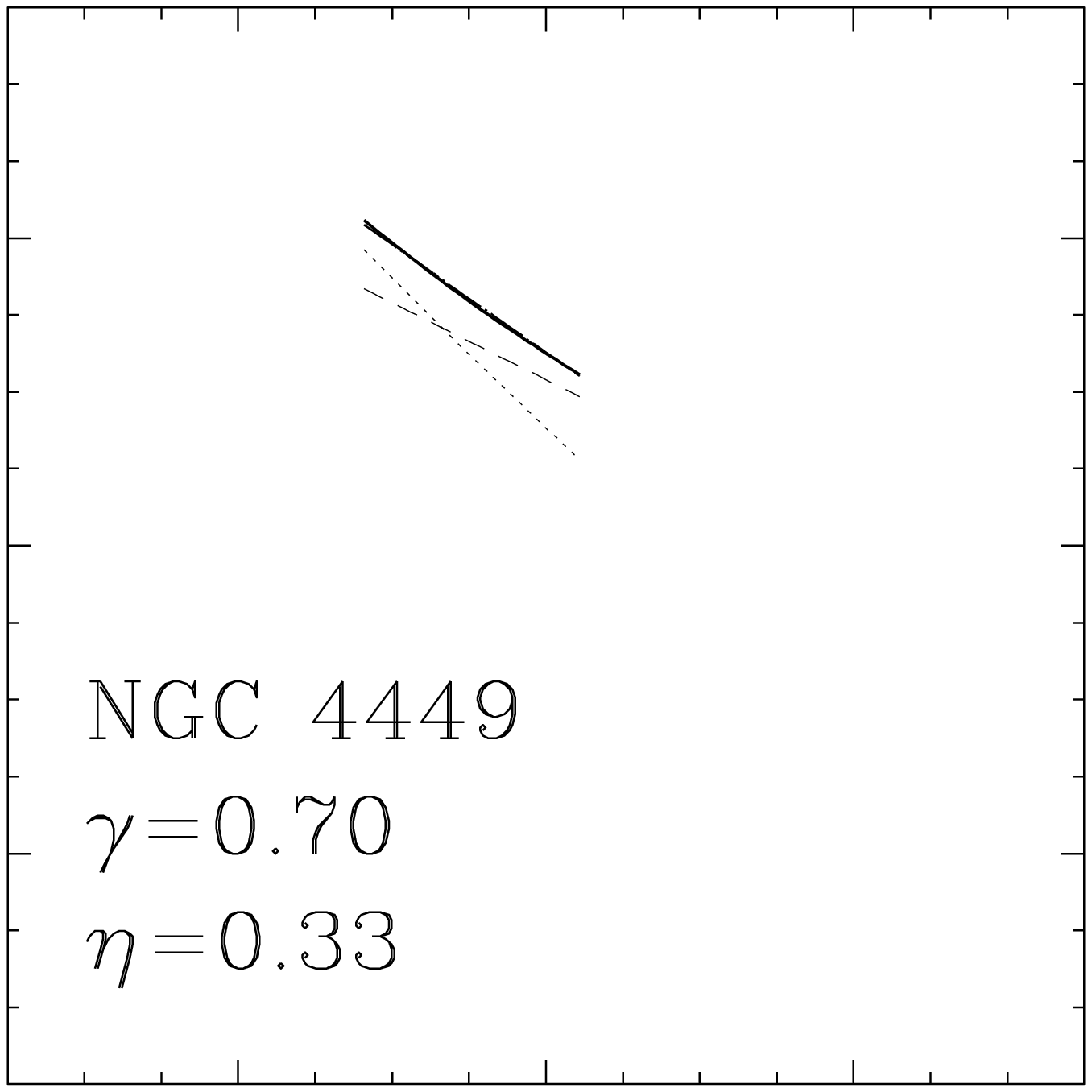}
\includegraphics{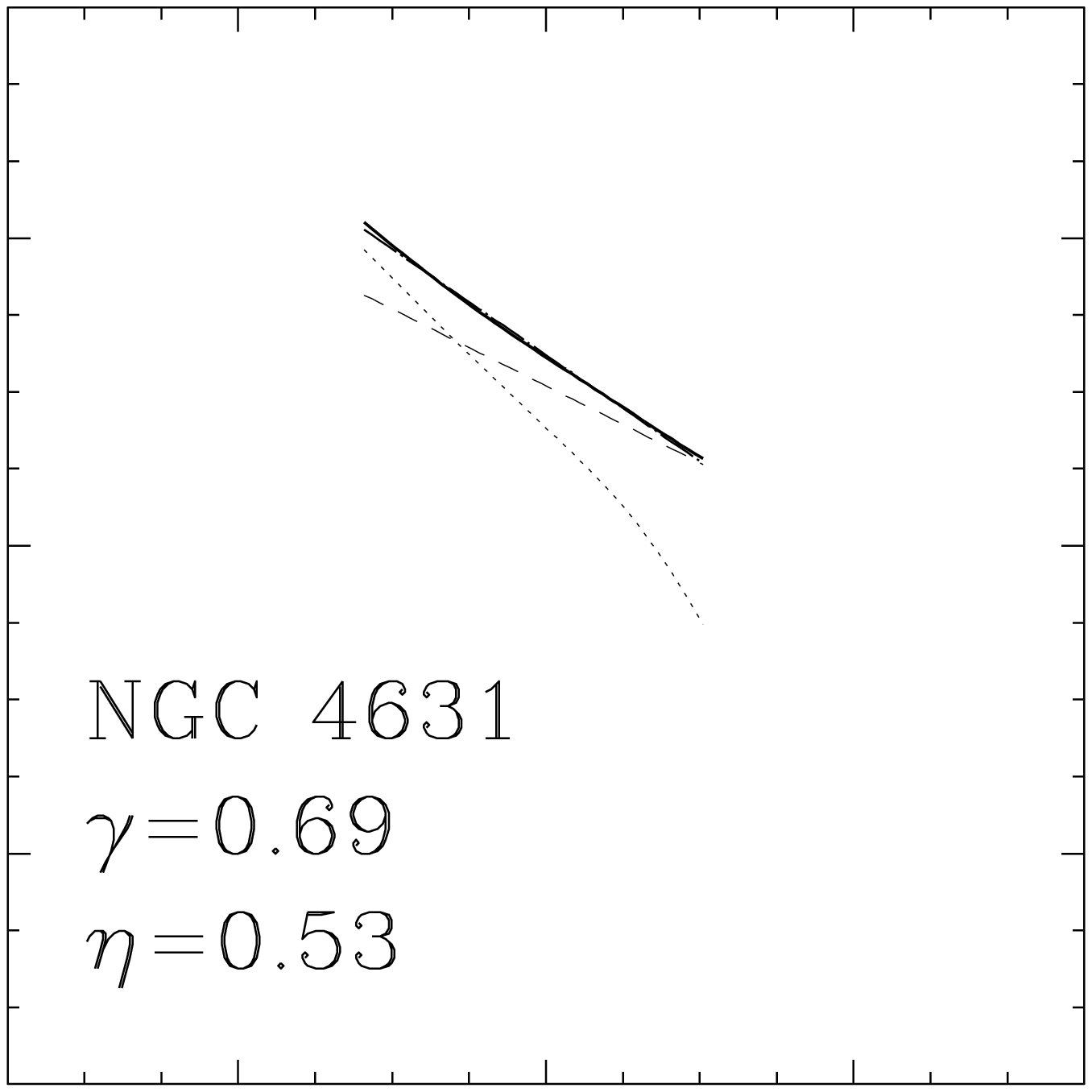}
\includegraphics{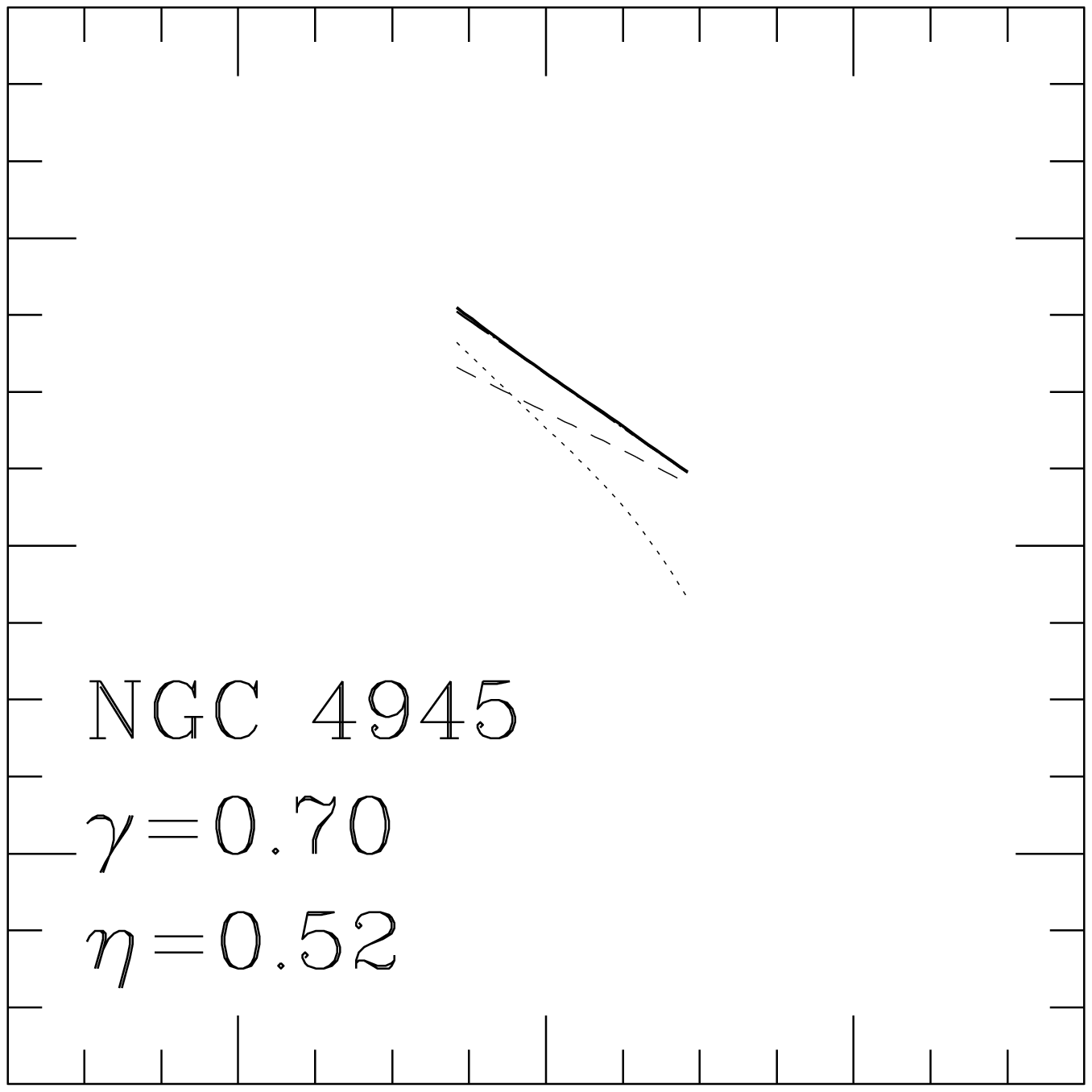}
\vspace{0.1cm}
\vspace{3.0cm}
\includegraphics{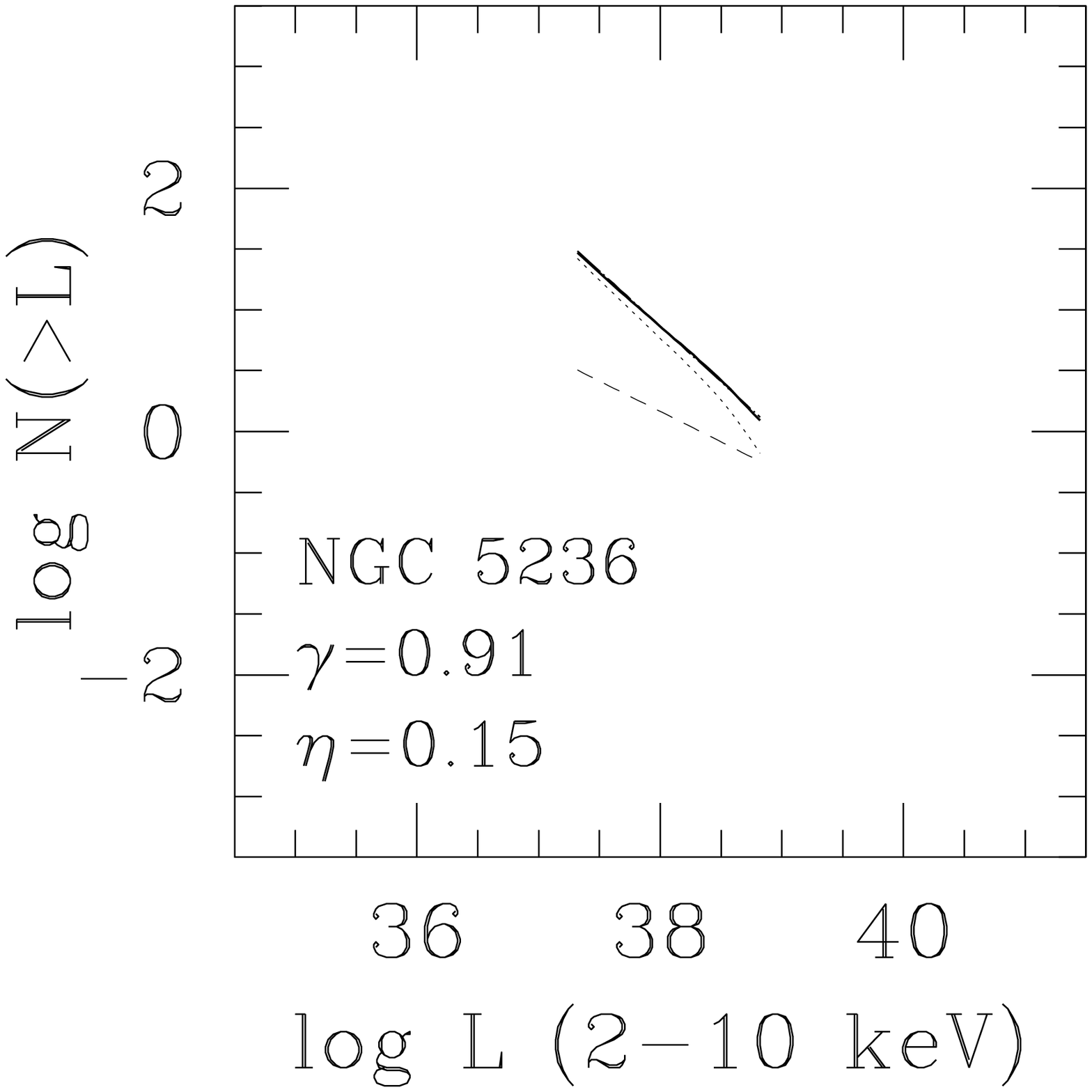}
\includegraphics{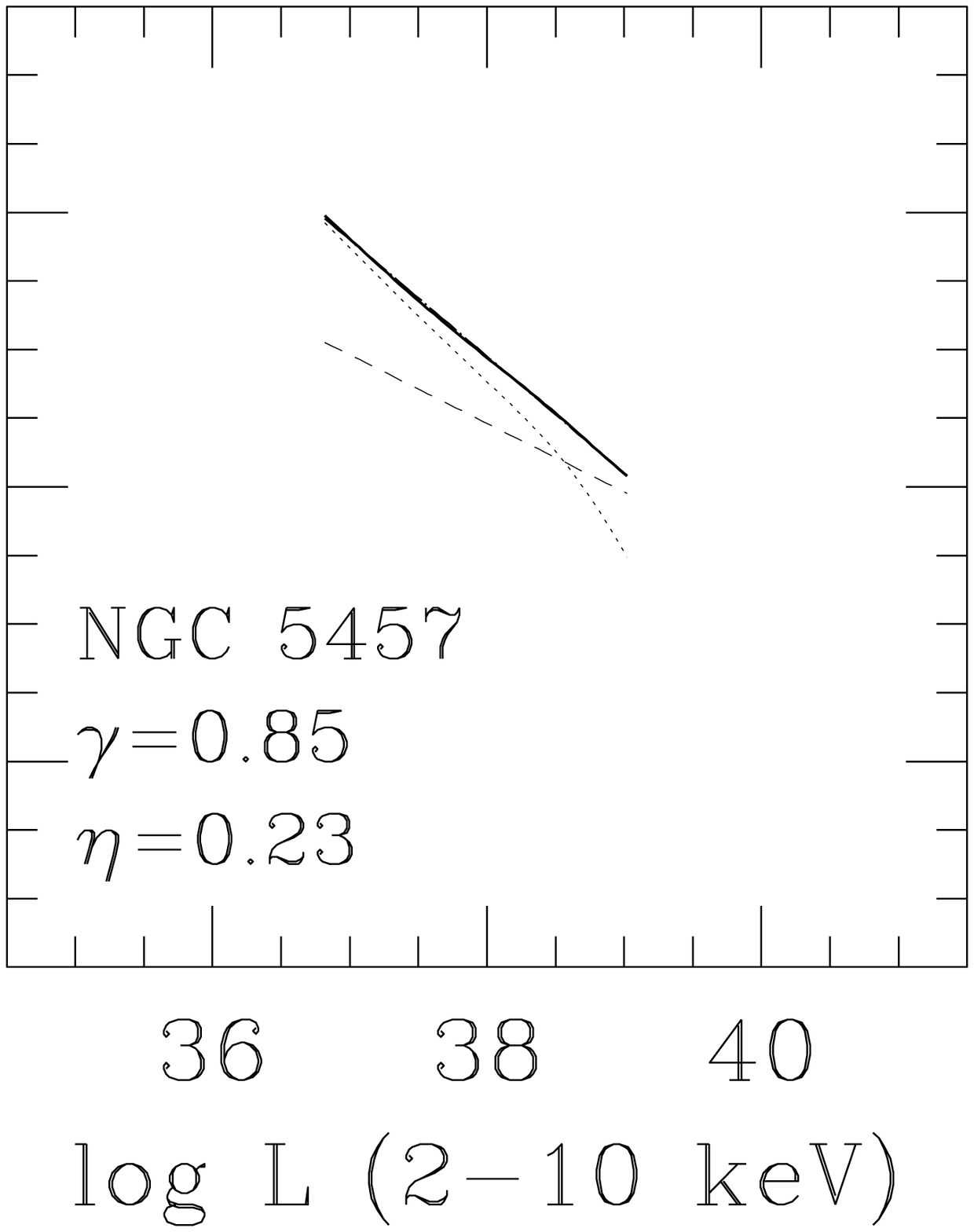}
\includegraphics{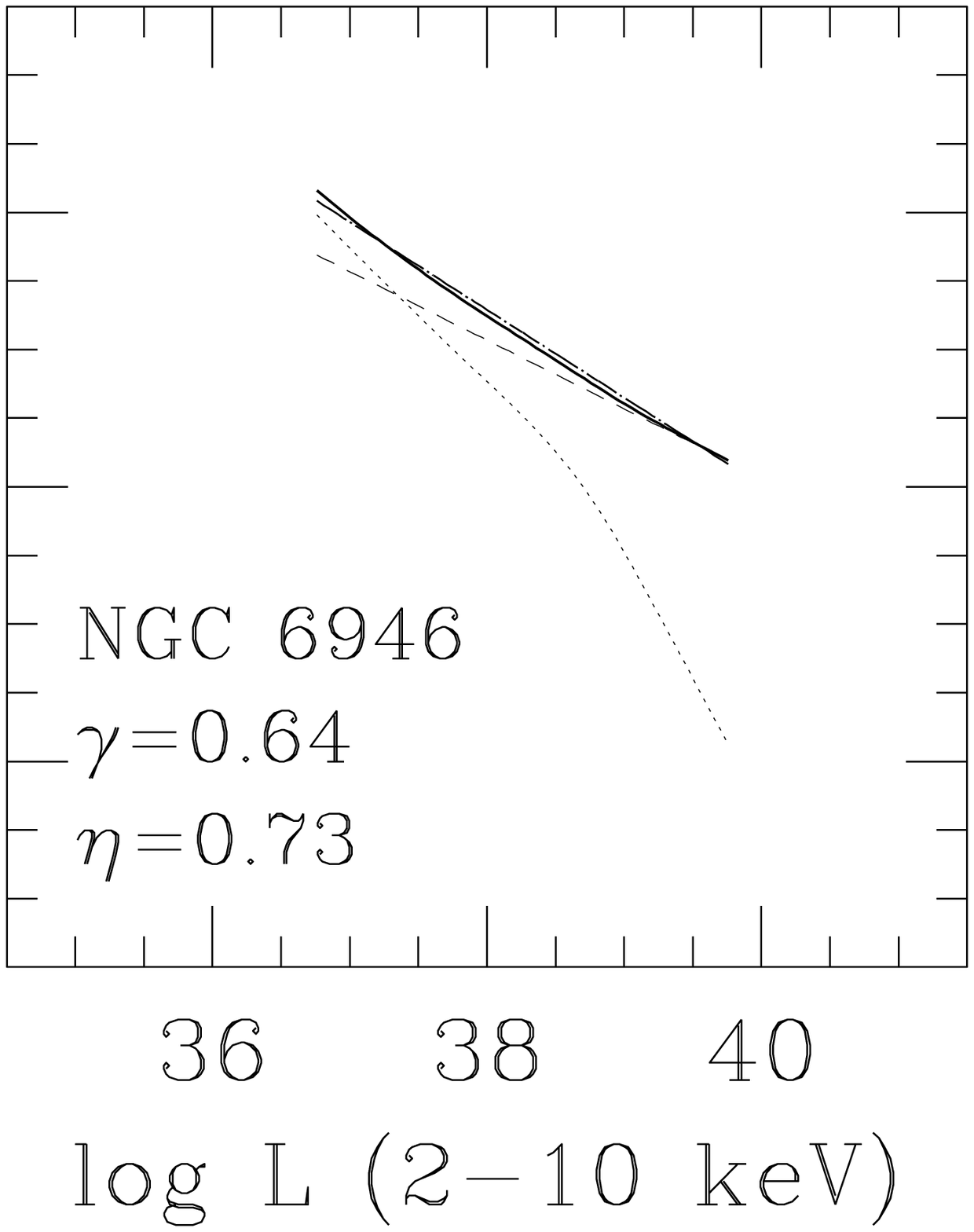}
\includegraphics{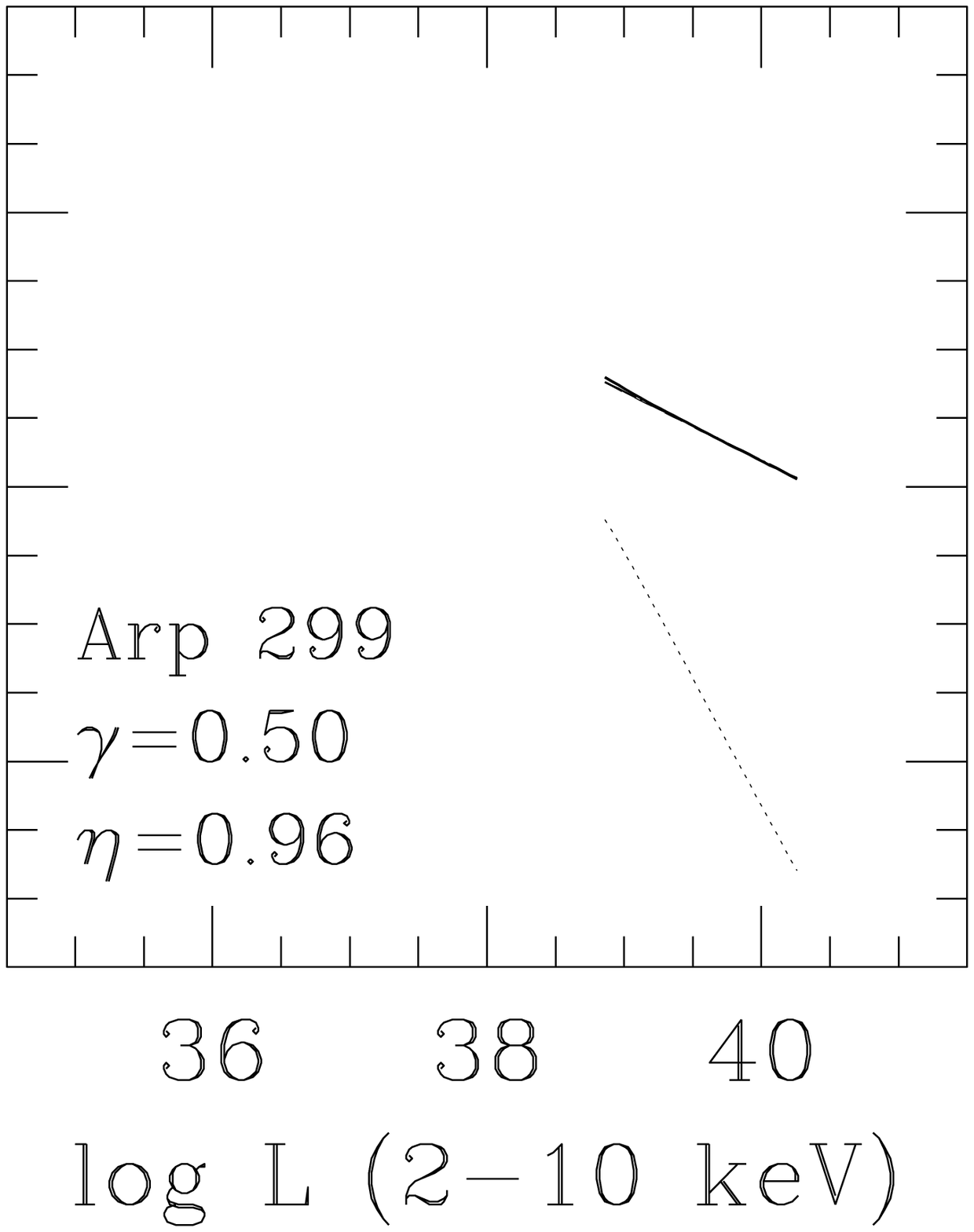}
\includegraphics{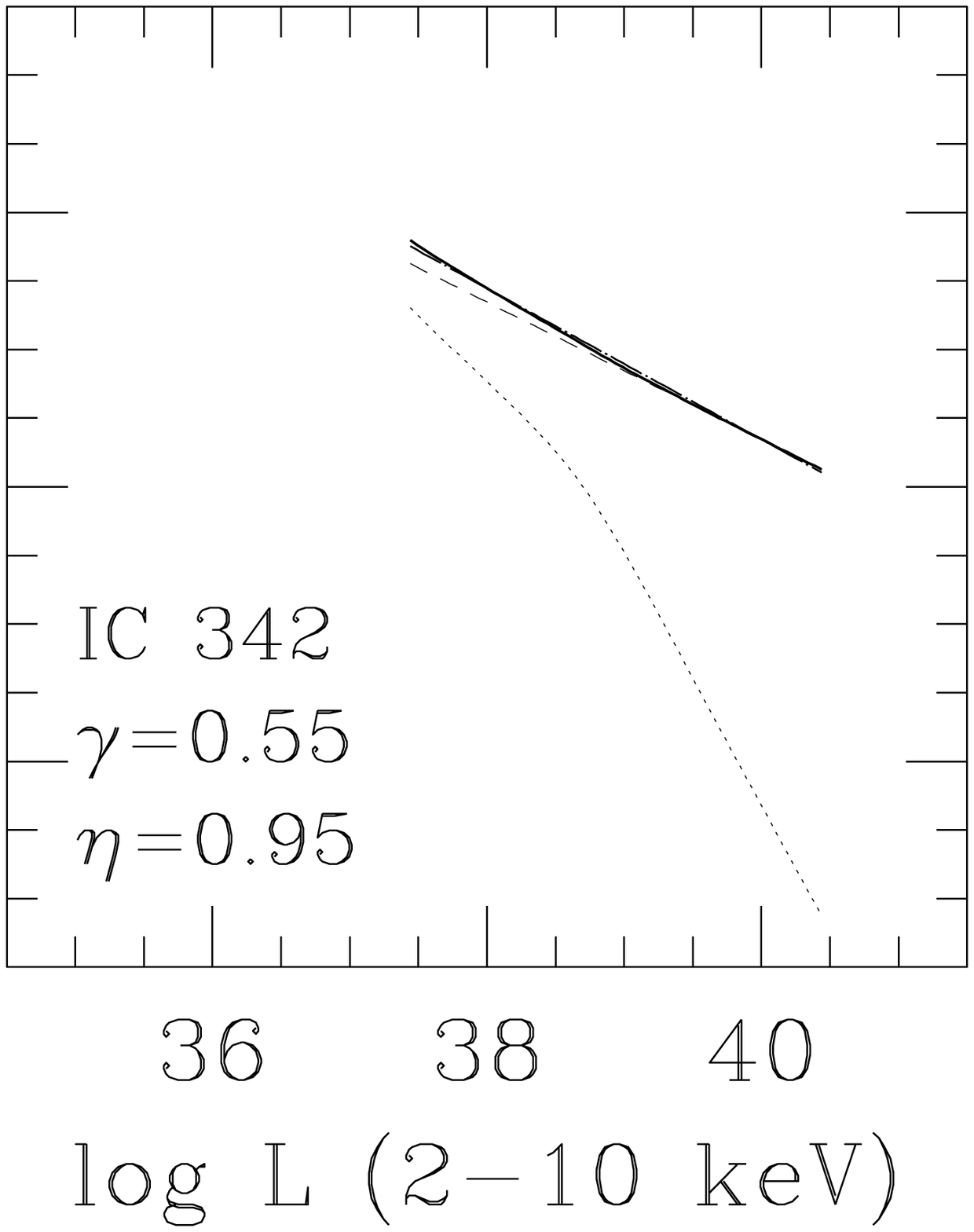}
\vspace{0.1cm}
\caption{
Published fits to measured XPLFs (dot-dashed lines) are modelled as the sum (solid 
line) of a 'young-XPLF' (dashed line; see eq.[1]) and an 'old-XPLF' (dotted line; see 
eq.[2]). In each case, the cumulative old-XPLF component is normalized at the 
break luminosity, $L_{\rm br}$$=$$5$$\times$$10^{38}$ erg s$^{-1}$, and 
its profile can be fitted with the following expression: $(1+x^{-1.02})/(1+x^{1.8})$ 
with $x$$=$$L/L_{\rm br}$. 
}
\end{figure*}

O stars are linked not only to the galactic thermal FIR emission through the
heating of their placental clouds, but also to the galactic nonthermal radio
emission through the acceleration of particles (which emit nonthermal synchrotron 
radiation) during their final SN explosion. This is reflected in a strong 
radio-FIR correlation which can be quantified (Helou et al. 1985) by a 
parameter, 
\begin{eqnarray} 
q_{\rm FIR} ~\equiv~ 
{\rm log} \, \biggl( { f_{\rm FIR} \over \nu_{60\mu{\rm m}} }\biggr) ~-~ 
{\rm log} \, \bigl( f_{\rm 1.4\, GHz} \bigr)   
\end{eqnarray}
(with $f_{\rm FIR}$ in W m$^{-2}$, $\nu_{60\mu{\rm m}}$$=$$3.75$$\times$$
10^{12}$ Hz, and $f_{\rm 1.4\, GHz}$ in W m$^{-2}$ Hz$^{-1}$), that 
turns out to have a value  
\begin{eqnarray}
q_{\rm FIR} ~ \simeq~ 2.35 \pm 0.02
\end{eqnarray}
for local samples (Condon et al. 1991a; Yun et al. 2001; Bell 2003). In 
principle then, the nonthermal radio emission provides us with another 
sensitive tracer of the massive stellar population and hence of the 
instantaneous SFR (Condon 1992). For local SFGs, we adopt the calibration 
between 1.4 GHz luminosity and SFR, 
\begin{eqnarray}
{\rm SFR} ~=~ { L_{1.4} \over 1.61 \times 10^{28} {\rm erg~s^{-1}Hz^{-1}} }
\end{eqnarray}
derived by Schmitt et al. (2006) assuming a Salpeter stellar IMF with mass 
limits 0.1-100 $M_\odot$. Once adjusted for the same mass interval, this 
calibration produces SFRs a factor of $\sim$2 lower than the calibration of 
Condon (1992), which was based on the Galactic relation between nonthermal 
1.4 GHz luminosity and SN rate; however, it is very similar to the more recent 
calibrations of Yun et al. (2001) and Bell (2003). (The above relations in 
eqs.(3)-(7) are, of course, self-consistent.)

\section{Analysis and results}

To evaluate $L_{x}^{\rm yXP}$, we model the measured XPLFs as linear 
combinations of the 'universal' young and old XPLFs (see eqs.[1],[2]). 
The normalization ratio of the young to old differential XPLFs can be 
expressed as a function of the fractional young-XP luminosity, $\eta$, 
according to: 
\begin{eqnarray}
\lefteqn{ { n_{\rm y,0} \over n_{\rm v,0}} ~=~ {\eta \over 1-\eta}  ~
\biggl[ {L_{\rm br}^{2-\beta_{{\rm v,}1}} - L_{\rm min}^{2-\beta_{{\rm v,}1}}
\over 2-\beta_{{\rm v,}1}} ~+~
\biggl( {L_{\rm br} \over L_{\rm min}} \biggr)^{-\beta_{{\rm v,}1}} \times  {} }
                \nonumber\\
& & {} \times {L_{\rm max}^{2-\beta_{{\rm v,}2}} - L_{\rm br}^{2-\beta_{{\rm v,}2}}
\over 2-\beta_{{\rm v,}2}}
\biggr] ~{\bigg /}~\biggl[{ L_{\rm max}^{2-\beta_{\rm y}} - 
L_{\rm min}^{2-\beta_{\rm y}} \over 2-\beta_{\rm y}}\biggr]
\,.
\end{eqnarray}
Since going back to the original XPLF data that are scattered in the 
literature is beyond the scope of the present work, in this paper 
we shall apply the young/old-XP decomposition to the published {\it 
fits} of the observed XPLFs. Only in the cases of NGC~628 and NGC~1569 
did we choose to decompose the actual data. After performing random 
checks, we are confident that introducing this approximation has not 
biased the results of our modeling in any significant way. Our results 
are shown in Figs.2 and 3. 

One further step involves correcting $L_{\rm XP}$ for incompleteness 
of the corresponding XPLFs. In nearby galaxies the XPLFs can be measured 
down to $L_{\rm min}$$\sim$$10^{36}$ erg s$^{-1}$, while in more distant 
galaxies the XPLFs can only be measured down to higher limiting luminosities, 
hence $L_{\rm XP}$ are biased low with distance. We correct for this bias 
-- at least approximately -- by assuming that all XPLFs can be extrapolated 
with the same $\gamma$ down to $L_{\rm min}$$=$$10^{36}$ erg s$^{-1}$. 
This leads to a new, distance-bias-corrected $L_{\rm XP}$. The flat XPLFs 
($\gamma$$<$1, except for NGC~628) ensure that these extrapolations are 
quite reasonable. 

\begin{figure}

\vspace{3.0cm}
\includegraphics{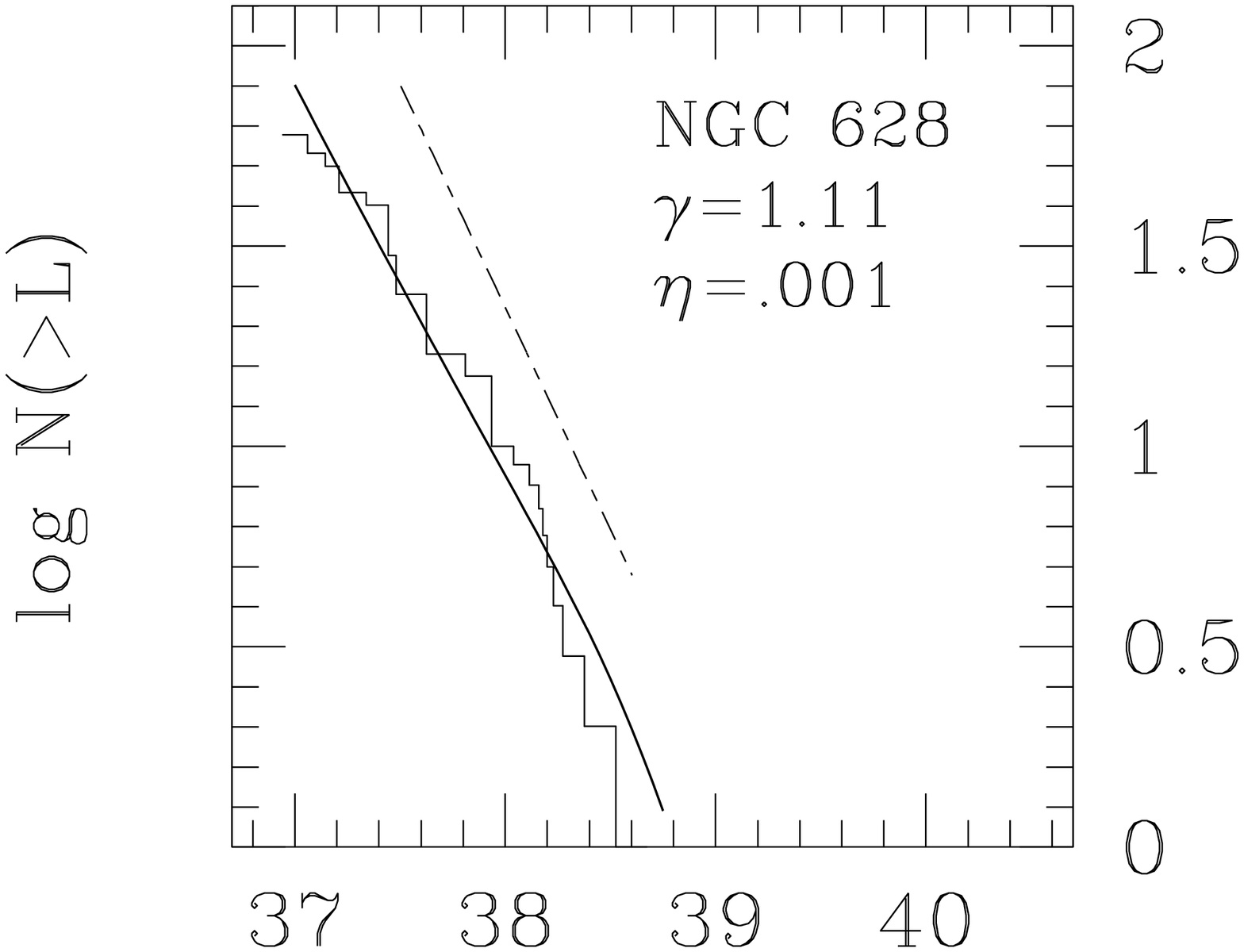}
\includegraphics{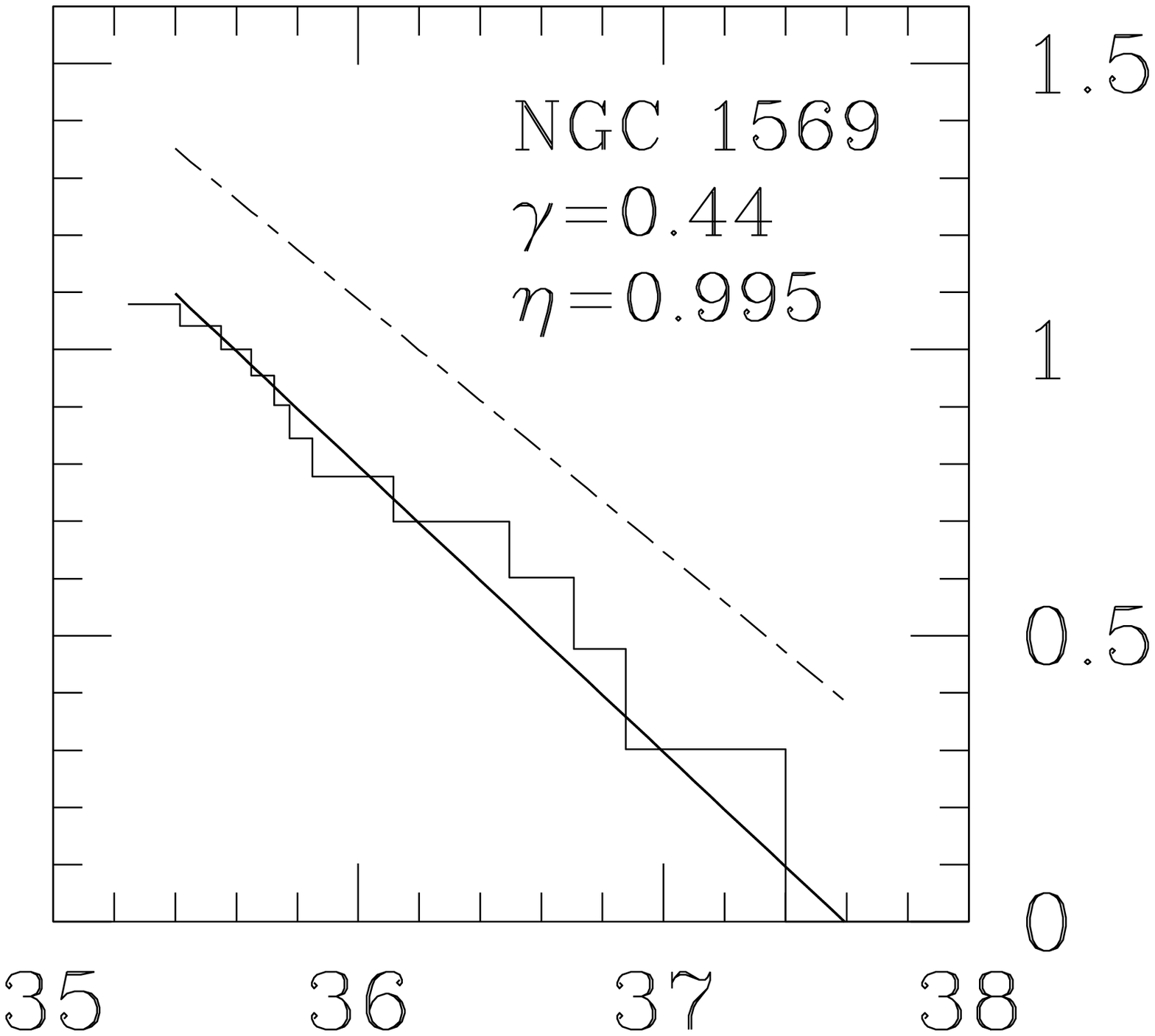}

\vspace{0.1cm}

\vspace{3.0cm}
\includegraphics{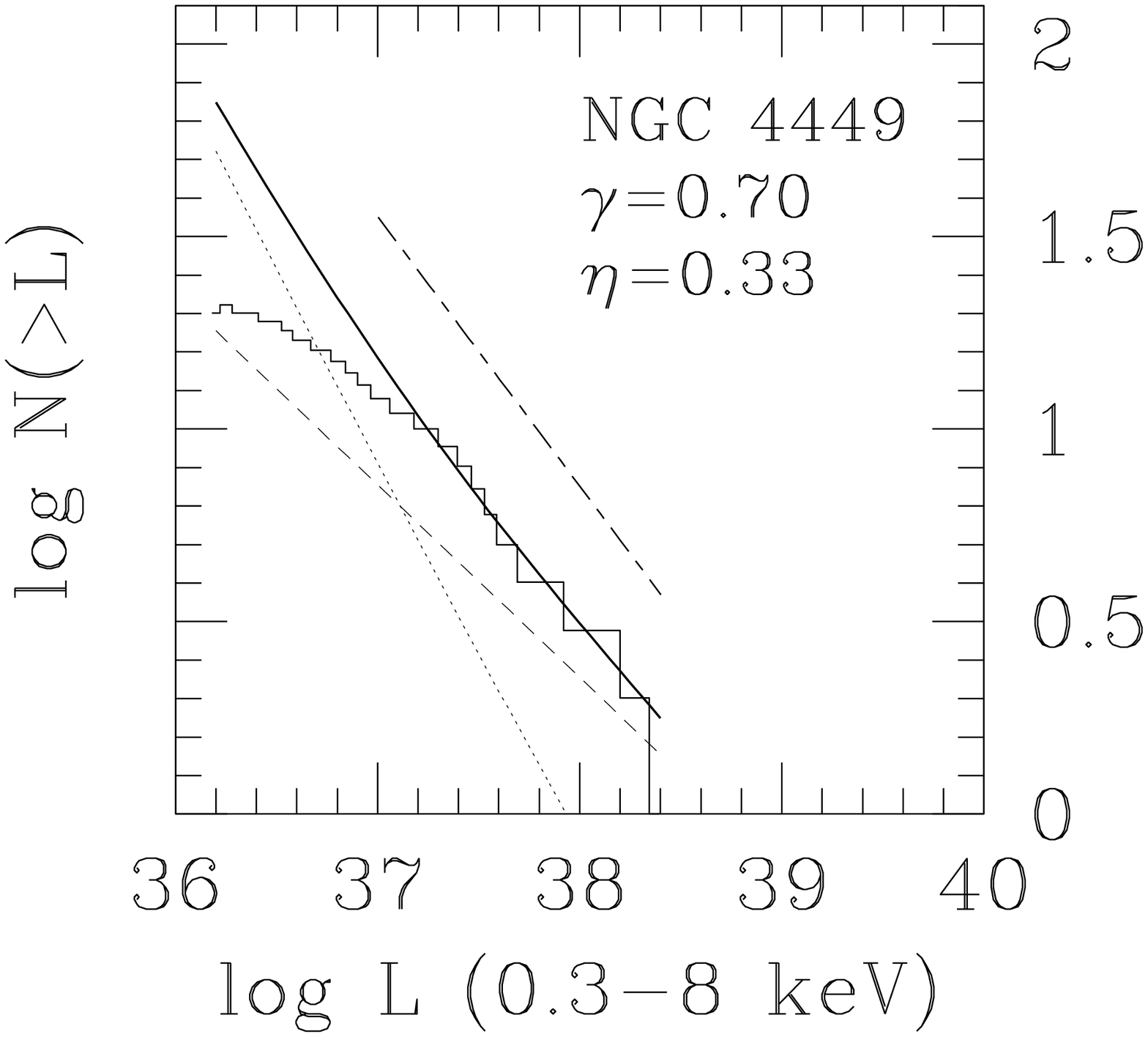}
\includegraphics{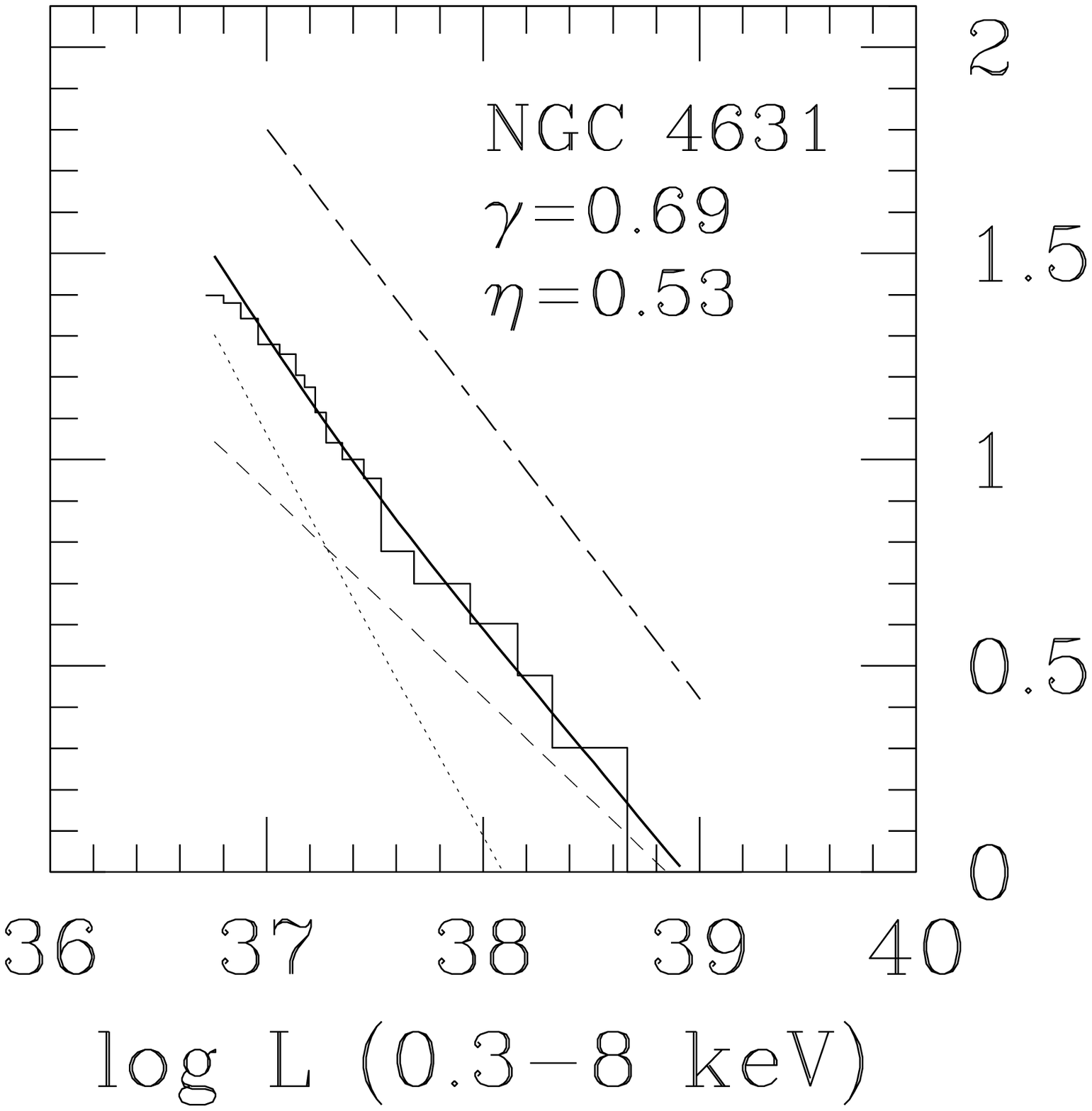}

\vspace{0.4cm}

\caption{ The measured XPLFs of NGC~628 and NGC~1569 (histograms), not shown in Fig.1, 
are modelled as the sum (solid line) of a 'young-XPLF' (dashed line; see eq.[1]) and an 
'old-XPLF' (dotted line; see eq.[2]); their published power-law fits (see Table 1) are 
also shown, as arbitrarily shifted long-short--dashed lines. 
Similar information is displayed for NGC~4449 and NGC~4631 to exemplify our decomposition 
method on original XPLF data (as opposed to the corresponding published XPLF fits, shown 
in Fig.1). In all cases, the power-law fits are displayed over the luminosity ranges where 
they were originally derived (see Table 1). NGC~4449 shows a clear case of incompleteness 
in source counts at low luminosities. All data are from Colbert et al. (2004).
}
\end{figure}

In Fig.4-{\it left} the FIR-deduced SFRs are plotted versus $L_{x}^{\rm yXP}$. 
The data clearly suggest a linear correlation,  
\begin{eqnarray}
{\rm SFR}(>0.1\,M_\odot) ~ = 
~ {L_x^{\rm yXP} \over 0.75 \times 10^{39}{\rm erg\, s}^{-1} } 
~~M_\odot{\rm yr}^{-1}
\end{eqnarray}
(with a $\sim$20\% statistical uncertainty on the calibration). Due to 
its normalization by a ratio of luminosities, the above relation is 
independent of the adopted cosmological model.

\begin{figure}

\vspace{15.0cm}
\includegraphics{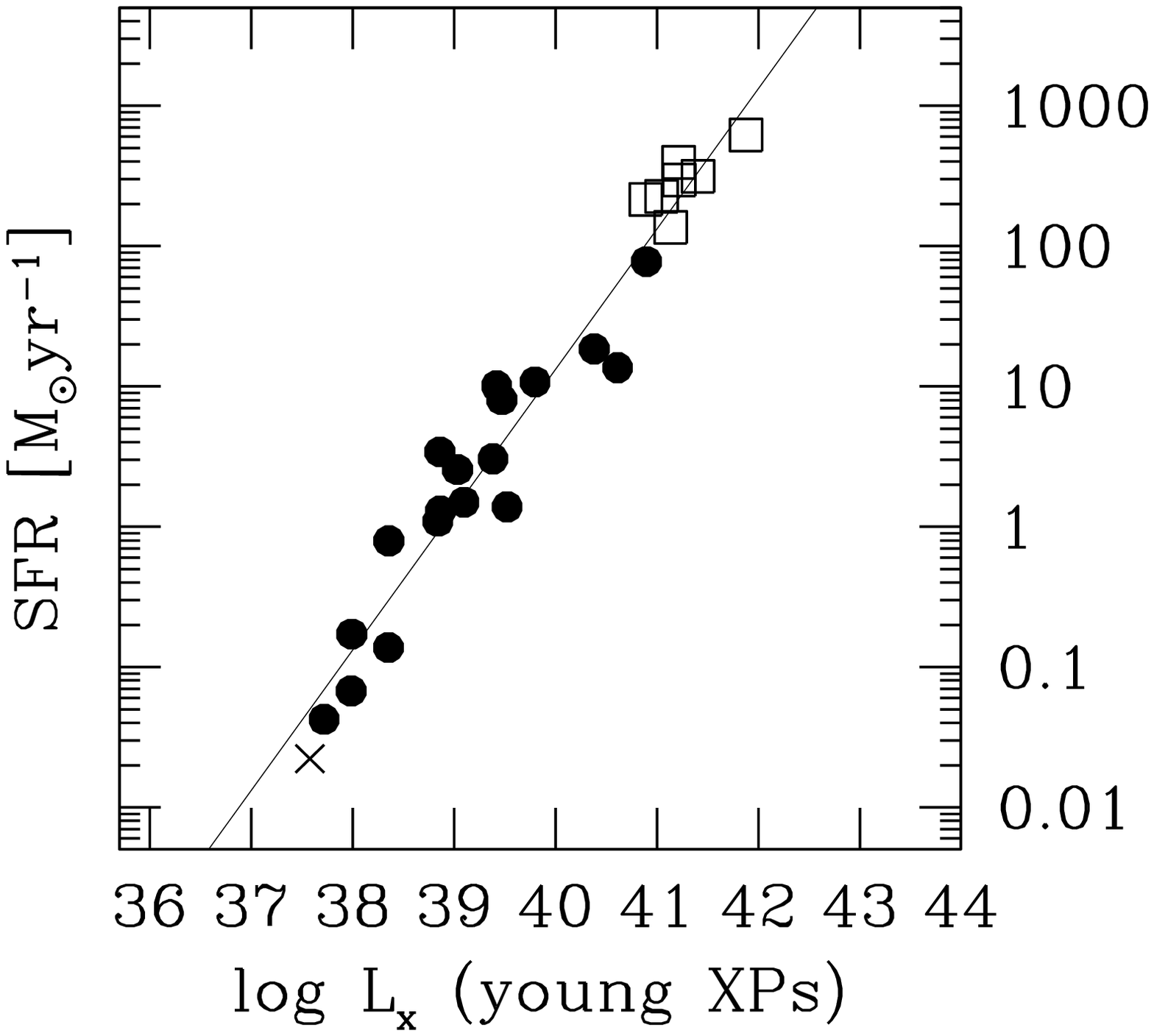}
\hspace{.1cm}
\includegraphics{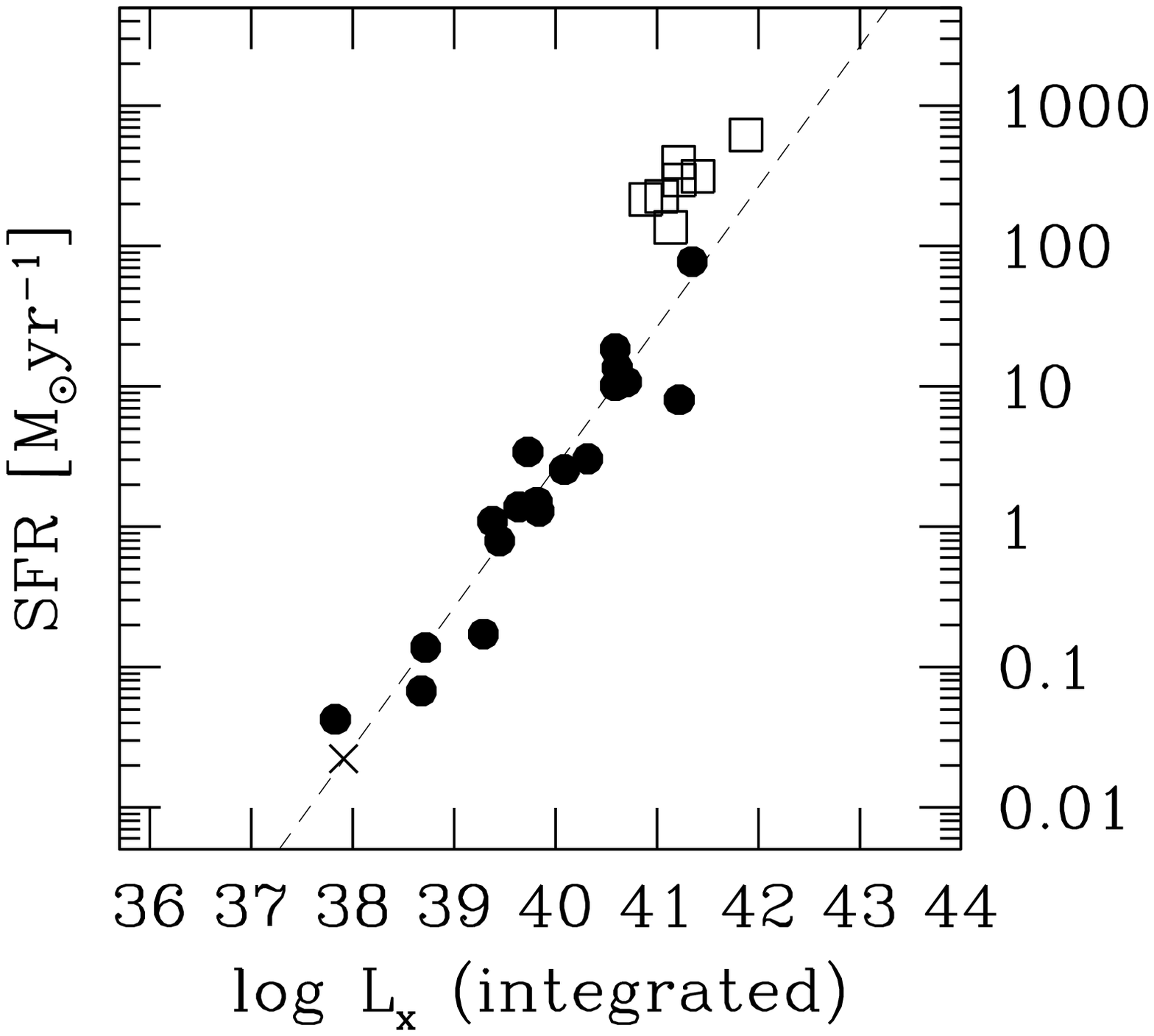}
\caption{ 
The FIR-based SFR versus 2-10 keV luminosity relation, using the total luminosity ({\it bottom}) 
and the collective luminosity of young XPs ({\it top}) for the local sample of star-forming 
galaxies (filled circles: see Tables 1,2; NGC~3077 is represented by a cross) and the more 
distant sample of starburst-ULIRGs (empty squares: see Table 3). In the latter set of 
high-SFR galaxies the emission is plausibly due to young XPs, hence the total 2-10 keV 
luminosities, $L_x$, are used in both panels. The solid line in the top panel shows the 
relation in eq.(9); the dashed line in the bottom panel shows the relation in eq.(10). 
}

\end{figure}

\section{Discussion}

In this section we discuss some issues of relevance to the viability and use 
of $L_{x}^{\rm yXP}$ as a SFR estimator. Specifically, we 
discuss properties of shape and calibration, and some underlying 
uncertainties, of the SFR--$L_{x}^{\rm yXP}$ relation.

\subsection{Linearity}

Our SFR--$L_{x}^{\rm yXP}$ relation appears to be linear over about 5 
decades in luminosity and SFR. This result is consistent with that of Colbert 
et al. (2004) who, using a sample of 32 elliptical and spiral galaxies, 
found that $L_{\rm XP}$ was linear in both galaxy stellar mass and SFR 
(measured by {\it K}-band and FIR+UV luminosities, respectively). 
In our sample, however, poor statistics prevent a full assessment of the 
relation in the very-low-SFR regime. 

In an attempt to gain furher insight, we examined the behaviour of the 
very-low-SFR galaxy, NGC~3077, in the $L_{x}^{\rm yXP}$--SFR plane. No 
published XPLF is available for this galaxy, but the measured spectra of 
its 6 detected XPs (Ott et al. 2003) can be used to single out young sources. 
The very soft sources S1, S5, and S6 are proposed by Ott et al. to be SNRs, 
and the hard source S3, speculated by Ott et al. to be an accreting binary, 
is spectrally consistent ($\Gamma$$\sim$1) with being a HMXB. 
It was suggested by Ott et al. that S2 is an 
accreting binary or a background AGN: in the former case, its $\Gamma$$\sim
$1.65 slope may suggest a LMXB interpretation. The supersoft source S4 is 
argued by 
Ott et al. to be either a hydrogen-burning white dwarf or an isolated NS. 
If the identification of 4 young XPs is correct, then NGC~3077 
approximately agrees with the SFR--$L_{x}^{\rm yXP}$ relation defined for 
our composite sample (see cross in Fig.2-{\it left}). 

\begin{table*}
\caption[] {Data II: starburst ULIRGs.}
\begin{flushleft}
\begin{tabular}{ l l l l l l l l l l l l l l }
\noalign{\smallskip}
\hline
\hline
\noalign{\smallskip}
Source & $~~z$ & ~~~D$^{(a)}$ & $~~~~~~f_{x}^{(b)}$ & $~~~ L_{x}^{(c)}$ &$~f_{12\mu}^{(d)}$ & $~f_{25\mu}^{(d)}$ & 
$~~f_{60\mu}^{(d)}$ & $~~f_{100\mu}^{(d)}$ & 
$L_{\rm FIR}^{(e)}$ & $L_{\rm IR}^{(f)}$ & SFR$^{(g)}$ & $f_{1.4}^{(h)}$ & $L_{1.4}^{(i)}$\\
\noalign{\smallskip}
\hline
\noalign{\smallskip}
IRAS~12112+0305 & 0.072 & 293.0 & $-13.79$ & 41.22 & 0.08  & 0.66 &  ~~8.18 & ~~~9.46 & 45.60 & 45.81 & 297 &     &     \\
IRAS~14348-1447 & 0.082 & 334.5 & $-13.72$ & 41.41 & 0.07  & 0.55 &  ~~6.82 & ~~~7.31 & 45.62 & 45.84 & 315 &     &     \\
IRAS~15250+3609 & 0.055 & 222.9 & $-13.64$ & 41.13 & 0.16  & 1.31 &  ~~7.10 & ~~~5.93 & 45.26 & 45.48 & 136 &0.013& 29.89\\
Arp~220         & 0.018 &~~72.3 & $-12.74$ & 41.06 & 0.61  & 8.00 &  104.0  & ~112.0  & 45.48 & 45.69 & 225 & 0.30& 30.27\\
IRAS~17208-0014 & 0.043 & 173.8 & $-13.34$ & 41.22 & 0.20  & 1.61 &  ~~9.53 & ~11.05  & 45.21 & 45.43 & 122 & 0.10& 30.56\\
IRAS~20100-4156 & 0.129 & 531.6 & $-13.65$ & 41.88 & 0.1   & 0.57 &  ~~5.44 & ~~~5.41 & 45.90 & 46.12 & 595 & 0.02& 30.83\\
IRAS~22491-1808 & 0.078 & 317.6 & $-14.19$ & 40.89 & 0.06  & 0.54 &  ~~5.54 & ~~~4.64 & 45.46 & 45.68 & 216 &     &     \\
\noalign{\smallskip}
\hline
\hline
\end{tabular}
\end{flushleft}
\smallskip

$^{(a)}$ Luminosity distances (in Mpc), computed from redshifts according to 
$D_{\rm L}$$=$$(2c/{\rm H}_0 \Omega_0^2)$ $[\Omega_0 z$$+$$(\Omega_0$$-$$2)\,$$
(\sqrt{1+\Omega_0 z}$$-$$1)]$ (Peacock 1999). 

$^{(b)}$ 2-10 keV fluxes (in erg cm$^{-2}$s$^{-1}$ and given in log form), 
from Iwasawa et al. (2001) for Arp~220 and from Franceschini et al. (2003) 
for the remaining sources. 

$^{(c)}$ 2-10 keV luminosities (in erg s$^{-1}$), in log form.

$^{(d)}$ {\it IRAS} flux densities (in Jy), from Genzel et al. (1998) and 
Sanders et al. (1995) for 20100-4156 and from Sanders et al. (2003) for 
the remaining sources. 

$^{(e)}$ FIR luminosities (in erg s$^{-1}$), in log form. 

$^{(f)}$ IR luminosities ($L_{\rm IR}$$\equiv$$1.65$$\times$$L_{\rm FIR}$; in 
erg s$^{-1}$), in log form. 

$^{(g)}$ Star-formation rates (from eq.3), in $M_\odot$ yr$^{-1}$. 

$^{(h)}$ 1.4~GHz flux densities (in Jy), from White \& Becker (1992) for 
IRAS~15250+3609 and from Condon et al. (1996) for the remaining sources.

$^{(i)}$ 1.4~GHz luminosities (in erg s$^{-1}$ Hz$^{-1}$), in log form.

\end{table*}

NGC~2403 is another galaxy with very low SFR, potentially useful to 
investigate the low-$L$ behavior of the SFR--$L_{x}^{\rm yXP}$ relation. 
However, it seems to be a problematic object. As noted by Schlegel \& 
Pannuti (2003) and discussed also by Fabbiano (2005), NGC~2403 is X-ray 
overluminous for its FIR luminosity and apparently violates Kilgard 
et al.'s (2002) correlation between XPLF-slope and FIR-based SFR. These 
apparent contradictions may be resolved assuming that SF in NGC~2403 
turned off a few $10^6$ yr ago: the FIR emission in the star-forming 
clouds, powered by OB stars, would be drastically reduced by now, 
whereas the HMXBs would still be shining (Schlegel \& Pannuti 2003). We 
suggest an alternative possibility to reconcile NGC~2403's X-ray and 
FIR properties. Spectral modelling of the 4 brightest XPs of NGC~2403 
(Schlegel \& Pannuti 2003) suggests a LMXB nature for two of these 
(sources 1 and 28), and a BHXB nature for the other two (sources 20 and 
21). For the latter we also suggest a low-mass donor interpretation, 
based on the galaxy's very low SFR and the lack of spatial correlation 
with likely SF sites (see Schlegel \& Pannuti 2003). If LMXB-related, 
the 4 brightest sources of NGC~2403 would no longer be unclassified and 
hence should be removed from the unclassified (as young or old) list of 
XPs. (Being old sources, these four sources would of course be unsuitable 
to trace the ongoing SF.) Of the remaining 37 less luminous unclassified 
XPs, distributed now in a truncated XPLF, 12 are estimated to be interlopers 
(apparently distributed randomly with luminosity, see Schlegel \& Pannuti 
2003). Finally, 25 sources are left as the galaxy's XP population out of 
which young sources have to be extracted by XPLF modelling. The young-XP 
luminosity estimated with our approach does comply with the SFR--$L_{x}^
{\rm yXP}$ relation (see Fig.2-{\it left}). Further detailed study of 
NGC~2403 and its XP population is clearly much needed, both for its 
intrinsic interest, and for investigating the lowest-$L$ reaches of the 
SFR--$L_x^{\rm yXP}$ relation.

An analogous study of the use of young XPs (specifically: HMXBs) as SFR 
indicators led Grimm et al. (2003) to suggest a non-linear regime, SFR 
$\propto$ (luminosity)$^{0.6}$ for SFR$\mincir$$4.5\,M_\odot$ yr$^{-1}$,
in the SFR--X-ray-luminosity relation. This behaviour was attributed by 
Gilfanov et al. (2004b) to effects of low-numbers statistics in the 
distribution of XPs in low-$L_x$, low-SFR galaxies. However, we suggest 
that differences in our sample selection and that of Grimm et al. may 
also play a role. The procedure of Grimm et al. involved {\it a priori} 
selection of galaxies with high SFR--to--stellar-mass ratios (based on 
dynamical estimates for the stellar masses of galaxies, and SFRs derived 
from a variety of indicators) to ensure that $L_x$ would be HMXB-dominated 
and hence a tracer of the ongoing SFR. In contrast, we use all galaxies 
with measured source counts: by decomposing their XPLFs into young and 
old component, we {\it a posteriori} obtain the young-XP luminosity. 

Given this uncertain situation, the faint limit of the SFR--$L_{x}^{\rm 
yXP}$ relation clearly needs further investigation.

\subsection{Calibration}

The calibration of our SFR--$L_{x}^{\rm yXP}$ relation, 
(0.75$\pm$0.15)$\times$10$^{39}$ erg s$^{-1}$ per $M_\odot$yr$^{-1}$, is 
compatible with that, $\sim$$10^{39}$ erg s$^{-1}$ per $M_\odot$yr$^{-1}$,
derived by Grimm et al. (2003) using a variety of SFR estimators. Grimm et 
al. used $L_x$ in their relation, but for their sample, which was selected 
such that $L_x$ would be largely HMXB-dominated (see section 6.1), 
$L_x$$\sim$$L_{x}^{\rm yXP}$ by construction. 

Our calibration agrees also with the corresponding calibration of Colbert 
et al.'s (2004) bilinear correlation between $L_{\rm XP}$ and host-galaxy 
SFR and stellar mass. To see this we should account for the different 
definition of the variables, SFR and X-ray luminosity, in Colbert et al.'s 
eq.(7) and in our eqs.(3),(4). Specifically, we convert their $L_{\rm 
FIR+UV}$-based definition of SFR into our adopted $L_{\rm IR}$-based 
definition, which gives SFR$_{\rm FIR+UV}$$=$$1.067\,$SFR$_{\rm IR}$, and 
transform their 0.3-8 keV luminosities into our 2-10 keV luminosities 
(assuming with Colbert et al. $\Gamma=1.7$ PL spectra), which gives 
$L_{0.3-8}$$=$$1.529\,$$L_{2-10}$. In terms of our variables, Colbert 
et al.'s bivariate relation can be rewritten as 
$$
L_{\rm XP} - (0.85\pm 0.13) \times 10^{29} \,M = 
(0.49 \pm 0.21) \times 10^{39} \, {\rm SFR} 
$$
(with 2-10 keV XP luminosities in erg s$^{-1}$, masses in $M_\odot$, and 
IR-derived SFR in $M_\odot$ yr$^{-1}$). Comparing this expression with 
eq.(9) one sees that: 

\noindent
{\it (i)} our young-XP luminosity corresponds to Colbert et al.'s XP 
luminosity {\it minus} a quantity, proportional to the galaxy's stellar 
mass, that represents the old-XP luminosity: 
$L_x^{\rm yXP} = L_{\rm XP} - (0.85 \pm 0.13) \times 10^{29} M$; and 

\noindent
{\it (ii)} once adjusted for the same definitions of SFR and luminosity, 
Colbert et al.'s calibration, $(0.49 \pm 0.21) \times 10^{39}$ erg s$^{-1}$ 
per $M_\odot$yr$^{-1}$, is consistent with ours.

\subsection{Uncertainties}

The main uncertainties concern the precision with which our adopted 
young-XPLF and old-XPLF have been detemined. 
\medskip

\noindent
{\it Old-XPLF.} The old-XPLF proposed by Kim \& Fabbiano (2004), though the 
most updated and reliable available, extends in luminosity down to only 
$\sim$5$\times$10$^{37}$ erg s$^{-1}$, whereas measured XPLFs often reach 
down to $\sim$$10^{36}$ erg s$^{-1}$. In our analysis we chose 
to extrapolate Kim \& Fabbiano's function down to $\sim$$10^{36}$ 
erg s$^{-1}$, but this assumption may not be fully realistic. 
A flattening of the old-XPLF at $L \mincir 10^{37}$ erg s$^{-1}$, with a 
differential slope of $\sim$1, is suggested by counts of bulge LMXBs of 
some nearby spirals (Gilfanov 2004) and of XPs in NGC~5128 (Voss \& 
Gilfanov 2005). However, the situation concerning the low-$L$ XPLFs may 
be rather complicated. For example, in M~31's very well studied LMXB 
population clear differences are seen between bulge and globular-cluster 
XPLFs, and different low-$L$ breaks appear in inner-bulge, outer-bulge, 
and globular-cluster XPLFs (see Fabbiano 2005 and references therein). 

Our choice is motivated by concerns of unexplored complexities in the 
low-$L$ old-XPLF, as well as by the consideration that in SFGs the 
details of the low-$L$ old-XPLF may not be crucial, if $L_{\rm XP}$ is 
dominated by bright sources -- i.e. when the cumulative XPLF index is 
$\gamma$$<$1, as is the case for the SFGs in Table 1 (except for NGC~628, 
whose measured XPLF however extends over a range where Kim \& Fabbiano's 
function is defined).

In conclusion, use of Kim \& Fabbiano's (2004) {\it combined} broken-PL 
old-XPLF probably represents a significant improvement in accuracy 
over the use of a single-PL with (cumulative) slope $\magcir$$1$ as 
suggested by {\it individual} XPLFs (e.g., Kim \& Fabbiano 2004; Colbert 
et al. 2004). We consider this assumption to be adequate for modeling 
XLPFs that do not extend much lower than $L$$\sim$$10^{37}$ 
erg s$^{-1}$. But accurate modeling of deeper XLPFs (either measured 
or extrapolated) will require better knowledge of the low-$L$ old-XPLF. 
\medskip

\noindent
{\it Young-XPLF.} Either measured from very active global starburst 
galaxies (e.g.: NGC~4038/9, Zezas \& Fabbiano 2002) or from spatially 
resolved young XP population (e.g.: M~81, Tennant et al. 2001; M~83: 
Soria \& Wu 2003), young-XPLFs turn out as single PLs with (cumulative) 
slopes $-$0.5$\pm$0.1 -- the flatter slopes were measured in more 
intensely star-forming galaxies (Kilgard et al. 2002). Grimm et al. 
(2003) suggested the existence of a 'universal' young-XPLF described 
as a single-PL, with cumulative slope $-$0.6 and normalization 
proportional to the SFR, in the luminosity interval 
$\sim$4$\times$$10^{36}-$10$^{40}$ erg s$^{-1}$. However, as discussed 
in Fabbiano (2005), a significant scatter of individual XPLF 
behaviors is observed. In our analysis we adopted Grimm 
et al.'s suggested regularity of a universal single-PL XPLF, but we chose 
the slightly flatter slope $-$0.5 which we feel (following Kilgard et al. 
2002) to better represent homogeneous young-XP populations in high-SFR 
environments.

A further complication is possible. If in a high-SFR environment the 
stellar IMF is top-heavy (e.g., Doane \& Mathews 1993; Rieke et al. 
1993), the resulting correlation between SFR and stellar IMF implies 
a proportionally higher number of massive stars, and hence a flatter 
young-XPLF, for higher SFRs. This would in principle challenge the 
concept of a 'universal' young-XPLF. If so, using one same young-XPLF 
to model the XPLFs of galaxies with very different SF activities would 
be incorrect and would lead to a systematic bias in the analysis. In 
this scenario, the assumption of 'universal' young-XPLF implies that 
SF sites in galaxies should have very similar characteristics everywhere 
within individual galaxies and in different galaxies, the main difference 
between high- and low-SFR galaxies being the number and sizes, not the 
physical properties, of such SF sites.

Given the known uncertainties, we checked that our main results are not 
significantly altered by changing our adopted young-XPLF slope by $\pm$0.1, 
which is believed to represent a reasonable uncertainty in the 'universal' 
young-XPLF slope.

\begin{table}
\caption[] {Data III: {\it Hubble} Deep Field North galaxies.}
\begin{flushleft}
\begin{tabular}{ l l l l l l l }
\noalign{\smallskip}
\hline
\hline
\noalign{\smallskip}
Source &  $~~z$  & $f_{1.4}$ & $~L_{1.4}$ & ~~~~~~~~~SFR & $~~~~~~f_{x}$ & $~~~L_{x}$ \\
\noalign{\smallskip}
\hline
\noalign{\smallskip}
134& 0.456& 210& 30.00&~~~~62.1/ 112.0 & $-15.55$& 41.13\\
136& 1.219& 180& 30.87& ~~455.5/ 821.3 & $-15.72$& 41.96\\
188& 0.410&~~83& 29.50&~~~~19.6/~~~35.3 & $-16.24$& 40.34\\
194& 1.275& ~60& 30.43& ~~167.8/ 302.5 & $-15.70$& 41.95\\
246& 0.423& ~36& 29.16&~~~~~~9.1/~~~16.3 & $-16.12$& 40.49\\
278& 0.232& 160& 29.26&~~~~11.3/~~~20.4 & $-15.80$& 40.26\\
\noalign{\smallskip}
\hline
\hline
\end{tabular}
\end{flushleft}
\smallskip

All fluxes (2-10 keV: erg cm$^{-2}$ s$^{-1}$; $1.4\,$GHz: $\mu$Jy) are rest-frame and 
are taken from Ranalli et al. (2003). Radio/X-ray luminosities (measured, respectively, 
in erg s$^{-1}$Hz$^{-1}$ and erg s$^{-1}$), as well as X-ray fluxes, are given in log 
form. X-ray fluxes and luminosities were derived by modeling {\it Chandra} 0.5-8 keV 
counts with a power-law model (Ranalli et al. 2003). Star-formation rates, derived from 
the 1.4~GHz luminosity by means of eq.(7)/eq.(11), are expressed in $M_\odot$ yr$^{-1}$.
\end{table}


\section{Star formation in the nearby Universe}

To sample SF more completely in the nearby Universe, we extend our 
analysis to a sample of starburst-ULIRGs 
\footnote{By this definition we mean ULIRGs that show no 
	  obvious X-ray spectral evidence of harboring a 
	  central AGN (e.g., Franceschini et al. 2003).}.
Their very-high-SFRs ($\magcir$$100$$\,M_\odot$yr$^{-1}$), and the 
apparent flatness of their 2-10 keV spectra ($\Gamma$$\sim$$1.2$: 
e.g., Franceschini et al. 2003) reminiscent of Galactic HMXB spectra, 
suggest that in these objects $L_x$ mainly originates from young XPs. 
Then $L_x^{\rm yXP}$$\sim$$L_x$, and the current SFR of these 
galaxies is effectively traced by $L_{x}$. Starburst-ULIRGs are 
natural calibrators for the SFR--X-ray-luminosity relation. From 
Fig.4-{\it left} we see that, using $L_{x}$, our ULIRGs (empty 
squares) do lie on the extrapolation of the SFR-$L_{x}^{\rm yXP}$ 
relation defined (at lower SFRs) by the SFG sample. 

\begin{figure}

\vspace{4.0 cm}
\includegraphics{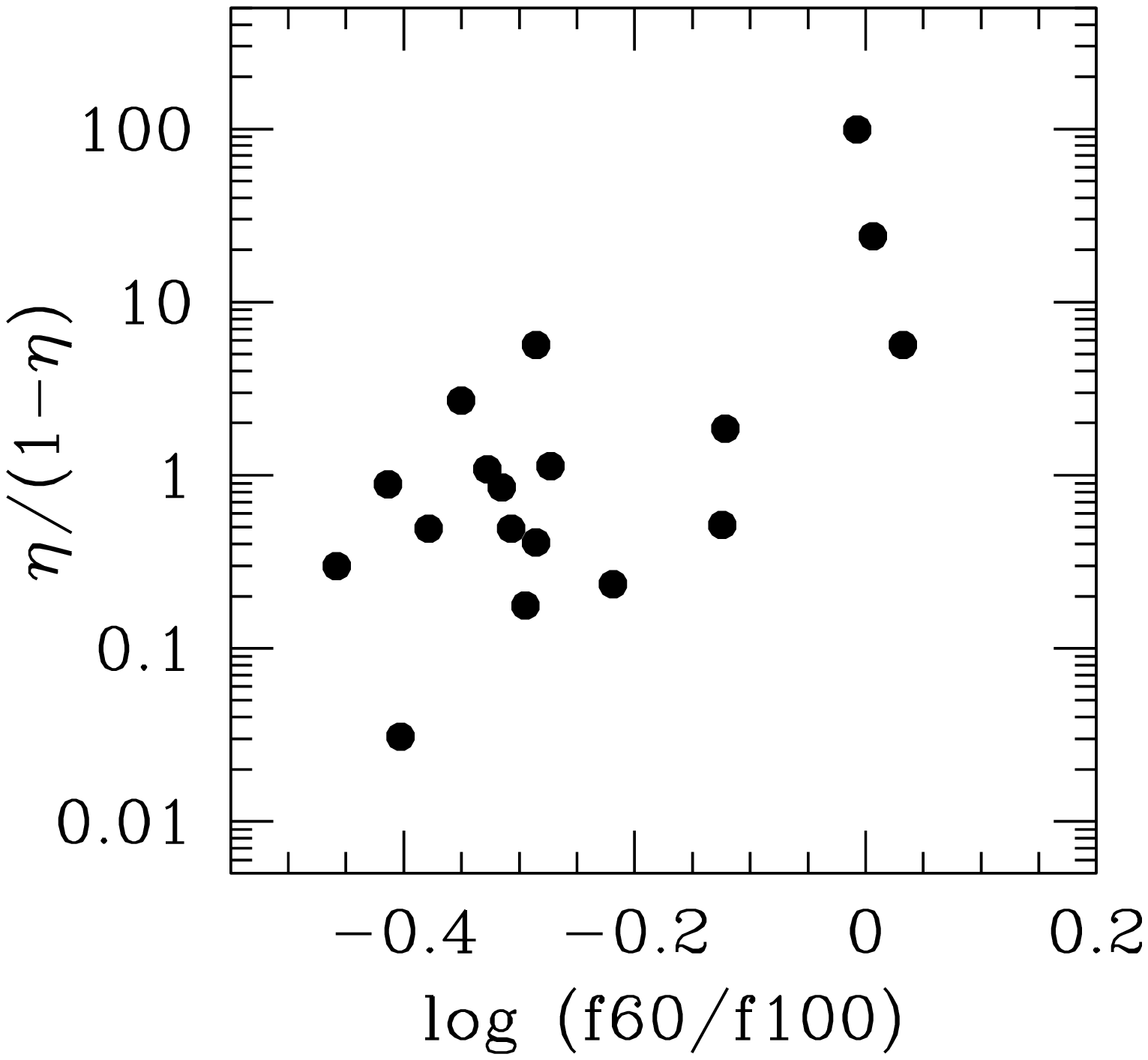}
\includegraphics{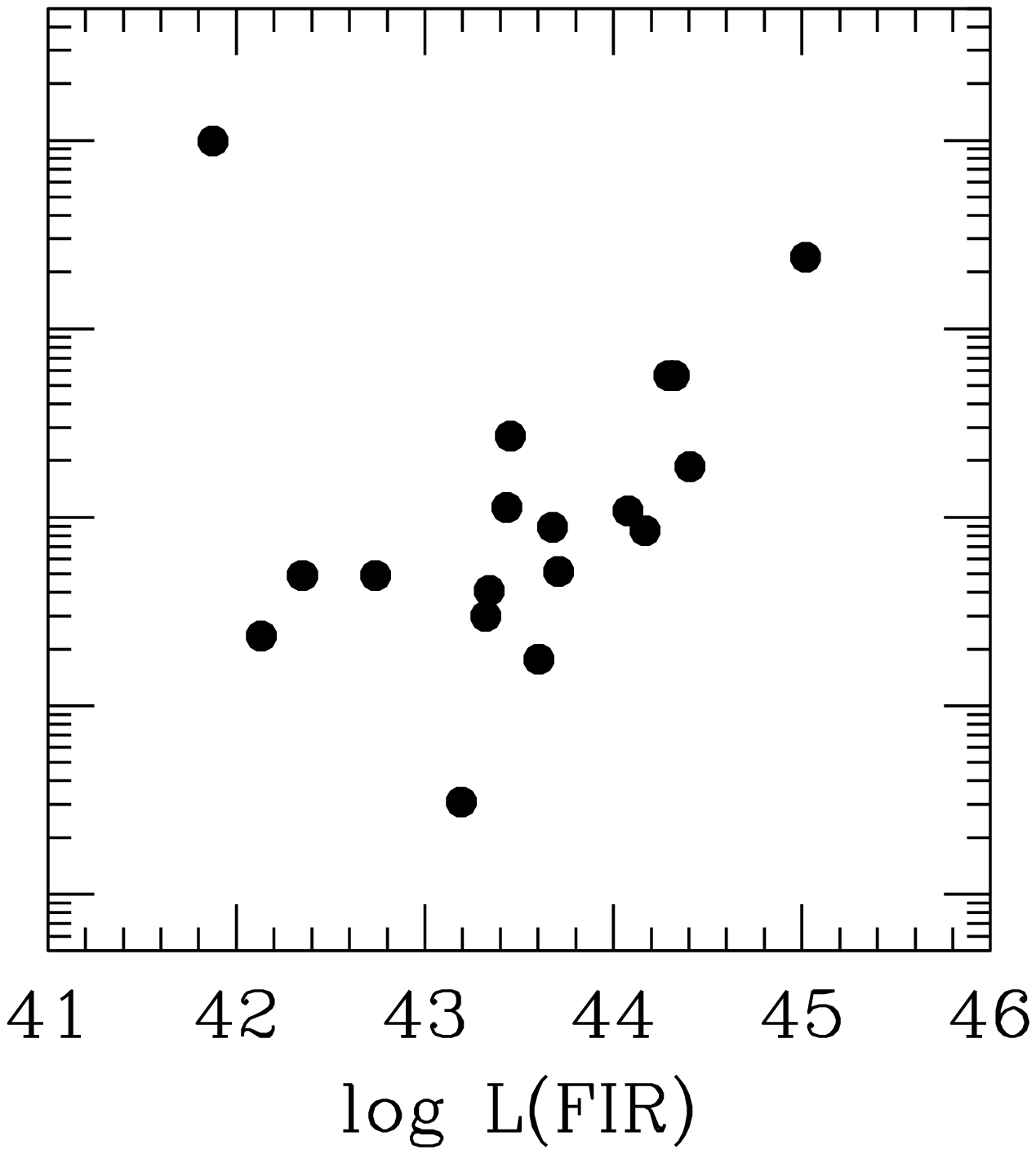}
\caption{ 
The young- to old-XP 2-10 keV luminosity ratio plotted versus the 
temperature index $f_{60}/f_{100}$ ({\it left}) and the FIR luminosity 
({\it right}) for the galaxies in sample 1. The young-XP fraction is 
generally higher in more FIR-luminous galaxies, but the correlation 
appears to be tightest with the starburst phase (indicated by $f_{60}
/f_{100}$): it is highest in peak-starbursts and gets progressively 
lower in evolved-starbursts and post-starbursts (see also Fig.1). The 
outlier in the right panel is NGC~1569 whose {\it IRAS} colors are 
suggestive of a peak-starburst phase in spite of its low luminosity. 
}
\end{figure}

For a comparison with the SFR--$L_x^{\rm yXP}$ relation in eq.(9), 
we now consider the SFR--$L_{x}$ correlation. For sample 1 this is 
reproduced by (see Fig.2-{\it right}, filled circles): 
\begin{eqnarray}
{\rm SFR}(>0.1\,M_\odot) ~ = ~ {L_{x}  \over  3.8 \times 10^{39}{\rm 
erg\, s}^{-1} }  ~~M_\odot{\rm yr}^{-1}
\end{eqnarray}
(with a $\sim$$20$$\%$ statistical uncertainty on the calibration). 
This expression is consistent with an analogous expression by 
Ranalli et al. (2003), based on nearly the same data, once their 
different definition of SFR, involving $L_{\rm FIR}$ instead of 
$L_{\rm IR}$, is accounted for.) Inspecting the two relations in 
Fig.4, it is clear that sample 1 complies with both, whereas 
sample 2 does not. 

The mismatch between the SFG sample and the ULIRG sample in the 
SFR--$L_{x}$ plane is so considerable that no simple function can 
adequately characterize our full combined sample over the SFR 
range ($-2$$\mincir$log[SFR/($M_\odot$yr$^{-1}$)]$\mincir$$2.5$) 
considered here. Since there is nothing special about the way our 
star-forming objects were selected, we suggest that the discrepancy 
is real and does not stem from a known bias. Our interpretation of 
the discrepancy is as follows. In all star-forming galaxies virtually 
the total $L_{\rm FIR}$ traces the istantaneous SFR, whereas $L_x$, 
that is emitted partly by LMXBs (which represent the SFR of previous 
epochs) and partly by young sources (SNRs, HMXBs which trace the 
ongoing SFR), traces the integrated (from some previous epoch to the 
present) SFR. In those galaxies, therefore, only a fraction of $L_x$ 
is related to the instantaneous SFR -- a fraction which is large in 
SF-dominated galaxies (like our ULIRGs) but can be quite small in more 
quiescent spirals (e.g., the Galaxy) (see Fig.5)

The simultaneous validity of both the SFR--$L_{x}^{\rm yXP}$ and 
SFR--$L_{x}$ relations for our local SFGs (with SFR$\leq$$50$$\, 
M_\odot\, $yr$^{-1}$) suggests that in these galaxies the SF activity 
has remained essentially constant over the past several $10^8$ yr, so 
the corresponding SFR (traced by $L_x$) is, for most galaxies of 
sample 1, approximately the same multiple of the instantaneous SFR 
(traced by $L_x^{\rm yXP}$). 

\begin{figure}

\vspace{3.0cm}
\includegraphics{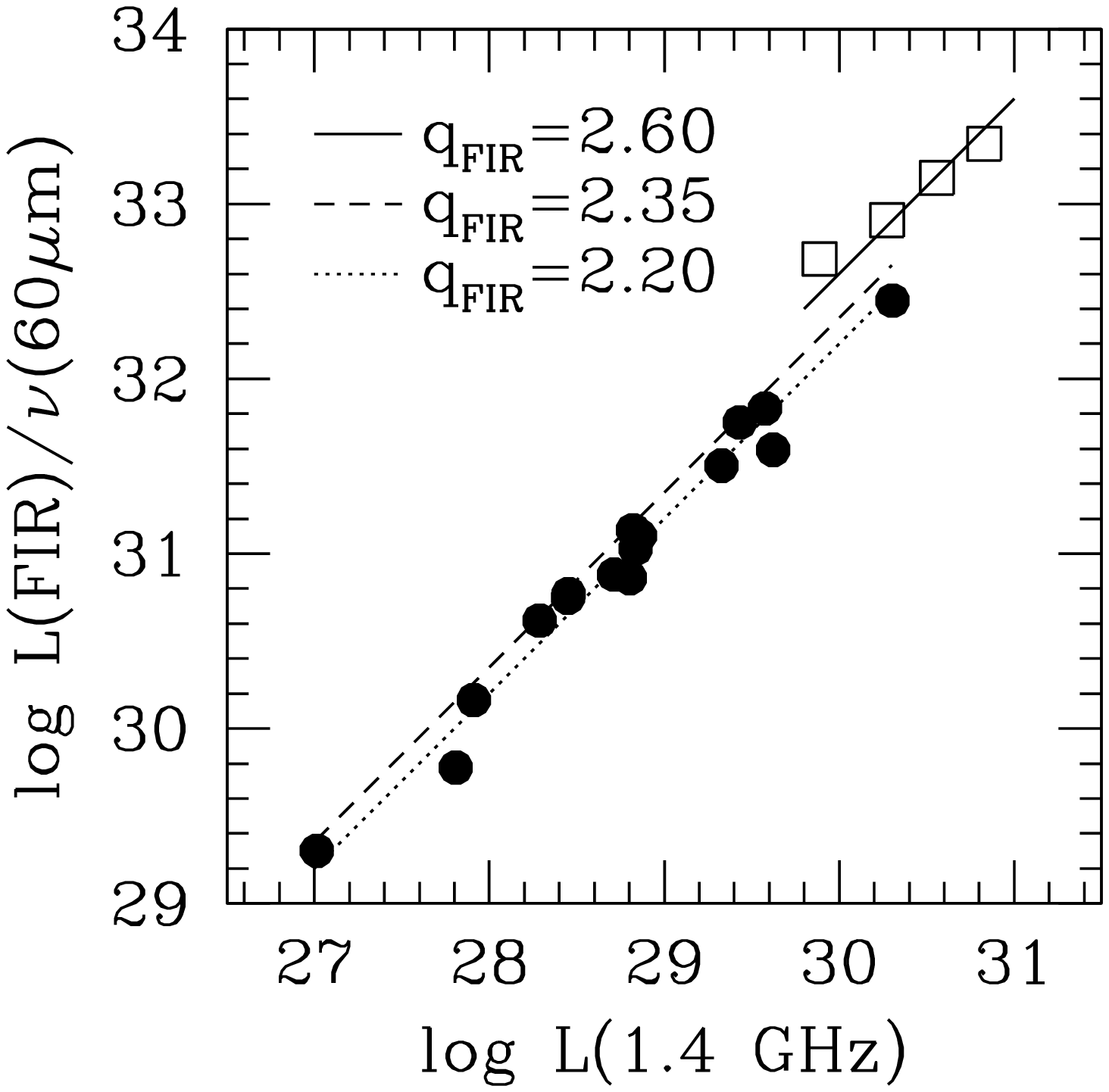}
\includegraphics{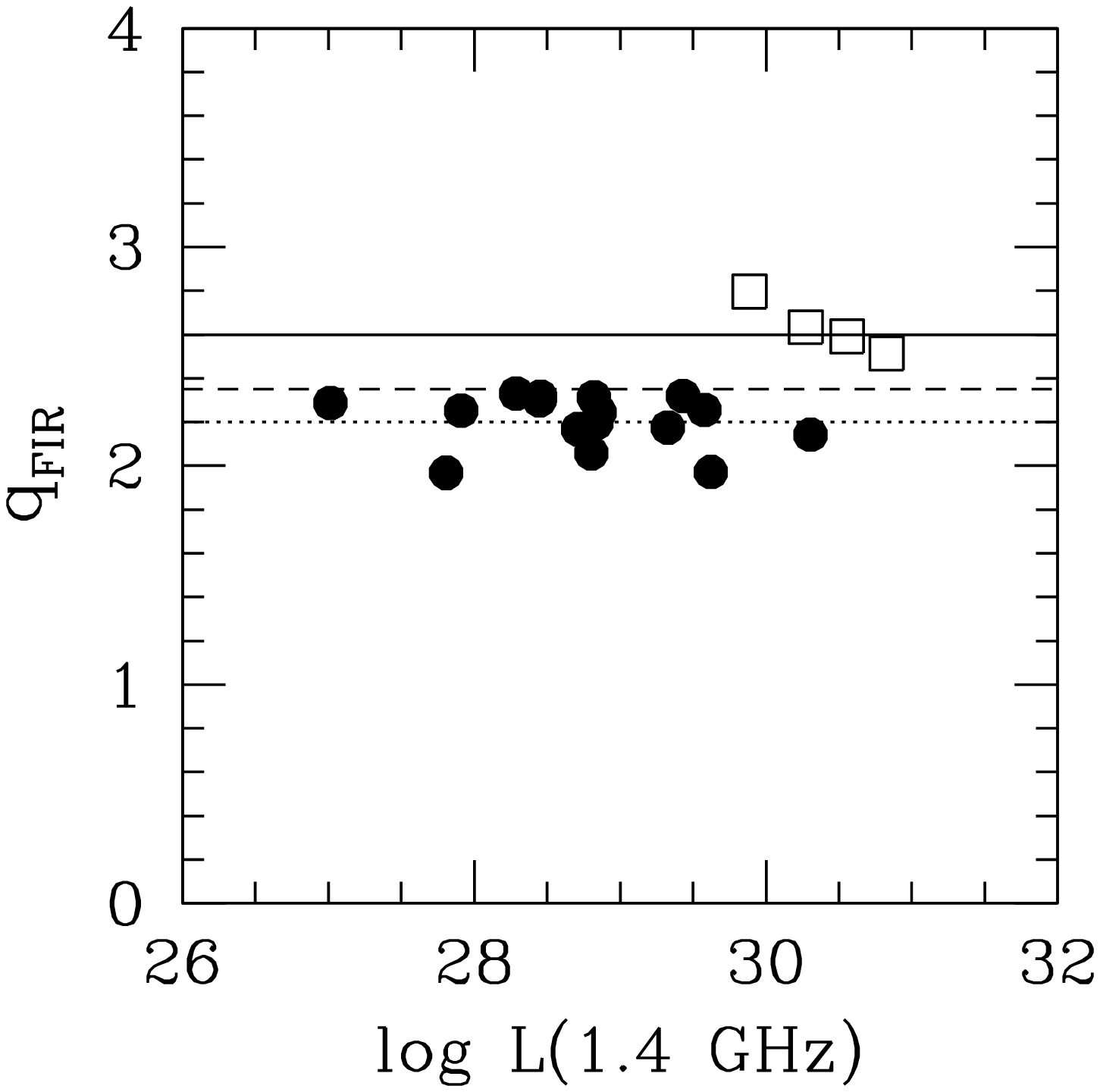}

\vspace{3.375cm}
\includegraphics{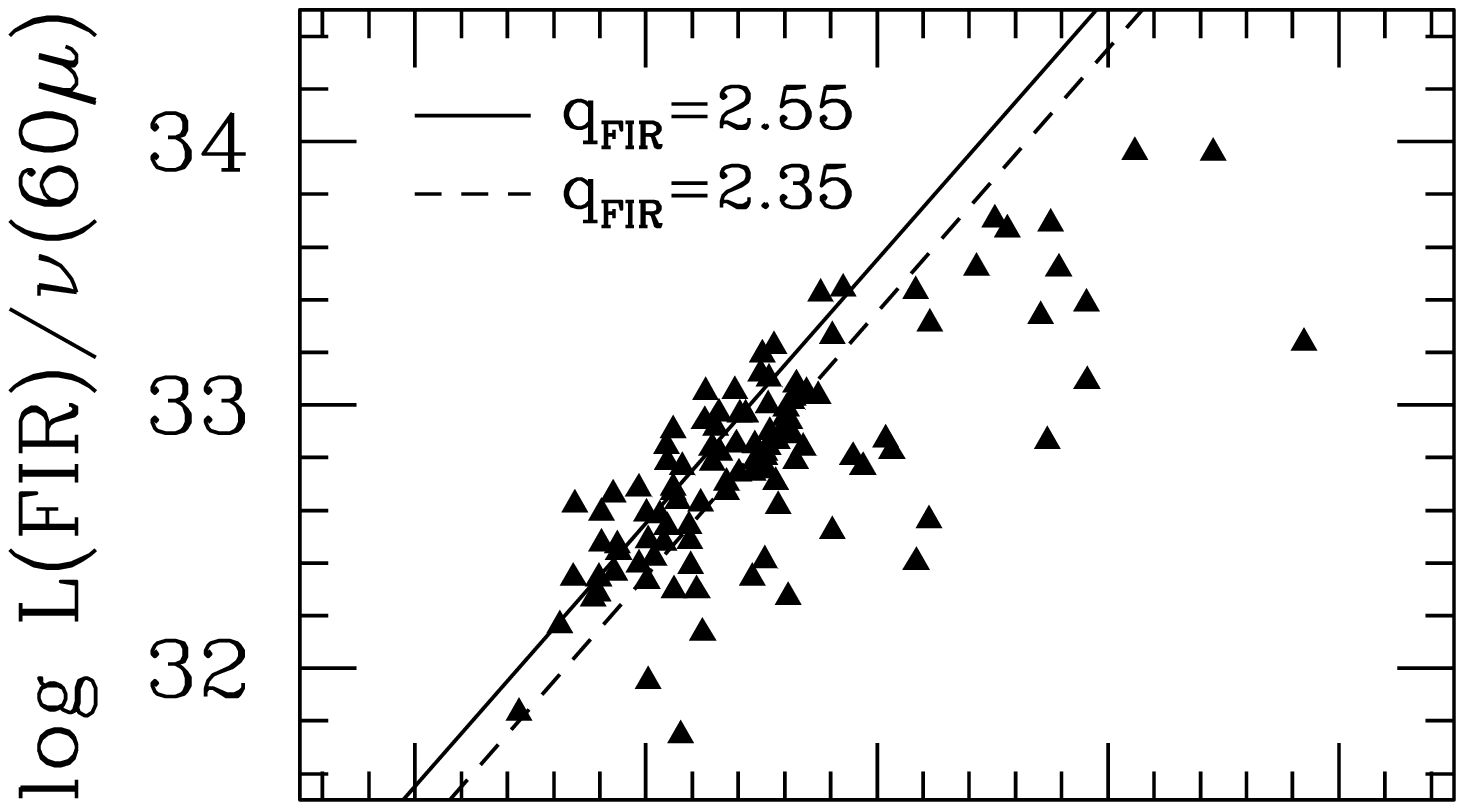}
\includegraphics{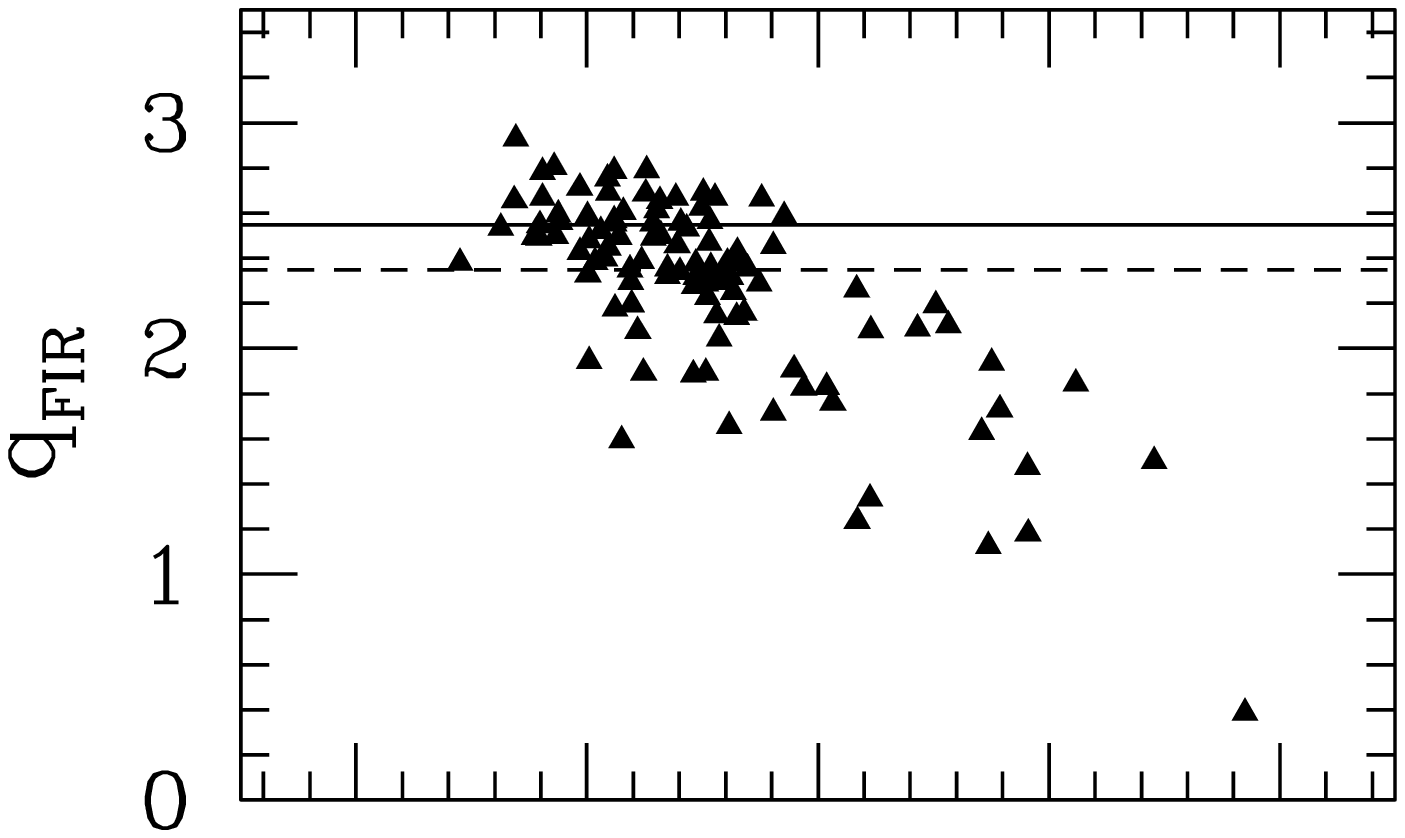}

\vspace{3.375cm}
\includegraphics{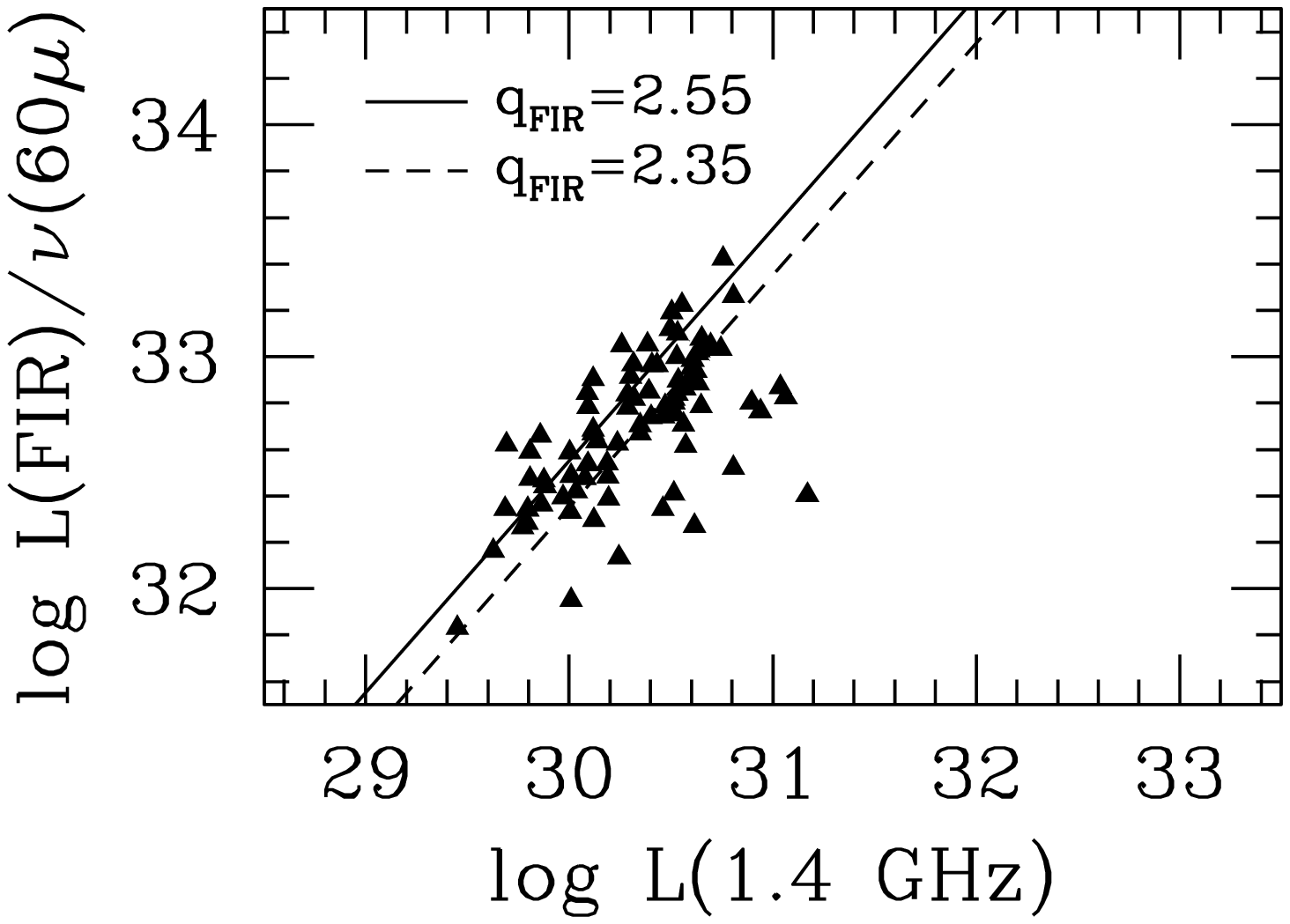}
\includegraphics{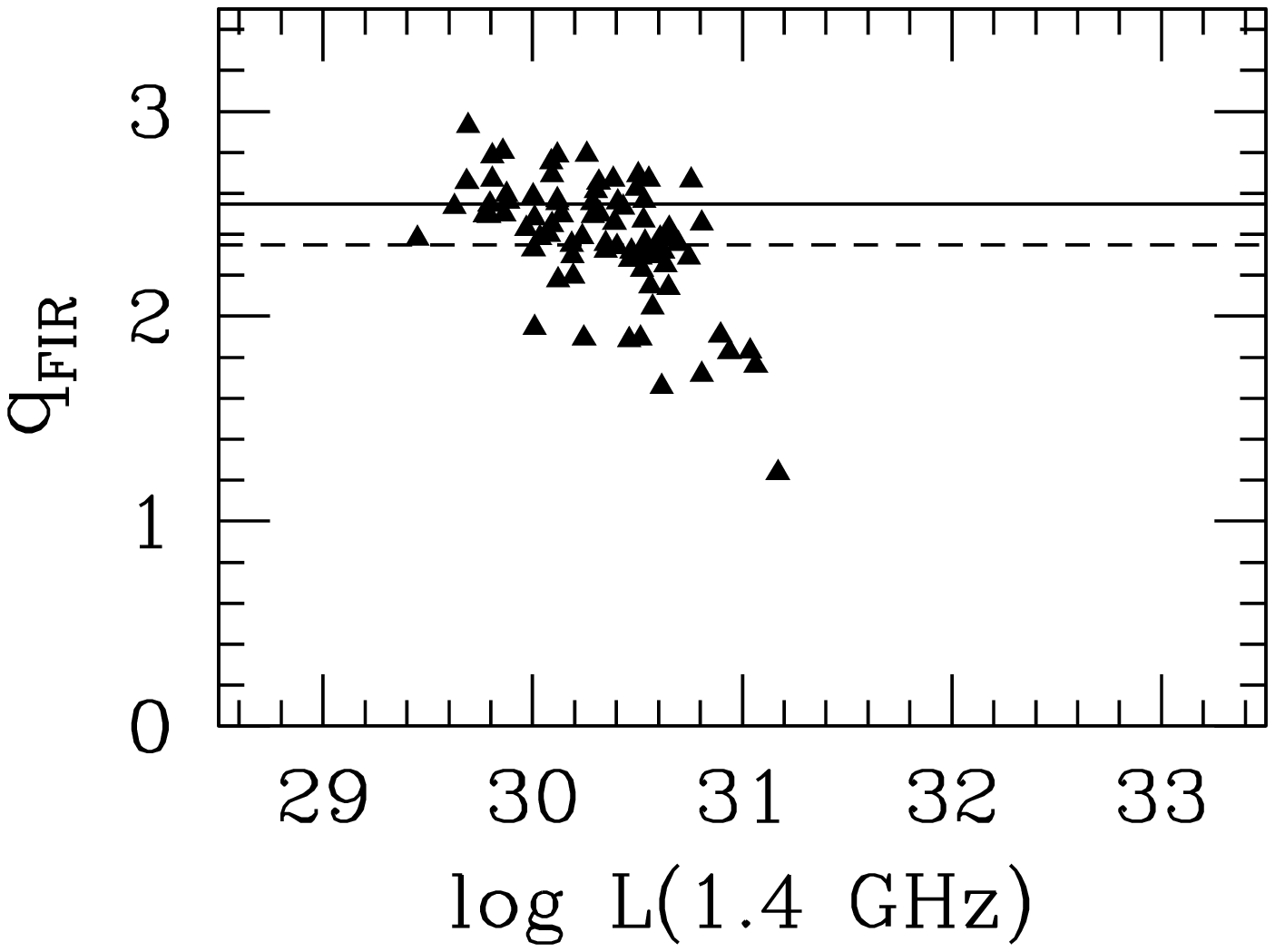}

\caption{ 
The FIR-radio correlation ({\it left}) and, equivalently, the FIR/radio
parameter $q_{\rm FIR}$ ({\it left}) for several samples of star-forming 
galaxies. 
{\it Top:} The local galaxies of sample 1 (filled circles, dotted line) 
and the relatively faraway ULIRGs of sample 2 (empty squares, solid line). 
The dashed line denotes the characteristic relation for local field galaxies 
according to Condon et al. (1991a), Yun et al. (2001), and Bell (2003). 
{\it Middle:} The high-redshift ULIRG sample of Stanford et al. (2000). All 
data are used; when 100$\mu$m detections could not be obtained, 1$\sigma$ 
upper limits were used to compute $L_{\rm FIR}$.
{\it Bottom:} Stanford et al.'s (2000) sample, with non-detections and 
sources with $z$$>$$0.45$ removed.
}
\end{figure}

It would be misleading, however, to infer the ratio of the integrated to 
instantaneous SFR from the normalization ratio ($\sim$$5$) of eq.(10) to 
eq.(9). We emphasize, in fact, that in sample 1 there is a systematic 
discrepancy between the XP luminosity (used to build $L_x^{\rm yXP}$$
\equiv$$\eta$$L_{\rm XP}$) and the integrated luminosity: 
$<$$L_{\rm XP}$$>$$\simeq$$0.40_{-0.08}^{+0.11}$$<$$L_x$$>$. This 
discrepancy may originate from inaccurate modeling of the individual XP 
spectra (affecting $L_{\rm XP}$; e.g., Schlegel \& Pannuti 2003) and/or 
of the integrated galaxy spectra (affecting $L_x$; e.g. Dahlem et al. 
2000), or from the presence of a deeply buried AGN (e.g.: Della Ceca 
et al. 2002; Komossa et al. 2003; Ballo et al. 2004) or of truly diffuse 
emission (e.g., Griffith et al. 2000), or from a combination of all 
these. Whatever its origin, the discrepancy between $L_x$ and $L_{\rm 
XP}$ contributes significantly to the difference between the 
normalizations of eqs.(9) and (10). As a comparison, from Table 2 we 
derive $<$$\eta$$>$$\simeq$$0.50$$\pm$$0.07$ for sample 1. Hence 
$<$$L_x^{\rm yXP}$$>$$\sim$$0.2$$\,$$<$$L_x$$>$, as implied by the 
respective normalizations. (Incompleteness corrections, being relatively 
modest, do not substantially alter this result.) 


\section{Star formation at high redshift}

Knowledge of the cosmological SF history is crucial to constrain models of 
galaxy evolution. One key step forward in this direction is developing our 
ability to measure the ongoing SFR in galaxies at cosmological distances, 
in a way that is mostly unaffected by absorption. 

Sample 3 is a set of distant ($z$$\sim$1) {\it Hubble} Deep Field North galaxies 
(HDFNGs) with available 1.4 GHz flux densities and {\it Chandra}-based 2-10 keV 
fluxes (see Table 4). No FIR data are available. In Table 4 we report (from 
Ranalli et al. 2003) their (k-corrected: e.g., Bauer et al. 2002) rest-frame 
2-10 keV luminosities and radio luminosity densities. The former are computed from 
{\it Chandra} counts in the soft (0.5--2 keV) and hard (2--8 keV) band, assuming 
a power-law (PL) model. If a more realistic model is adopted, e.g. a sub-keV thermal 
plus a hard PL model (see Dahlem et al. 1998), the resulting 2-10 keV luminosities 
would be slightly ($\sim$10$\%$) lower. 

The SFRs of our HDFNGs, estimated from $L_{\rm 1.4}$ using the (locally derived) 
conversion in eq.(7), are high (SFR$\sim$$100$$-$$1000$$\,M_\odot$yr$^{-1}$: see 
Table 4), suggesting that these galaxies are SF-dominated (see also Cohen 2003). 
This, and the consideration that at the epoch corresponding to $z$$\sim$$1$ 
($\sim$$6$ Gyr for our adopted cosmology) there had been no time for LMXBs to form, 
lead to the expectation that the $L_x$ of our HDFNGs are largely dominated by young 
XPs. We then assume $L_{x}^{\rm yXP}$$\sim$$L_x$. In the context of our current 
analysis, HDFNGs appear similar to ULIRGs. 

The link between a starburst's instantaneous SFR and thermal FIR emission is 
well established, rendering $L_{\rm FIR}$ a relatively accurate estimator of 
the instantaneous SFR in starburst galaxies, most notably in their peak phase 
(see section 4). This is not the case for the non-thermal radio emission, whose 
calibration with the SN rate is not known precisely (see Condon 1992; Condon et 
al. 2002) and whose characteristic synchrotron loss timescale strongly depends 
on the magnetic field -- which can be very different in galaxies of (e.g.) very 
different SFR. This means that, in principle, in a given sample we should check 
the cross-correlation of the FIR-based and radio-based SFR indicators to ensure 
that the two sets of SFRs derived for that sample are mutually consistent. For 
a sample for which the radio SFR indicator is not known directly, combining the 
FIR SFR indicator in eq.(3) with the observed FIR-radio correlation for that 
sample will yield a suitable radio-based SFR indicator. 

\begin{figure*}

\vspace{4.8cm}
\includegraphics{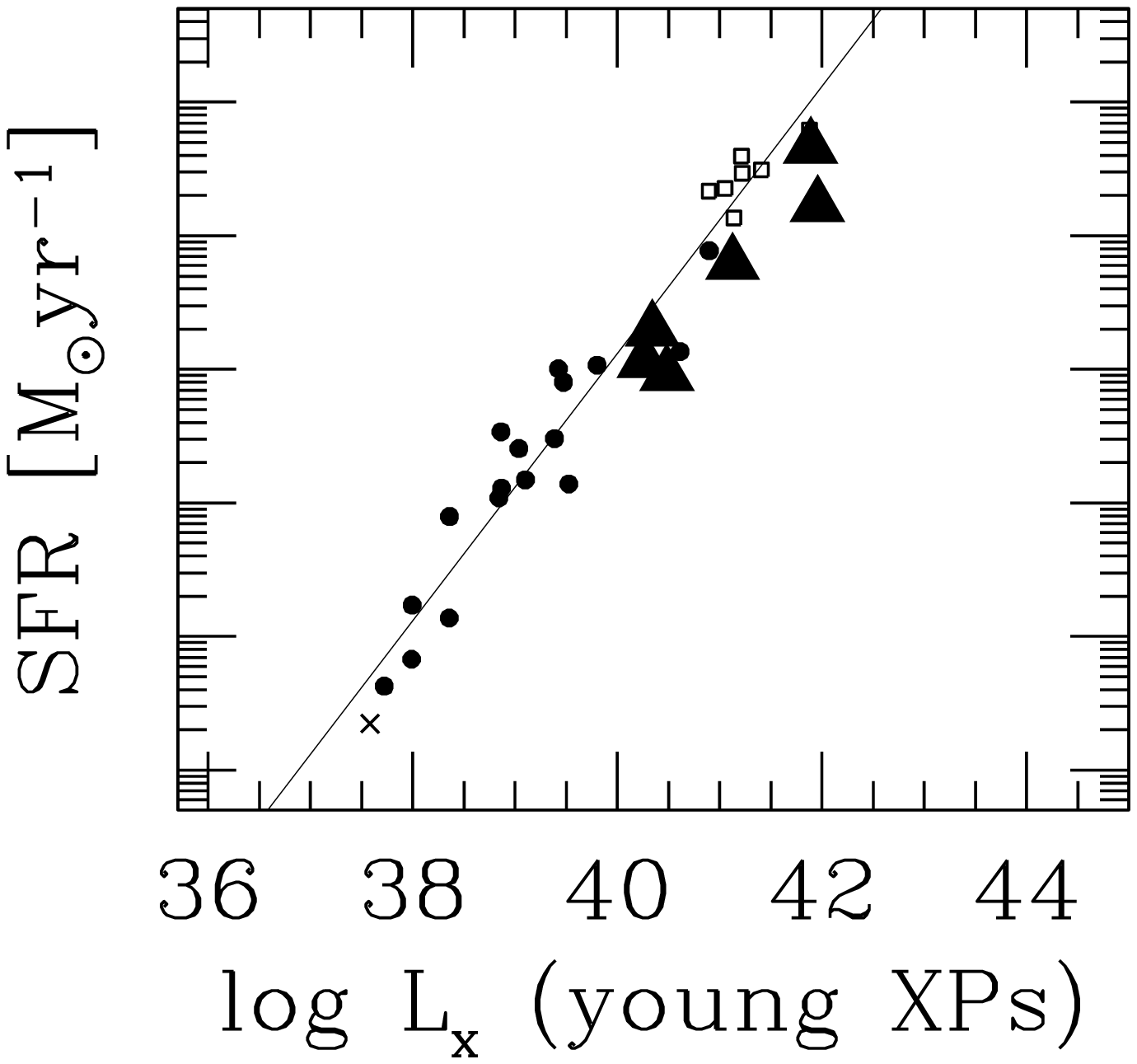}
\includegraphics{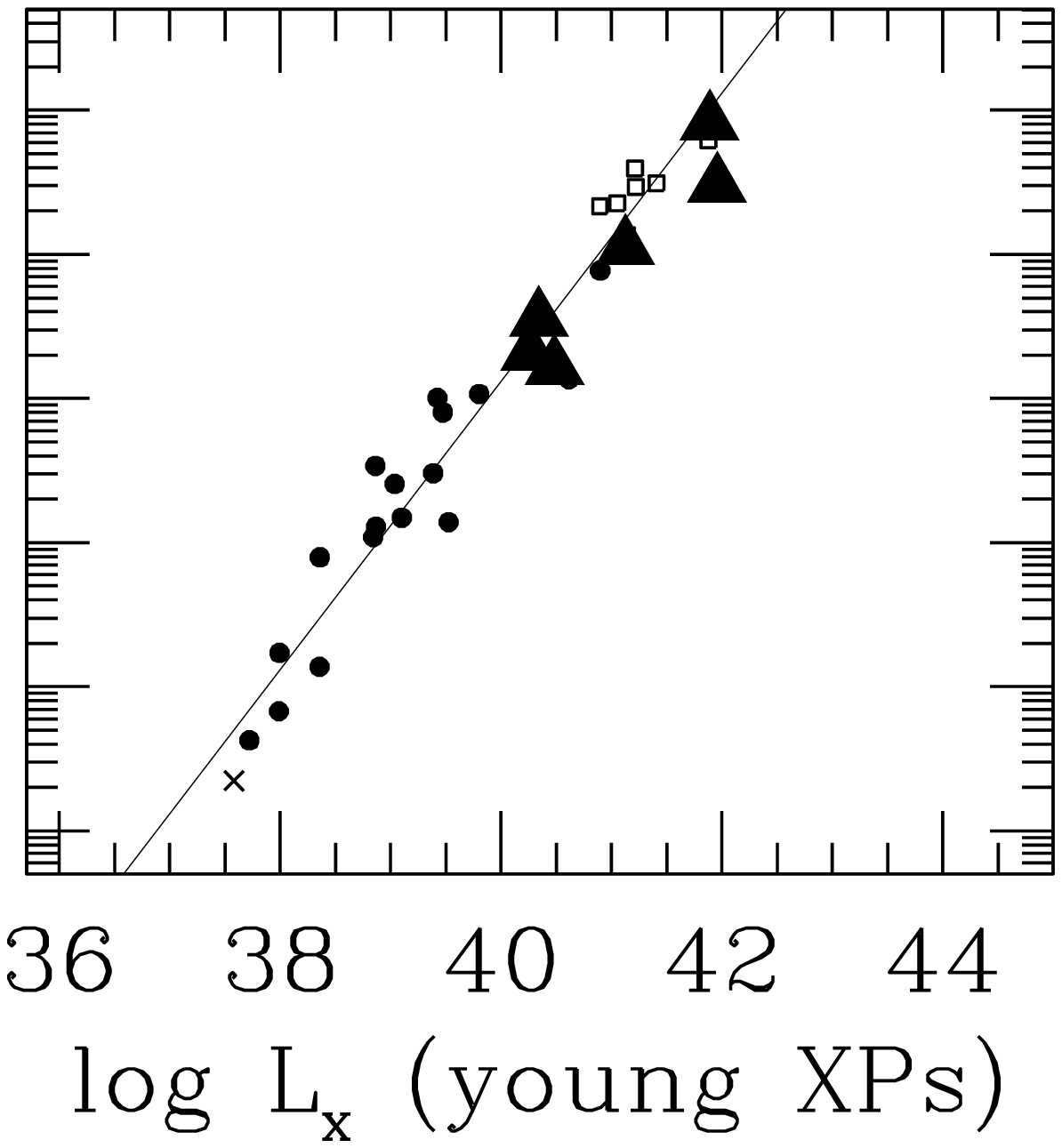}
\includegraphics{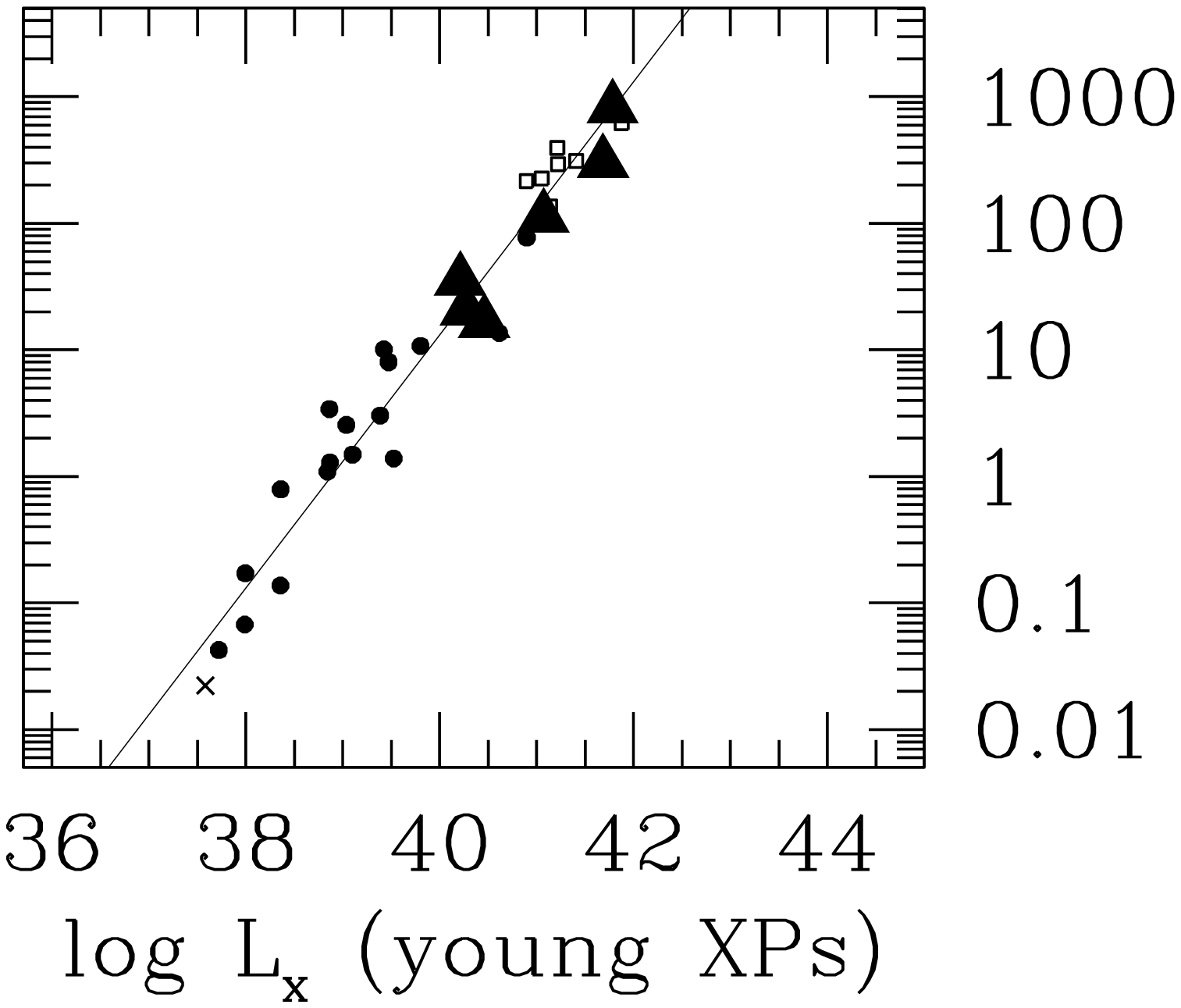}

\vspace{0.3cm}

\vspace{4.8cm}
\includegraphics{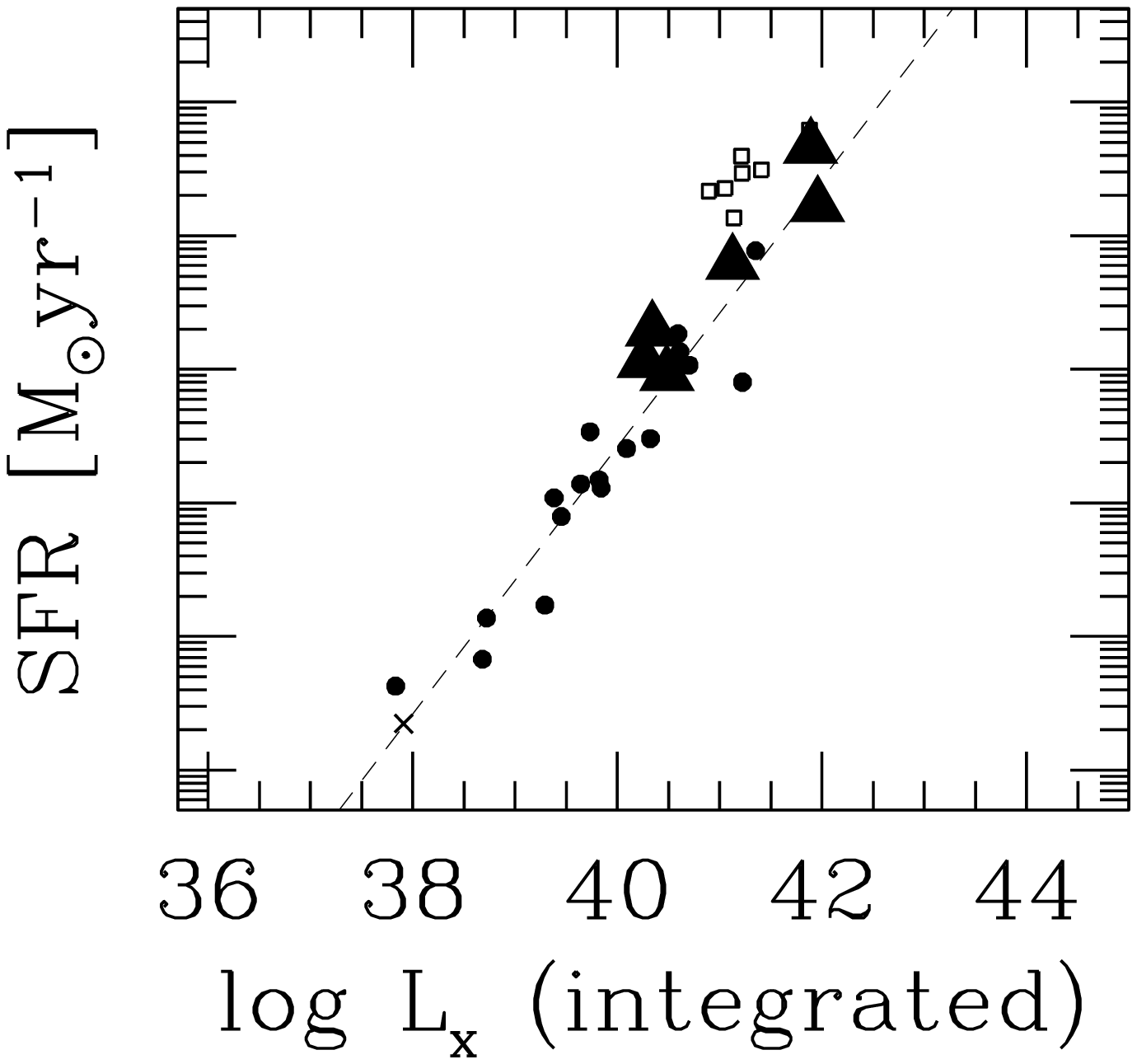}
\includegraphics{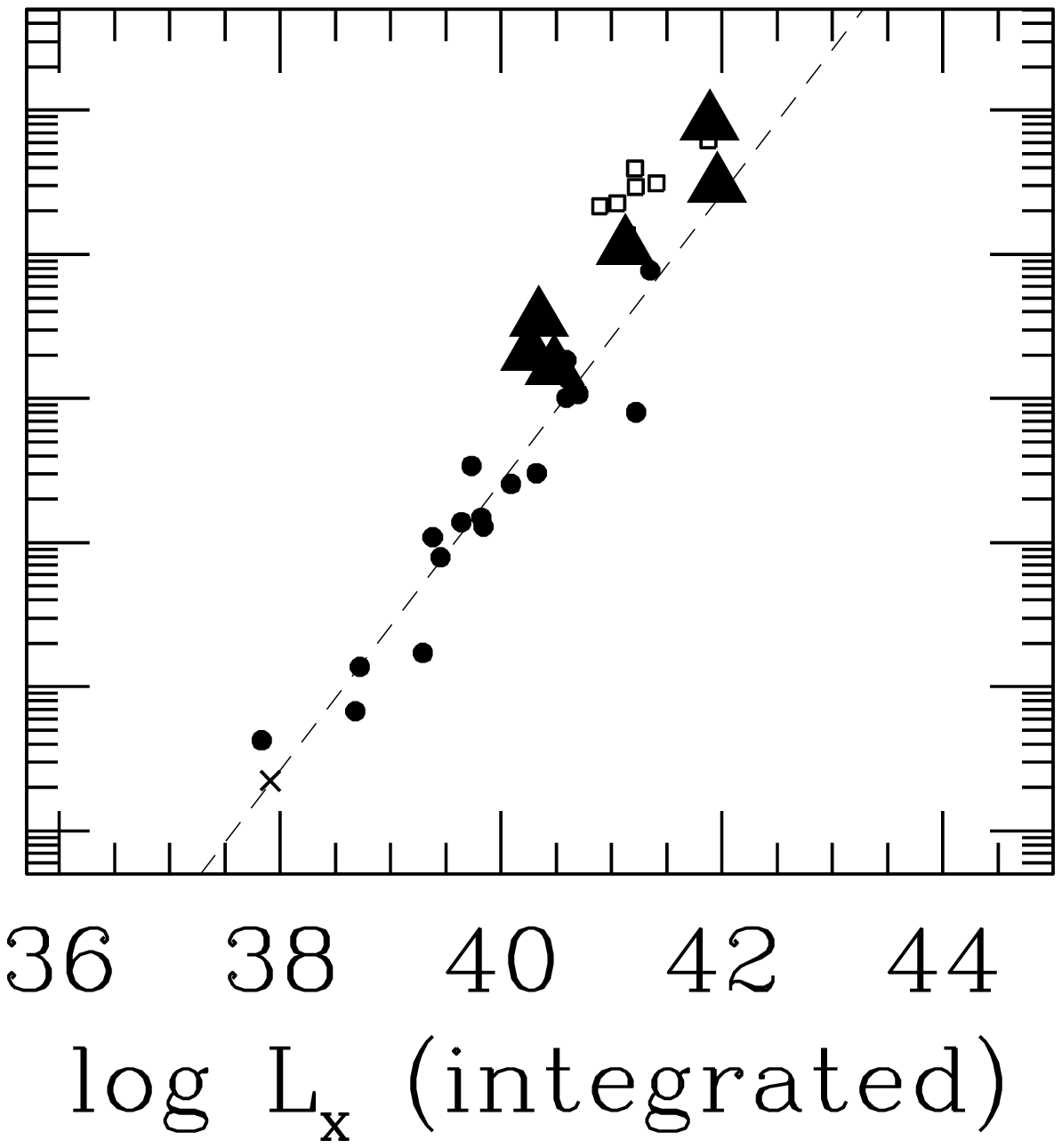}
\includegraphics{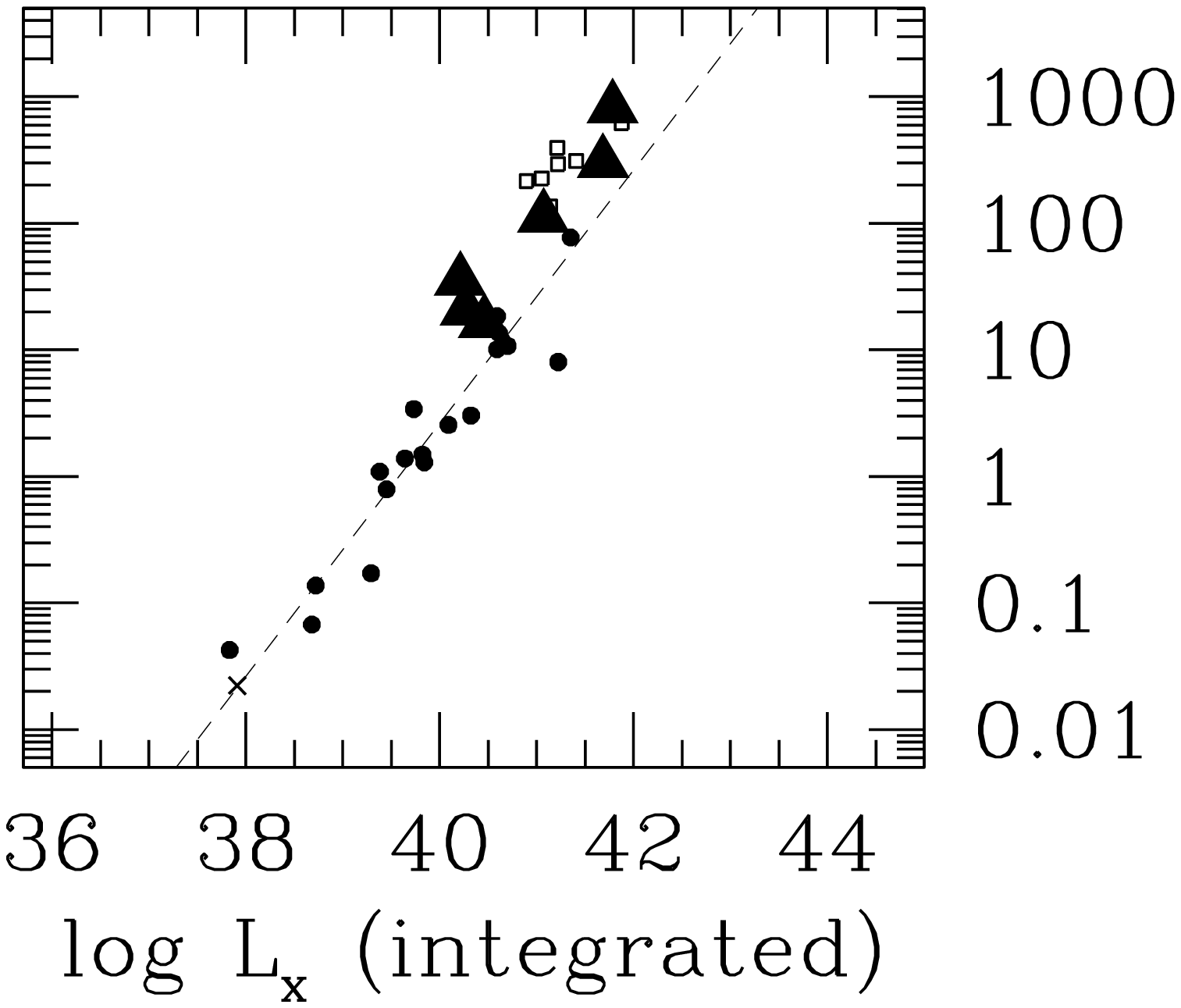}

\vspace{0.3cm}

\caption{ 
The SFR versus 2-10 keV luminosity relations, using the collective luminosity of young 
XPs ({\it top}) and the total luminosity ({\it bottom}), for the distant sample of HDFNGs 
in sample 3 (large filled triangles) and, as a comparison, for the local SFGs in sample 1 
(small filled circles) and the nearby ULIRGs in sample 2 (small empty squares). Because 
the emission of the HDFNGs is arguably due to young XPs, their total 2-10 keV luminosities, 
$L_x$, are used in both panels. These are computed from {\it Chandra} counts assuming a 
single-PL model ({\it left}, {\it middle}) and, alternatively, a soft-thermal plus hard-PL 
model ({\it right}). In the latter case, in order to maximize the effect we assumed the 
thermal component to have $0.2\,Z_\odot$ (e.g., Tsuru et al. 1997) and the PL component 
to have $\Gamma$$=$$1.2$ (e.g., Franceschini et al. 2003). (All new luminosities were 
k-corrected accordingly.) HDFNG SFRs were computed from 1.4~GHz luminosity densities using 
Schmitt et al.'s (2006) locally-calibrated conversion in eq.(7) ({\it left}) and our 
proposed 'ideal starburst' conversion in eq.(11) ({\it middle}, {\it right}). Shown are 
also the relations in eq.(9) ({\it top}, solid line) and in eq.(10) ({\it bottom}, dashed 
line).
}
\end{figure*}

Our direct knowledge of the FIR-radio correlation for deep samples is still quite 
limited. Garrett (2002), analyzing a sample of distant ($z$$\mincir$1.3) HDFNGs 
with $28.5$$\mincir {\rm log}[L_{\rm 1.4\, GHz}/({\rm erg~s}^{-1}{\rm Hz}^{-1})]
$$\mincir$$32$ (i.e., overlapping in luminosity with sample 2), could reach no 
conclusive results because the FIR fluxes of those galaxies, which were not 
directly accessible, had to be extrapolated from available {\it ISO} 15$\mu$m 
data assuming a starburst template: the resulting $q_{\rm FIR}$ depended crucially 
on the adopted template, hence no information could be obtained on the actual value 
of $q_{\rm FIR}$ for that sample. Appleton et al. (2004), based on {\it Spitzer} 
70$\mu$m data for a distant galaxy sample in the luminosity range $10^{27}$$\mincir 
L_{\rm 20\, cm}/({\rm W~Hz}^{-1})$$\mincir$$10^{30}$ (i.e., overlapping in 
luminosity with sample 1), concluded that $q_{\rm 70}$$\simeq$$2.15$ out to 
$z$$\sim$$2$. 

Attempting to circumvent any lack of direct knowledge, and pushing further our 
assumption of similarity between HDFNGs and ULIRGs, we {\it assume} that any 
FIR-radio correlation observed for the latter will also be representative of the 
former. Our own data suggest an ULIRG value of $q_{\rm FIR}$$\simeq$$2.6$ 
(Fig.6-{\it top}). 
\footnote{ 
	Also from Fig.6, we point out that sample 1 has 
	$q_{\rm FIR}$$\sim$$2.2$, hence it offers an 
	essentially unbiased representation of the local 
	SFG population. }
The smallness of our ULIRG sample is somewhat compensated for by its homogeneity, 
as it includes only peak-phase starbursts (i.e., objects with $f_{60}/f_{100}$$
\sim$$1$). This possibly is the reason for the negligible scatter of our ULIRGs 
around $q_{\rm FIR}$$=$$2.6$ in Fig.6-{\it top}. A likely confirmation comes 
from an analysis of Stanford et al.'s (2000) sample of high-$z$ ULIRGs (with 
$29.5$$\mincir {\rm log}[L_{\rm 1.4\, GHz}/({\rm erg~s}^{-1}{\rm Hz}^{-1})]$$
\mincir$$31$, i.e. spanning the same luminosity range as our ULIRGs): although 
we find $2$$\mincir$$q_{\rm FIR}$$\mincir$$2.6$, in agreement with Condon et 
al.'s (1991b) earlier result for a flux-limited {\it IRAS} sample of starbursts 
and ULIRGs, nevertheless, when obvious outliers and {\it IRAS} non-detections 
are removed, we find that most of Stanford et al.'s (2000) data are consistent 
with $q_{\rm FIR}$$\sim$$2.5-2.6$, definitely higher than the 'canonical' local 
value of $\sim$2.35 (see Figs.6-{\it middle}, 6-{\it bottom}).

Based on the above considerations, we {\it tentatively} propose $q_{\rm FIR}$$
=$$2.6$ for the 'pure starbursts' represented by the ULIRGs in sample 2. If, 
according to our assumption, this value is also representative of our HDFNGs, 
then by combining it with eqs.(3)-(5) we obtain a consistent radio SFR 
indicator, 
\begin{eqnarray}
{\rm SFR} ~=~  { L_{1.4} \over 8.93 \times 10^{27} {\rm erg~s^{-1}Hz^{-1}} }
\,,
\end{eqnarray}
which we suggest may apply to ideal starbursts. The proposed conversion happens 
to be close to that of Condon (1992), who assumed a Galactic calibration of the 
non-thermal radio luminosity with SN rate, and to be higher by a factor of 1.8 
than that of Schmitt et al.'s (2006). We shall use eq.(11) alongside eq.(7) to 
estimate SFRs from 1.4 GHz luminosities for our sample 3 galaxies.

Concerning the 'nonthermal-luminosity versus SN-rate' calibration issue, we point 
out that Yun \& Carilli (2002), using a FIR-radio spectral template in which the 
normalization of the non-thermal radio continuum had been left free to vary in 
order to determine a normalization most suitable for starburst galaxies, found that 
a local sample of FIR-luminous ($>$$10^{11}L_\odot$) galaxies were best fit by the 
Galactic normalization adopted by Condon (1992); and that the spectral template, 
incorporating such Galactic normalization, could fit the observed SEDs of some 
distant ($z$$\sim$$1$), intensely star-forming (SFR$\sim$$200$$-$$\,1000 M_\odot$ 
yr$^{-1}$) galaxies. So Yun \& Carilli's (2002) results may provide circumstantial 
evidence that in very-high-SFR evironments the calibration of the radio-SFR 
conversion may be quite similar to that adopted by Condon (1992) -- and hence in 
implicit agreement with our proposed conversion in eq.(11). 

Using the data in Table 4, we plot in Fig.7 the HDFNGs on the SFR versus 
X-ray luminosity plane (large filled triangles), for both $L_{x}^{\rm yXP}$ 
({\it top}) and $L_{x}$ ({\it bottom}). The left panels use Schmitt et al.'s 
(2006) conversion in eq.(7), whereas the middle panels use the our proposed 
empirical conversion in eq.(11). The right panels use eq.(11) and a two-component 
(soft-thermal plus hard-PL) model, which are probably more realistic for 
starburst galaxies than Ranalli et al.'s (2003) simple PL model upon which 
the values of $L_{\rm x}$ in Table 4 are based. Inspection of Fig.7 leads us 
to conclude that: 
{\it (a)} if Schmitt et al.'s (2006) local calibration holds also at high 
redshifts (or, alternatively, for strong starbursts), then our distant HDFNGs 
don't really seem to strictly follow either relationship, falling somewhere 
in between the two; 
{\it (b)} if our HDFNGs comply with eq.(11), then they match the SFR--$L_{x}^
{\rm yXP}$ relation but not the SFR--$L_{x}$ relation; 
{\it (c)} the previous point is further strengthened if the HDFNG 0.5-10 keV 
spectra are similar to those of local starburst galaxies (e.g., Dahlem et al. 
1998).

Therefore, the issue of where HDFNGs are located in the X-ray versus SFR 
plane is still unsettled. More investigations are needed to effectively 
measure the ongoing SFR in these galaxies. Broad-band IR photometry would 
clearly prove crucial given the established role and effectiveness of the 
8-1000$\,\mu$m luminosity as a SFR indicator. Once calibrated -- using 
IR-derived SFRs -- on distant HDFNGs, the X-ray--based SFR indicator could 
be used to gauge ongoing SFRs in yet more distant galaxy samples.

\section{X-ray versus radio SFR indicators}

Finally, we plot all the galaxies of our combined samples on the 
X-ray versus radio luminosity plane (Fig.8): the correlation involving 
$L_{\rm x}$ ({\it left}) is strong, whereas the one involving $L_x^{\rm 
yXP}$ ({\it right}) is weak -- a piecewise behavior is discernible, 
with a break at $L_x^{\rm yXP}$$\sim$$5$$\times$$10^{39}$ erg s$^{-1}$. 
We interpret the different behavior as follows. 

\noindent
{\it (a)}
The radio luminosity is a function both of the SN rate (hence the massive SFR) 
and of the average magnetic field strength. The syncrotron loss timescale is 
$\tau_{\rm s}$$=$$({4\over 3}\,$${\sigma_{\rm T}\over m_e c}\,$$\gamma_e\,$${B_
{\rm \mu G}^2 \over 8\,\pi })^{-1}$$\simeq$$1.25$$\times$$10^{10}$$(B_{\rm \mu 
G}^{-2}$ $E_{\rm GeV}^{-1}$ yr (where $B_{\rm \mu G}$ is the magnetic field 
strength measured in $\mu$G, and $E_{\rm GeV}^2$ is the elecron kinetic energy 
measured in GeV). Assuming $E$$\sim$$5$ GeV (typical of electrons radiating at 
$\sim$1.4 GHz) and $B_{\rm \mu G}$$\sim$$1$--$5$ gives $\tau_{\rm s}$$\sim$$10^8
$--$2.5$$\times$$10^9$ yr. Synchrotron loss timescales are then typically (much) 
longer than the typical SF timescale, $\tau_{\rm SF}$$\sim$$10^8$ yr. This means 
that the current radio emission of a local SFG may trace the SFR integrated over 
the last $\mincir$$10^9$ yr. As remarked in section 7, a similar consideration 
holds for the total 2-10 keV emission, $L_x$. 
\smallskip 

\noindent
{\it (b)}
During a strong starburst phase we expect the prompt radio emission to be 
dramatically enhanced due to the more intense particle accleration (and 
perhaps also increased magnetic field, e.g. Hirashita \& Hunt 2006), with 
a corresponding decrease in the synchrotron loss time. Clearly, the FIR 
energy density is much higher then (perhaps as high as $U_{\rm FIR}$$\
sim$$10^{-8}$ erg cm$^{-3}$ in the starburst region; Condon et al. 1991b). 
Relativistic electron Compton energy losses are then important (with very 
short loss timescales, $\tau_{\rm IC}$$=$$( {4\over 3}\, {\sigma_{\rm 
T}\over m_e c}\, \gamma_e \,{U_{\rm FIR} \over 10^{-8} {erg\,cm}^{-3}})^
{-1}$$\mincir$$5$$\times$$10^4$ yr, for the same parameters as in the 
previous paragraph). [Of course, Compton losses of synchro-emitting 
electrons off CMB photons increase with redshift as $(1+z)^4$.] These 
three factors (enhanced fresh radio emission and shorter synchrotron- 
and Compton-loss timescales) concur in making the observed radio emission 
of strong starbursts an accurate measure of their instantaneous SFR.

The varying ability of synchrotron radio emission to trace the 
instantaneous SFR, as a function of the different physical conditions 
prevailing in different galaxies, may play some role in the discussion 
on whether the value of $q_{\rm FIR}$ is 'universal' and on the real 
meaning and calibration of the radio SFR indicator (see section 8). 

In either case, the galactic radio and (total) X-ray luminosities should 
measure, at any given phase of a galaxy's SF history, essentially the same 
(definition of) SFR (i.e., instantaneous in strong starbursts, and integrated 
over the past $\mincir$$10^9$ yr in more quiescent disks). We thus expect a 
linear $L_{\rm 1.4\,GHz}$--$L_x$ correlation. 

\begin{figure}

\vspace{4.0cm}
\includegraphics{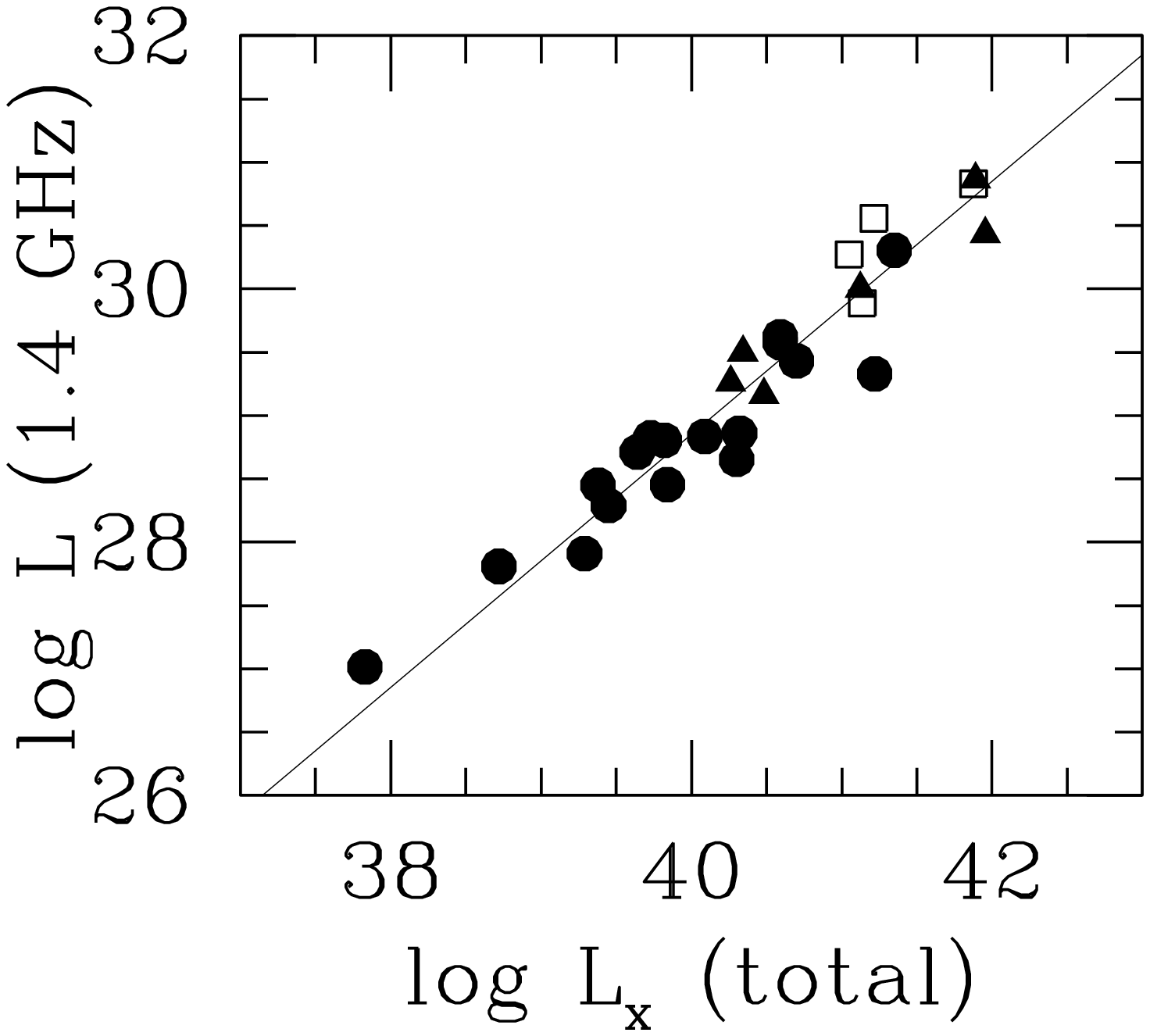}
\includegraphics{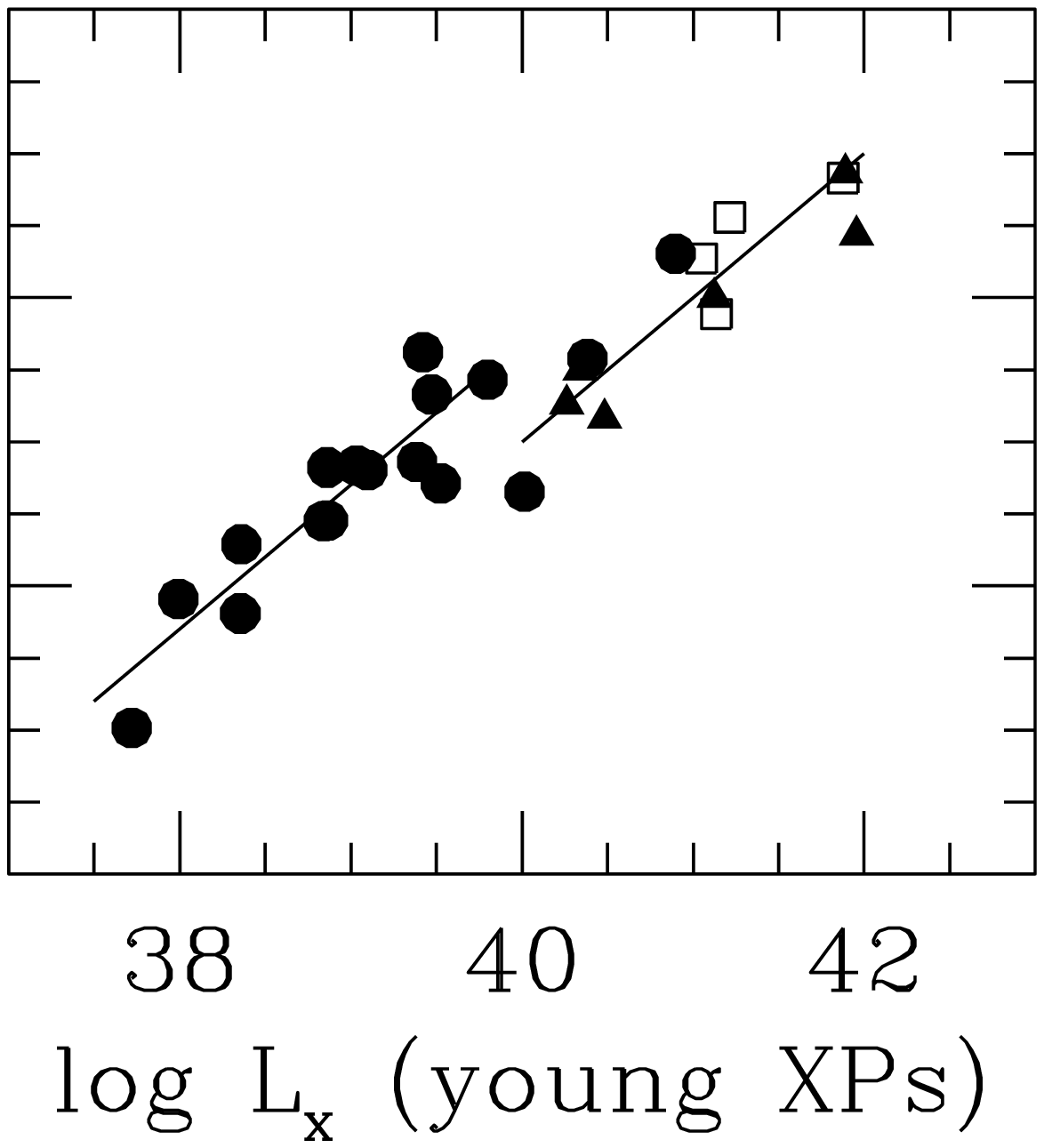}
\caption{ 
1.4 GHz luminosity densities are plotted versus 2-10 keV luminosities, 
using total luminosities ({\it left}) and collective young-XP luminosities 
({\it right}), for the combined samples of local SFGs (filled circles), 
nearby ULIRGs (empty squares), and distant HDFNGs (filled triangles). The 
lines are solely meant to guide the eye. 
}
\end{figure}

The integrated past SFR may in principle be unrelated to the current SFR, traced 
by $L_x^{\rm yXP}$. In our local SFGs galaxies, however, if the SF activity has 
remained essentially constant over the past several $\sim$$10^8$ yr (as argued 
above), the past integral SFR traces (to a multiplicative factor) the current SFR. 
This makes the integrated vs instantaneous SFR correlation possible, in the form 
of a linear correlation in the $L_{\rm 1.4~GHz}$$,$$L_x^{\rm yXP}$ plane 
(Fig.8-{\it right}, filled circles). For strong starbursts a linear correlation 
exists, too, but it is displaced from the SFG one given the different definition 
of the SFR being measured by the radio emission in the two cases (instantaneous 
for starbursts, integrated for SFGs). Therefore the linear correlation between 
young-XP emission and radio emission would exist separately for mildly star-forming 
galaxies and for extreme starbursts. Its piecewise linearity, as well as the 
meaning of measuring the current versus integrated SFR, makes $L_x^{\rm 
yXP}$--$L_{\rm 1.4~GHz}$ a companion relation to SFR--$L_{\rm x}$.

\section{Summary and conclusion}

In this paper we examined the issue of the collective 2-10 keV emission of 
young XPs and its role as a SFR indicator in star-forming galaxies. For 
a sample of local star-forming galaxies with available data, we estimated 
the young-XP luminosity by modelling the observed XPLFs in terms of 
'universal' young- and old-XPLFs. 

For SFR$\mincir$$50$$\,M_\odot$yr$^{-1}$ galaxies, the collective emission 
of young XPs turns out to correlate linearly with the FIR-based SFR. In this 
SFR range both the SFR--$L_{x}^{\rm yXP}$ relation, presented here, and the 
SFR--$L_{x}$ relation, proposed by Ranalli et al. (2003), are valid.

The relation is extended to higher SFRs by using a sample of extremely 
starburst-dominated ULIRGs. Since their $L_x$ is arguably dominated by young 
XPs and hence by the instantaneous SFR, when $L_{x}^{\rm yXP}$$=$$L_{x}$ 
these ULIRGs comply with the SFR--$L_{x}^{\rm yXP}$ relation. Similar 
considerations may hold for a sample of $z$$\sim$$1$, intensely star-forming 
{\it Hubble} Deep Field North galaxies -- especially so if their radio SFR 
indicator is calibrated slightly higher than for local galaxies, as suggested 
by the FIR-radio relation for starburst-dominated ULIRGs.

Overall, the SFR--$L_{x}^{\rm yXP}$ relation is suggested to hold over the 
broad SFR range $0.01$$\mincir$SFR/($M_\odot$yr$^{-1}$)$\mincir$$1000$, 
according to:
$$
{\rm SFR}(>0.1\, M_\odot) ~=~ 
{L_x^{\rm yXP} \over (0.75 \pm 0.15) \times 10^{39} {\rm erg \,s}^{-1}} ~
M_\odot {\rm yr}^{-1}\,.
$$

For galaxies of very different SFRs, the same SFR--X-ray-relation holds only 
when the young-XP emission, {\it not} the total luminosity, is used. This is so 
because the FIR emission and the young-XP emission both trace the instantaneous 
SFR, whereas the total X-ray luminosity traces the instantaneous SFR in extreme 
starburst galaxies and the integrated SFR in more quiescent ones. Consequently, 
a SFR--$L_{\rm x}$ relation exists separately for very-low and very-high SFR 
galaxies (e.g., our local SFGs and distant ULIRGs, respectively). In particular, 
the existence of the SFR--$L_{\rm x}$ relation for our local, low-SFR galaxies  
suggests that SF has not changed dramatically over the last $\sim$$10^9$ yr.

The SFR--$L_{x}^{\rm yXP}$ relation represents the most adequate X-ray estimator 
of the instantaneous SFR in galaxies. Its reflects the equivalence of two 
complementary measures of the current SFR, based on the observed manifestations 
of massive stars at their birth (short-lived FIR emission from placental dust 
clouds) and near their death as compact remnants (short-lived X-ray emission from 
close binary accretors), respectively.
\medskip

\noindent
{\it Acknowledgement.} 
This research has made use of the NASA/IPAC Extragalactic Database (NED), which 
is operated by the Jet Propulsion Lab, CalTech, under contract with NASA. We 
acknowledge constructive remarks from an anonymous referee and useful exchanges 
with several colleagues, including Ed Colbert, Pepi Fabbiano, Jimmy Irwin, Jean 
Swank, Rowan Temple, and Min Yun. This paper is fondly dedicated to the memory of 
MP's father, who died shortly before the inception of the work reported here.
\bigskip

\def\ref{\par\noindent\hangindent 20pt}

\noindent
{\bf References}
\vglue 0.2truecm

\ref{\small Appleton, P.N., Fadda, D.T., Marleau, F.R., et al. 2004, ApJS, 154, 147}
\ref{\small Ballo, L., Braito, V., Della Ceca, R., et al. 2004, ApJ, 600, 634}
\ref{\small Bauer, F.E., Alexander, D.M., Brandt, W.N., et al. 2002, AJ, 124, 2351}
\ref{\small Bell, E. 2003, ApJ, 586, 794}
\ref{\small Burstein, D., \& Heiles, C. 1982, AJ, 87, 1165}
\ref{\small Calzetti, D., Armus, L., Bohlin, R.C., et al. 2000, ApJ, 533, 682}
\ref{\small Cohen, J.G. 2003, ApJ, 598, 288}
\ref{\small Colbert, E.J.M., Heckman, T.M., Ptak, A.F., et al. 2004, ApJ, 602, 231}
\ref{\small Condon, J.J., Helou, G., Sanders, D.B., \& Soifer, B.T. 1996, ApJS, 103, 81}
\ref{\small Condon, J.J., Cotton, W.D., Greisen, E.W., et al. 1998, AJ, 115, 1693}
\ref{\small Condon, J.J., Anderson, M.L., \& Helou, G. 1991a, ApJ, 376, 95}
\ref{\small Condon, J.J., Huang, Z.-P., Yin, Q.F., \& Thuan, T.X. 1991b, ApJ, 378, 65}
\ref{\small Condon, J.J. 1992, ARA\&A, 30, 575}
\ref{\small Condon, J.J., Cotton, W.D., \& Broderick, J.J. 2002, AJ, 124, 675}
\ref{\small Dahlem, M., Parmar, A., Oosterbroek, T., et al. 2000, ApJ, 538, 555}
\ref{\small Dahlem, M., Weaver, K.A., \& Heckman, T.M. 1998, ApJS, 118, 401}
\ref{\small Dale, D.A., Helou, G., Contursi, A., et al. 2001, ApJ, 549, 215}
\ref{\small David, L.P., Jones, C., \& Forman, W. 1992, ApJ, 338, 82}
\ref{\small de Vaucouleurs, G., de Vaucouleurs, A., Corwin, H.G. Jr., et al. 1991,
        The Third Reference Catalogue of Bright Galaxies, Univ. Texas Press, Austin (RC3)}
\ref{\small Della Ceca, R., Ballo, L., Tavecchio, F., et al. 2002, ApJ, 581, L9}
\ref{\small Devereux, N.A., \& Eales, S.A. 1989, ApJ, 340, 708}
\ref{\small Doane, J.S., \& Mathews, W.G. 1993, ApJ, 419, 573}
\ref{\small Fabbiano, G. 2005, ARA\&A, submitted (astro-ph/05111481}
\ref{\small Fabbiano, G., \& Trinchieri, G. 1985, ApJ, 296, 430}
\ref{\small Franceschini, A., Braito, V., Persic, M., et al. 2003, MNRAS, 343, 1181}
\ref{\small Garrett, M.A. 2002, A\&A, 384, L19}
\ref{\small Genzel, R., Lutz, D., \& Sturn, E., et al. 1998, ApJ, 498, 579}
\ref{\small Gilfanov, M. 2004, MNRAS, 349, 146}
\ref{\small Gilfanov, M., Grimm, H.-J., \& Sunyaev, R. 2004a, MNRAS, 347, L57}
\ref{\small Gilfanov, M., Grimm, H.-J., \& Sunyaev, R. 2004b, MNRAS, 351, 1365}
\ref{\small Griffiths, R.E., Ptak, A., Feigelson, E.D., et al. 2000, Science, 250, 1325}
\ref{\small Grimm, H.-J., Gilfanov, M., \& Sunyaev, R. 2003, MNRAS, 339, 793}
\ref{\small Guainazzi, M., Matt, G., Brandt, W.N., et al. 2000, A\&A, 356, 463}
\ref{\small Hartwell, J.M., Stevens, I.R., Strickland, D.K., et al. 2004, MNRAS, 348, 406}
\ref{\small Helou, G., Khan, I.R., Malek, L, \& Boehmer, L. 1988, ApJS, 68, 151}
\ref{\small Helou, G., Soifer, B.T., \& Rowan-Robinson, M. 1985, ApJ, 298, L7}
\ref{\small Heeschen, D.S., \& Wade, C.M. 1964, AJ, 69, 277}
\ref{\small Hirashita, H., \& Hunt, L.K. 2006, A\&A in press (astro-ph/0609733)}
\ref{\small Holt, S.S., Schlegel, E.M., Hwang, U., \& Petre, R. 2003, ApJ, 588, 792}
\ref{\small Hopkins, A.M., Miller, C.J., Nichol, R.C., et al. 2003, ApJ, 599, 971}
\ref{\small Hunter, D.A., Gillett, V.C., Gallagher, III J.S., et al. 1986, ApJ, 303, 171}
\ref{\small Inui, T., Matsumoto, H., Tsuru, T.G., et al. 2005, PASJ, 57, 135}
\ref{\small Iwasawa, K., Matt, G., Guainazzi, M., \& Fabian, A.C. 2001, MNRAS, 326, 894}
\ref{\small Jenkins, L.P., Roberts, T.P., Ward, M.J., \& Zezas, A. 2004, MNRAS, 352, 1335}
\ref{\small Kennicutt, R.C.Jr. 1998a, ApJ, 498, 541}
\ref{\small Kennicutt, R.C.Jr. 1998b, ARAA, 36, 189}
\ref{\small Kewley, L.J., Geller, M.J., Jansen, R.A., \& Dopita, M.A. 2002, AJ, 124, 3135}
\ref{\small Kilgard, R.E., Kaaret, P., Krauss, M.I., et al. 2002, ApJ, 573, 138}
\ref{\small Kim, D.-W., \& Fabbiano, G. 2004, ApJ, 611, 846}
\ref{\small Komossa, S., Burwitz, V., Hasinger, G., et al. 2003, ApJ, 582, L15}
\ref{\small Kong, A.K.H. 2003, MNRAS, 346, 265}
\ref{\small K\"uhr, H., Witzel, A., Pauliny-Toth, I.I.K., \& Nauber, U. 1981, A\&AS, 45, 367}
\ref{\small Lira, P., Ward, M., Zezas, A., et al. 2002, MNRAS, 330, 259}
\ref{\small Liu, J.-F., Bregman, J.N., \& Seitzer, P. 2002, ApJ, 580, L31}
\ref{\small Ott, J., Martin, C.L., \& Walter, F. 2003, ApJ, 594, 776}
\ref{\small Peacock, J.A. 1999, Cosmological Physics (Cambridge: Cambridge University Press)}
\ref{\small Persic, M., \& Rephaeli, Y. 2002, A\&A, 382, 843}
\ref{\small Persic, M., Rephaeli, Y., Braito, V., et al. 2004a, A\&A, 419, 849}
\ref{\small Persic, M., Cappi, M., Rephaeli, Y., et al. 2004b, A\&A, 427, 35}
\ref{\small Ranalli, P., Comastri, A., \& Setti, G. 2003, A\&A, 399, 39}
\ref{\small Rephaeli, Y., Gruber, D., Persic, M., \& McDonald, D. 1991, ApJ, 380, L59}
\ref{\small Rephaeli, Y., Gruber, D., \& Persic, M. 1995, A\&A, 300, 91}
\ref{\small Rieke, G.H., Loken, K., Rieke, M.J., \& Tamblyn, P. 1993, ApJ, 412, 99}
\ref{\small Salpeter, E.E. 1955, ApJ, 121, 161}
\ref{\small Sanders, D.B., Egami, E., Lipari, S., et al. 1995, AJ, 110, 1993 }
\ref{\small Sanders, D.B., Mazzarella, J.M., Kim, D.-C., et al. 2003, AJ, 126, 1607}
\ref{\small Sanders, D.B., Scoville, N.Z., \& Soifer, B.T. 1991, ApJ, 370, 158}
\ref{\small Sarazin, C.L., Irwin, J.A., \& Bregman, J.N. 2000, ApJ, 544, L101}
\ref{\small Schlegel, D., Finkbeirer, D.P., \& Davis M. 1998, ApJ, 500, 525}
\ref{\small Schlegel, E.M., \& Pannuti, T.G. 2003, AJ, 125, 3025}
\ref{\small Schmitt, H.R., Calzetti, D., Armus, L., et al. 2006, ApJ, 643, 173}
\ref{\small Soria, R., \& Wu, K. 2003, A\&A, 410, 53}
\ref{\small Stanford, S.A., Stern, D., van Breugel, W., \& De Breuck, C. 2000, ApJS, 131, 185}
\ref{\small Summers, L.K., Stevens, I.R., Strickland, D.K., \& Heckman, T.M. 2004, MNRAS, 351, 1}
\ref{\small Swartz, D.A., Ghosh, K.K., Tennant, A.F., \& Wu, K. 2004, ApJS, 154, 519}
\ref{\small Temple, R., Raychaudhury, R., \& Stevens, I. 2005, MNRAS, 362, 581}
\ref{\small Tennant, A.F., Wu, K., Ghosh, K.K., et al. 2001, ApJ, 549, L43}
\ref{\small Tsuru, T.G., Awaki, H., Koyama, K., \& Ptak, A. 1997, PASJ, 49, 619}
\ref{\small Tully R.B. 1988, Nearby Galaxies Catalog, Cambridge Univ. Press (Cambridge)}
\ref{\small Vega, O., Silva, L., Panuzzo, P., et al. 2005, MNRAS, 364, 1286}
\ref{\small Voss, R., \& Gilfanov, M. 2005, A\&A, in press (astro-ph/0505250}
\ref{\small White, R.L., \& Becker, R.H. 1992, ApJS, 79, 331}
\ref{\small Wright, A., \& Otrupcek, R. 1990, Parkes Catalogue, Australia Telescope National Facility}
\ref{\small Yun, M.S., \& Carilli, C.L. 2002, ApJ, 568, 88}
\ref{\small Yun, M.S., Reddy, N.S., \& Condon, J.J. 2001, ApJ, 554, 803}
\ref{\small Zezas, A., \& Fabbiano, G., 2002, ApJ, 577, 726}
\ref{\small Zezas, A., Fabbiano, G., Rots, A.H., \& Murray, S.S. 2002, ApJ, 577, 710}
\ref{\small Zezas, A., Ward, M.J., \& Murray, S.S. 2003, ApJ, 594, L31}

\end{document}